\documentclass[aps,prb,twocolumn,showpacs,eqsecnum]{revtex4-1}
\usepackage{amssymb}

\usepackage{graphicx}
\usepackage{amsmath}
\usepackage{verbatim}

\renewcommand{\bm}{\mathbf}

\newcommand{\mc}{\mathcal}

\begin{document}

\title{Spin susceptibility of interacting two-dimensional electrons\\
in the presence of spin-orbit coupling}
\author{Robert Andrzej \.Zak$^{1}$, Dmitrii L. Maslov$^{2}$, and Daniel Loss$^{1}$}

\begin{abstract}
A long-range interaction via virtual particle-hole pairs between Fermi-liquid quasiparticles leads to a nonanalytic behavior of the spin susceptibility $\chi$ as a function of the temperature ($T$), magnetic field ($\mathbf{B}$), and wavenumber. In this paper, we study the effect of the Rashba spin-orbit interaction (SOI) on the nonanalytic behavior of $\chi$ for a two-dimensional electron liquid. Although the SOI breaks the $SU(2)$ symmetry, it does not eliminate nonanalyticity but rather makes it anisotropic: while the linear scaling of $\chi_{zz}$ with $T$ and $|\mathbf{B}|$ saturates at the energy scale set by the SOI, that of $\chi_{xx}$ ($=\chi_{yy}$) continues through this energy scale, until renormalization of the electron-electron interaction in the Cooper channel becomes important.
We show that the Renormalization Group flow in the Cooper channel has a non-trivial fixed point, and study the consequences of this fixed point for the nonanalytic behavior of $\chi$.
An immediate consequence of SOI-induced anisotropy in the nonanalytic behavior of $\chi$ is a possible instability of a second-order ferromagnetic quantum phase transition with respect to a first-order transition to an $XY$ ferromagnetic state.
\end{abstract}

\affiliation{$^1$Department of Physics, University of Basel, Klingelbergstrasse
82, CH-4056 Basel, Switzerland}

\affiliation{$^2$Department of Physics, University of Florida, P. O. Box 118440,
Gainesville, FL 32611-8440}
\today


\maketitle

\section{\label{sec:Int}INTRODUCTION}

The issue of nonanalytic corrections to the Fermi liquid theory has
been studied extensively in recent years. \cite{belitz_RMP,woelfle_RMP}
The interest to this subject is stimulated by a~variety of topics, from
intrinsic instabilities of ferromagnetic quantum phase transitions\cite{belitz_RMP,woelfle_RMP,pepin06,maslov06_09,green09}
to enhancement of the
indirect exchange interaction
between nuclear spins in semiconductor heterostructures with potential
applications in quantum computing.\cite{PhysRevB.77.045108,PhysRevLett.98.156401,PhysRevB.79.115445}
The origin of the nonanalytic behavior can be traced to an effective long-range
interaction of fermions via virtual particle-hole pairs with small energies
and with momenta which are either small (compared to the Fermi momentum $%
k_{F}$) or near $2k_{F}$.~\cite{chubukov06,chubukov05a}
In 2D, this interaction leads to a linear scaling of $\chi $ with a characteristic energy scale
$E$, set  by either the temperature $T$ or the magnetic field $\left| \mathbf{B}\right| $, or else by the wavenumber of an
inhomogeneous magnetic field $\left| \mathbf{q}\right| $
 (whichever is larger when
measured in appropriate units).
\cite{PhysRevB.68.155113,PhysRevB.69.121102,hirashima98,JETPLett.58.709,PhysRevLett.86.5337,PhysRevB.64.054414,PhysRevB.72.115112,PhysRevB.77.220401(R)}
Higher-order scattering processes in the Cooper (particle-particle) channel result in
additional logarithmic renormalization of the result: at the lowest
energies, $\chi \propto E/\ln ^{2}E.$ \cite{aleiner06,PhysRevB.74.205122,ProcNatlAcadSci.103.15765,
schwiete06,maslov06_09,PhysRevB.77.045108,PhysRevLett.98.156401} The sign of the effect,
i.e., whether $\chi $ increases or decreases with $E$,
turns out to be non-universal, at least in a generic Fermi liquid regime,
i.e., away from the ferromagnetic instability: while the second-order perturbation theory predicts that
$\chi $ increases with $E,$ the sign of the effect may be reversed either due to a
proximity to the Kohn-Luttinger superconducting instability\cite{PhysRevB.74.205122,PhysRevB.77.045108,PhysRevLett.98.156401}
or higher-order processes in the particle-hole channel.\cite{ProcNatlAcadSci.103.15765,maslov06_09}

In this paper, we explore the effect of the spin-orbit interaction (SOI) on
the nonanalytic behavior of the spin susceptibility. The SOI is
important for practically all systems of current interest at low enough
energies; at the same time,  the nonanalytic behavior is also an inherently low-energy
phenomenon. A natural question to ask is: what is the interplay between these
two low-energy effects? At first sight, the SOI should regularize
nonanalyticities at energy scales below the scale set by this coupling. (For
a Rashba-type SOI, the relevant scale is given by the product $%
|\alpha| k_{F}$, where $\alpha $ is the coupling constant of the Rashba
Hamiltonian; here and in the rest of
the paper, we set $\hbar $ and $k_{B}$ to unity). Indeed, as we have
already mentioned, the origin of the nonanalyticity is
the long-range effective interaction
originating from the singularities in the particle-hole polarization
bubble. If, for instance, the temperature is the largest scale in the
problem, these singularities are smeared
by the temperature with an ensuing
nonanalytic dependence of $\chi $ on $T$. On the other hand, if the Zeeman energy $\mu _{B}|%
\mathbf{B}|$ is larger than $T$, it provides a more efficient mechanism
of regularization of the singularities and, as result, $\chi $ exhibits a
nonanalytic dependence on $|\mathbf{B}|$. The same argument applies also to the $%
|\mathbf{q}|$ dependence. It is often said that the SOI plays the
role of an effective magnetic field, which acts on electron spins. If so, one
should expect, for instance, a duality between the $T$ and $|\alpha| k_{F}$ scalings of $\chi$
(by analogy to a duality between $T$ and $\mu_B|\mathbf{B}|$ scalings of $\chi$),
i.e., in a system with fixed SOI,
the nonanalytic $T$ dependence of $\chi $ should saturate at $T\sim |\alpha| k_{F}$.
The main message of this paper is that such a duality
does not, in fact, exist; more precisely, not all components of the
susceptibility tensor exhibit the duality. In particular, the in-plane
component of $\chi $, $\chi _{xx}$, continues to scale linearly with $T$ and $\mu_B|\mathbf{B}|$, even if these energies
are smaller than $|\alpha| k_{F}$. On the other hand, the $T$ and $\mu_B|\mathbf{B}|$ dependences
of $\chi _{zz}$ do saturate at $T\sim |\alpha|k_{F}$.

\begin{figure}[t]
\includegraphics[width=.48\textwidth]{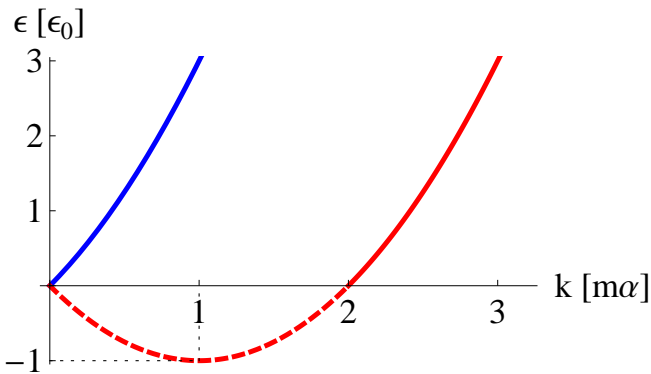}
\caption{(COLOR ONLINE) Rashba spectrum in zero magnetic field. The energy is measured in units of $\epsilon_0\equiv m\alpha^2/2$, the momentum is measured in units of $m\alpha$.}
\label{fig:Spectrum}
\end{figure}

\begin{figure}[t]
\includegraphics[width=.24\textwidth]{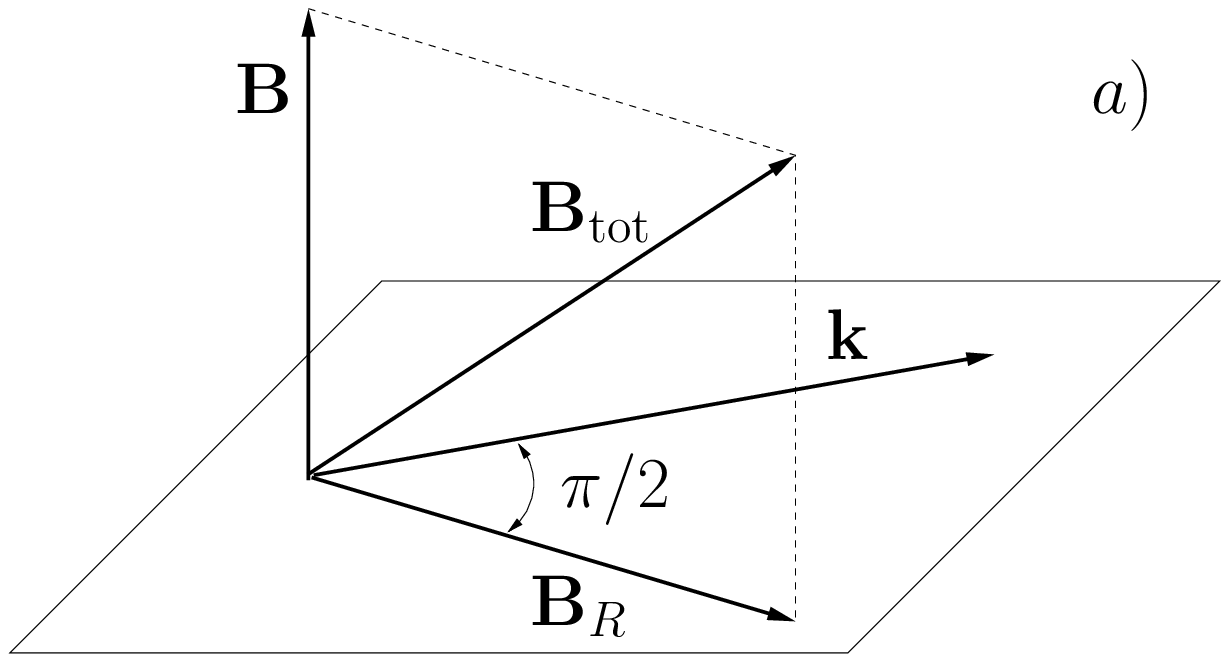}
\includegraphics[width=.16\textwidth]{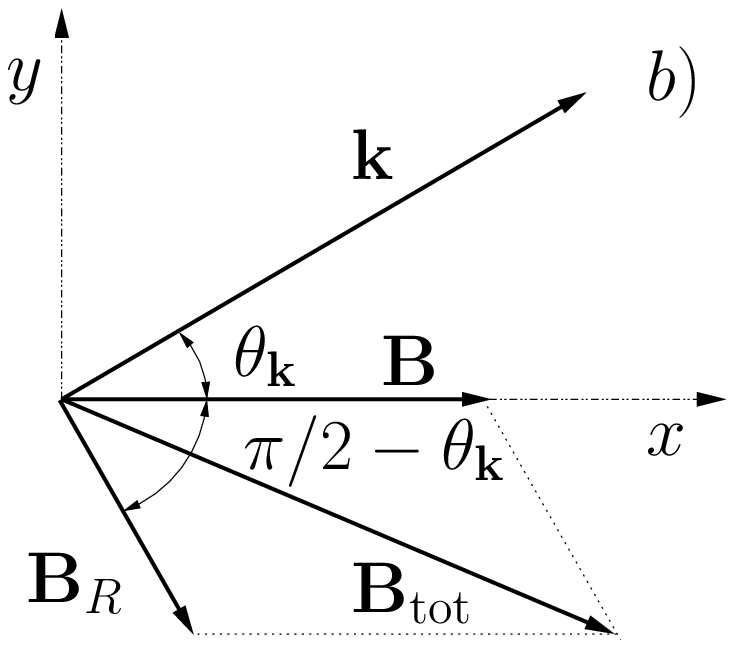}
\caption{Interplay between the external magnetic field $\mathbf{B}$ and
effective magnetic field $\mathbf{B}_{R}$ due to the Rashba spin-orbit
interaction. Left: the external field is perpendicular to the plane of
motion. Right: the external field is in the plane of motion.}
\label{fig:2DEG}
\end{figure}

The reason for such a behavior is that although the SOI does play
a role of the effective magnetic field, this field depends on the
electron momentum. To understand the importance of this fact, we consider
the Rashba Hamiltonian in the presence of an external magnetic field,\cite{bychkov84}
which couples only to the spins of the electrons but not to their orbital degrees of freedom
\begin{eqnarray}\label{eq:Ham}
H_{R} &=&\frac{k^{2}}{2m}{\hat{I}}+\alpha (\boldsymbol\sigma \times \mathbf{k%
})_{z}+\frac{g\mu _{B}\boldsymbol\sigma}{2} \cdot \mathbf{B}  \notag \\
&=&\frac{k^{2}}{2m}{\hat{I}}+\frac{g\mu _{B} \boldsymbol{\sigma}}{2}\cdot\left[
\mathbf{B}_{R}(\mathbf{k})+\mathbf{B}\right],
\end{eqnarray}
where $\alpha $ is the SOI, $\mathbf{k}$ is the electron momentum in
the plane of a two-dimensional electron gas (2DEG), $\protect\boldsymbol{\sigma}$ is
a vector of Pauli matrices, $\mathbf{e}_{z}$ is a normal to the plane, $g$
is the gyromagnetic ratio, $\mu _{B}$ is the Bohr magneton, and the
effective Rashba field, defined as $\mathbf{B}_{R}=(2\alpha /g\mu _{B})(%
\mathbf{k}\times \mathbf{e}_{z})$, is always in the 2DEG plane. The
effective Zeeman energy is determined by the total magnetic field $\mathbf{B}%
_{\text{tot}}=\mathbf{B}_{R} +\mathbf{B}$ as
\begin{equation}
\bar{\Delta}_{\mathbf{k}}\equiv \frac{g\mu _{B}\left| \mathbf{B}_{\text{tot}}\right|}{2} =\sqrt{\alpha ^{2}k^{2}+2\alpha \left( \boldsymbol{\Delta}\times
\mathbf{k}\right) _{z}+\Delta ^{2}},  \label{zeeman}
\end{equation}
where we introduced $\boldsymbol{\Delta}=g\mu _{B}\mathbf{B}/2$ for the
''Zeeman field'', such that $\Delta$ is the Zeeman energy of an electron spin in the external magnetic field ($\hbar=1$) and $2\Delta $ equals to the Zeeman splitting between spin-up and spin-down states. A combined effect of the Rashba and external magnetic fields gives rise to two branches of the electron spectrum (see Fig.~\ref{fig:Spectrum}) with dispersions
\begin{equation}
\varepsilon _{\mathbf{k}}^{\pm }=\frac{k^{2}}{2m}\pm \bar{\Delta}_{\mathbf{k}%
}.  \label{rsp}
\end{equation}
If $\mathbf{B}\parallel \mathbf{e}_{z}$, the external and effective
magnetic fields are perpendicular to each other, as shown in Fig. \ref{fig:2DEG}a,
so that the magnitude of the total magnetic field is $|\mathbf{B}_{\mathrm{%
tot}}|=|\mathbf{B}+\mathbf{B}_{R}|=\sqrt{B^{2}+B_{R}^{2}}$. This means that
the external and magnetic fields are totally interchangeable, and the $T$
dependence of the spin susceptibility is cut off by the largest of the two scales.
However, if the external field is in the plane (and defines the $x$ axis in
Fig. \ref{fig:2DEG}b), the magnitude of the total field depends on
the angle $\theta _{\mathbf{k}}$ between $\mathbf{k}$ and $\mathbf{B.}$ In
particular, for a weak external field,
\begin{equation}\label{eq:zeemaneff}
\bar{\Delta}_{\mathbf{k}}\approx
\left| \alpha \right| k+\Delta \sin \theta _{\mathbf{k}}.
\end{equation}
If the electron-electron interaction is weak, the nonanalytic behavior of the spin susceptibility
is due to particle-hole pairs with total momentum near $2k_F$, formed by
electron and holes moving in almost opposite directions. In this case,
the second term in Eq.~(\ref{eq:zeemaneff}) is of the opposite sign for electrons and holes.
The effective Zeeman energy of the whole pair, formed by fermions from Rashba branches $s$ and $s^{\prime},$
is
\begin{equation}
\Delta _{\text{pair}}=s\bar{\Delta}_{\mathbf{k}}-s^{\prime }\bar{\Delta}_{-%
\mathbf{k}}=\left( s-s^{\prime }\right) \left| \alpha \right| k+(s+s^{\prime
})\Delta \sin \theta _{\mathbf{k}}.  \label{zpair}
\end{equation}
Only those pairs which ``know'' about the external magnetic field
--via the second term in Eq.~(\ref{zpair})--
renormalize the spin susceptibility. According to Eq. (\ref{zpair}), such
pairs are formed by fermions from the same Rashba branch ($s=s^{\prime }$).
However, since the first term in Eq. (\ref{zpair}) vanishes in this case, such
pairs do not \lq\lq know \rq\rq\/ about the SOI, which means that the Rashba and
external magnetic fields are not interchangeable, and the SOI energy scale
does not provide a cutoff for the $T$ dependence of $\chi .$

We now briefly summarize the main results of the paper. We limit our
consideration to the 2D case and to the Rashba SO
coupling, present in any 2D system
with broken symmetry with respect to reversal of the normal to the plane. We focus on the
dependencies of $\chi $ on $T,$ $\alpha ,$ and $|B|,$ deferring a detailed discussion of the
dependence of $\chi $ on $\left| \mathbf{q}\right| $ to another occasion.
\cite{unpub:zak} Throughout the paper, we assume that the SO coupling and
electron-electron interaction, characterized by the coupling constant $U,$ are
weak, i.e., $\left| \alpha \right| k_{F}\ll E_{F}$ and $mU\ll 1.$
The latter condition implies that only $2k_F$ scattering processes are relevant for the nonanalytic behavior of the spin susceptibility. Accordingly, $U$ is the $2k_F$ Fourier transform of the interaction potential.
Renormalization provides another small energy scale at which the
product $(mU/2\pi )\ln \left( \Lambda /T_{C}\right) $ (where $\Lambda $ is
an ultraviolet cutoff of the theory) becomes comparable to unity,
i.e, $T_{C}\equiv \Lambda \exp \left( -2\pi /mU\right).$ Depending on the
ratio of the two small energy scales--$\left| \alpha \right| k_{F}$ and $T_{C}$--different behaviors are
possible.

\begin{figure}[t]
\includegraphics[width=.23\textwidth]{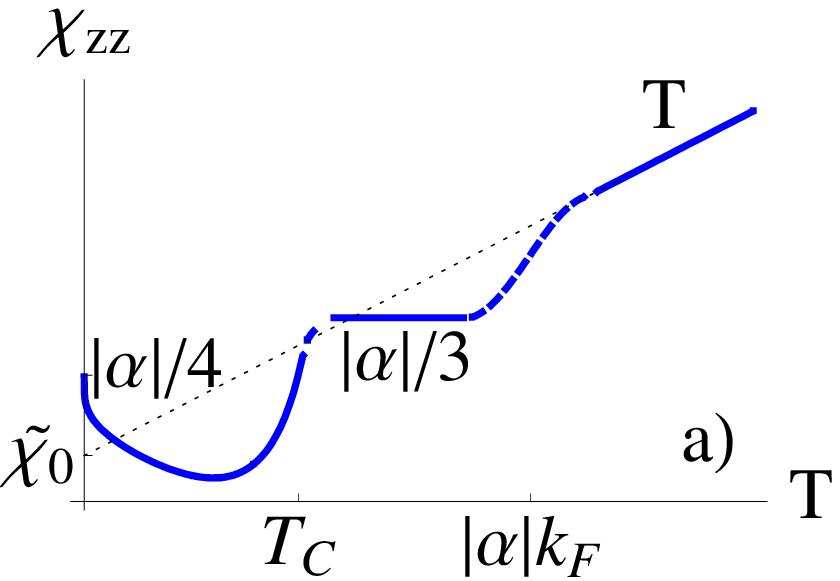}
\includegraphics[width=.23\textwidth]{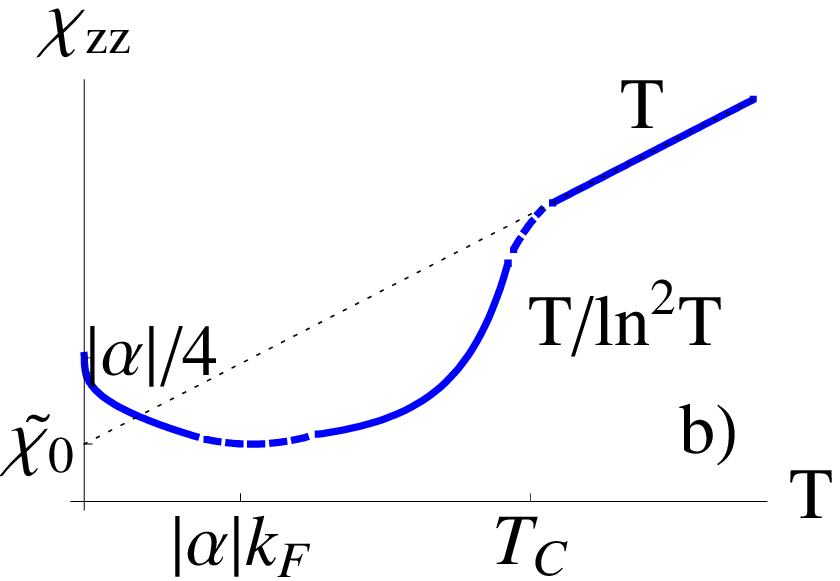}\\
\vspace*{2em}
\includegraphics[width=.23\textwidth]{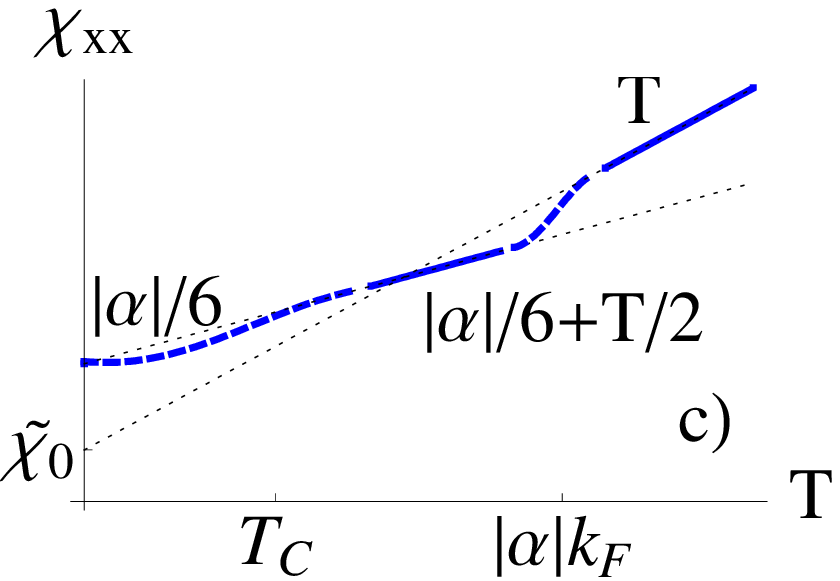}
\includegraphics[width=.23\textwidth]{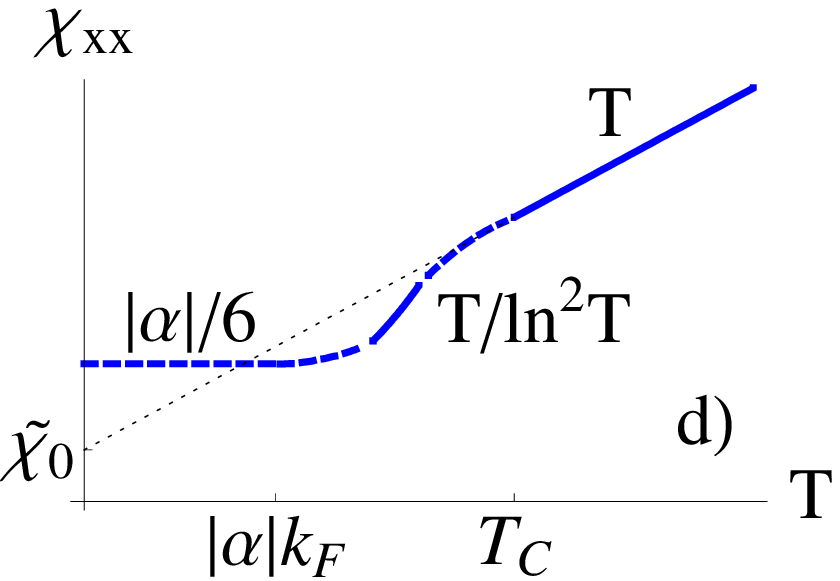}
\caption{
(COLOR ONLINE) A sketch of the temperature dependence of the spin susceptibility
for the transverse (top) and in-plane (bottom) magnetic field. Dashed segments in crossovers
between various asymptotic regimes do not represent results of actual
calculations. The left (right) panel is valid for
$T_C\ll|\alpha|k_F$ ($T_C\gg|\alpha|k_F$) where
$T_{C}\equiv \Lambda \exp \left( -2\pi /mU\right)$ is the temperature below which renormalization in the Cooper channel becomes significant. The zero temperature limit for interacting
electrons, denoted by $ \tilde\chi_0$, is given by Eq.~(\ref{stoner2}) in the Random Phase Approximation.}
\label{fig:Chi}
\end{figure}

In Fig.~\ref{fig:Chi}a, we sketch the $T$ dependence of $\chi _{zz}$ for the case of $%
T_{C}\ll \left| \alpha \right| k_{F}.$ For $T\gg \left| \alpha \right|
k_{F}, $ $\chi _{zz}$ scales linearly with $T,$ in agreement with previous
studies; a correction due to the SOI is on the order of $\left(
\alpha k_{F}/T\right) ^{2}$ [cf. Eq. (\ref{eq:Delta<<T})]. For $T_{C}\ll
T\ll \left| \alpha \right| k_{F},$ $\chi _{zz}$ saturates at a value
proportional to $\left| \alpha \right| ;$ the correction due to finite $T$ is
on the order of $\left( T/\left| \alpha \right| k_{F}\right) ^{3}$ [cf. Eq. (%
\ref{eq:Delta>>T})]. For $T\lesssim T_{C}$, renormalization in the Cooper
channel becomes important.
In the absence of the SOI, the coupling constant of the electron-electron interaction in the Cooper channel
flows to zero as $U/|\ln T|$. Consequently, the $T$ scaling of the spin susceptibility changes to $T/\ln ^{2}T$ for $T\ll T_C$.
In the presence of the SOI, the situation is different.
We show that the Renormalization Group (RG) flow of $U$ in this case has a non-trivial fixed point characterized
by finite value of the electron-electron coupling, which is only numerically smaller than its bare value. In between these two limits, the coupling constant
changes non-monotonically with $\ln T$, and so does $\chi_{zz}$.
In both high-and low $T$ limits (compared to $T_C$), however, $\chi_{zz}$ is almost $T$ independent, so Cooper renormalization affects
the $|\alpha|$ term in $\chi_{zz}$.

The $T$ dependence of $\chi _{zz}$ for $T_{C}\gg \left| \alpha
\right| k_{F}$ is sketched in Fig.~\ref{fig:Chi}b. In this case, the crossover between
the $T$ and $T/\ln ^{2}T$ forms occurs first, at $T\sim T_{C}$, while the $%
T/\ln ^{2}T$ form crosses over to $\left| \alpha \right|$ at $%
T\sim \left| \alpha \right| k_{F}.$

We now turn to $\chi _{xx}.$ For $T_{C}\ll \left| \alpha \right| k_{F},$ its
$T$ dependence is shown in Fig.~\ref{fig:Chi}c. The high-$T$ behavior is again linear
[also with a $\left( \alpha k_{F}/T\right) ^{2}$ correction, cf. Eq. (\ref{eq:ChiXX2ndT})]. For $T_{C}\ll T\ll \left| \alpha \right| k_{F},$ this behavior changes to $\chi _{xx}$ $\propto \left| \alpha \right| k_{F}/6+T/2$, which means that $\chi _{xx}$ continues to decrease with $T$ with a slope half of
that at higher $T$ [cf. Eq. (\ref{eq:ChiXX2ndAlpha})]. Finally, for $%
T\lesssim T_{C},$ Cooper renormalization leads to the same $\left| \alpha
\right|$ dependence as for $\chi _{zz}$ but with a different prefactor. The behavior of $\chi
_{xx}$ for $\left| \alpha \right| k_{F}\ll T_{C}$ is shown in Fig.~\ref{fig:Chi}d.
Apart from the numbers, this behavior is similar to that of $\chi _{zz}$
in that case.

The dependences of $\chi_{xx}$ and $\chi_{zz}$ on the external magnetic field can be obtained
(up to a numerical coefficient) simply by replacing $T$ by $\Delta$ in all formulas presented above.
In particular, if Cooper renormalization can be ignored, both $\chi_{xx}$ and $\chi_{zz}$ scale linearly with $\Delta$ for $\Delta\gg |\alpha|k_F$
but only $\chi_{xx}$ continues to scale with $\Delta$ for $\Delta\ll |\alpha|k_F$.
A linear scaling of $\chi_{ii}$ with $\Delta$ implies the presence of a nonanalytic, $|M_i|^3$ term in the free energy,
where  $\mathbf{M}$ is the magnetization and $i=x,y,z$. Consequently, while the cubic term is isotropic ($F\propto |M|^3$) for larger $M$ (so that the corresponding Zeeman energy is above $|\alpha| k_F$), it is anisotropic at smaller $|M|$: $F\propto |M_x|^3+|M_y|^3$.
A negative cubic term in $F$ implies metamagnetism and an instability of the second-order ferromagnetic quantum phase transition toward a first-order
one.\cite{belitz_RMP,maslov06_09} An anisotropic cubic term implies anisotropic metamagnetism, i.e., a phase transition in a finite magnetic field, if it is applied along the plane of motion, but no transition for a perpendicular field, and also that the first-order transition is into an $XY$ rather than Heisenberg ferromagnetic state. This issue is discussed in more detail in Sec.~\ref{sec:Sum}.

The rest of the paper is organized as follows. In Sec.~\ref{sec:Ham}, we
formulate the problem and discuss the $T$ dependence of the spin
susceptibility for free electrons with Rashba spectrum (more details on this subject are
given in Appendix \ref{app:0thOrder}). Section~\ref{sec:general} explains
the general strategy of extracting the nonanalytic behavior of $\chi $ from
the thermodynamic potential in the presence of the SOI. The
second-order perturbation theory for the temperature and magnetic-field
dependences of the transverse and in-plane susceptibilities is presented in Secs. ~%
\ref{sec:SS2zz} and~\ref{sec:SS2xx}, respectively. In Sec. \ref{sec:Diags},
we show that, as is also the case in the absence of the SOI, there
is no contribution to the nonanalytic behavior of the spin susceptibility
from processes with small momentum transfers, including the transfers
commensurate with (small) Rashba splitting of the free-electron spectrum
(more details on this issue are provided in Appendix \ref{app:smallq}).
Renormalization of spin susceptibility in the Cooper channel is considered in Sec.~\ref
{sec:Inf}. An explicit calculation of the third-order Cooper contribution to
$\chi _{zz}$ is shown in Sec. \ref{sec:cooper3}. In  Sec.~\ref{sec:Renormalization} we derive the RG flow equations
for the scattering amplitudes in the absence of the magnetic field; the effect of the finite field on the RG flow is discussed in Appendix \ref{app:RG}.
The effect of Cooper-channel renormalization on the nonanalytic behavior of $\chi_{zz}$ and $\chi_{xx}$ is discussed in Secs.~\ref{sec:InfZZ} and \ref{sec:InfXX}, correspondingly.
Implications of our results in the context of quantum phase transitions are discussed in Ref.~\ref{sec:Sum}, where we also give our conclusions.

\section{\label{sec:Ham}SPIN\ SUSCEPTIBILITY\ OF\ FREE RASHBA FERMIONS}

In this Section, we set the notations and discuss briefly the properties of
Rashba electrons in the absence of the electron-electron interaction. The
Hamiltonian describing a two-dimensional electron gas (2DEG) in the presence
of a Rashba SOI and an external magnetic field $\mathbf{B}$ is given
by Eq.~(\ref{eq:Ham}).

In the following, we consider two orientations of the magnetic field:
transverse ($\mathbf{B}=B\mathbf{e}_{z}$) and
parallel ($\mathbf{B}=B%
\mathbf{e}_{x}$) to the 2DEG plane.
It is important to emphasize that, when discussing the
perpendicular magnetic field, we neglect its orbital effect.
Certainly, if the spin
susceptibility is measured as a response to an external magnetic field, its
orbital and spin effects cannot be separated.
However, there are situations when the spin part of $\chi _{zz}$ is of primary
importance. For example, the Ruderman-Kittel-Kasuya-Yosida (RKKY)
interaction between the local moments located in the 2DEG plane arises only from the spin susceptibility of itinerant electrons, because the orbital effect of the dipolar magnetic field of such moments is negligible. In this case, $\chi _{xx}$ and $\chi _{zz}$ determine the strength of the RKKY interaction
between two moments aligned along the $x$ and $z$ axis, respectively. Also,
divergences of  $\chi _{zz}$ and $\chi _{xx}$ signal ferromagnetic
transitions into states with easy-axis and easy-plane anisotropies,
respectively. Since it is this kind of physical situations we are primarily interested  in this paper,
we will ignore the orbital effect of the field from now on.

The Green's function corresponding to the Hamiltonian (\ref{eq:Ham}) is
obtained by matrix inversion
\begin{equation}
\hat{G}_{K}\equiv \frac{1}{i\omega -\hat{H}}=\sum_{s=\pm }\hat{\Omega}_{s}(%
\mathbf{k})g_{s}(K),
\end{equation}
where we use the ``relativistic''\/ notation $K\equiv (\omega ,\mathbf{k})$
with $\omega $ being a fermionic Matsubara frequency, the matrix $\hat{\Omega}_{s}(\mathbf{k})$ is defined as
\begin{subequations}
\begin{eqnarray}
\hat{\Omega}_{s}(\mathbf{k}) &\equiv &\frac{1}{2}\left( \hat{I}+s\hat{\zeta}%
\right) ,  \label{eq:OmegaDef} \\
\hat{\zeta} &=&\frac{\alpha (k_{y}\hat{\sigma}_{x}-k_{x}\hat{\sigma}_{y})+%
\hat{\boldsymbol{\sigma}}\cdot\boldsymbol{\Delta}}{\bar{\Delta}_{\mathbf{k}}},
\end{eqnarray}
\end{subequations}
$g_{s}(K)=1/(i\omega -\xi _{\mathbf{k}}-s\bar{\Delta}_{\mathbf{k}})$ is the
single-electron Green's function, $\xi _{\mathbf{k}}\equiv \frac{k^{2}}{2m}-\mu $,
$\mu $ is the chemical potential, and $\bar{\Delta}_{\mathbf{k}}$ is given
by Eq. (\ref{zeeman}).

As we have already pointed out in the Introduction, an important difference
between the cases of transverse and in-plane magnetic field is the dependence of
the effective Zeeman energy [Eq. (\ref{zeeman})] on the electron
momentum. For the transverse magnetic field, ($\Delta _{x}=
\Delta _{y}=0,\Delta _{z}=\Delta $), the Zeeman energy is isotropic in the
momentum space and quadratic in $\Delta $ in the weak-field limit: $\bar{%
\Delta}_{\mathbf{k}}\approx |\alpha|k_{F}+\Delta ^{2}/2|\alpha|k_{F}.$
Correspondingly, the Fermi surfaces of Rashba branches are concentric circles
with slightly (in proportion to $\Delta ^{2}$) different radii. For the
in-plane magnetic field, ($\Delta _{x}=\Delta ,\Delta _{y}=
\Delta _{z}=0$), the effective Zeeman energy is anisotropic in the momentum
space, cf. Eq.~(\ref{eq:zeemaneff}).
Correspondingly, the Fermi surfaces of Rashba branches are also anisotropic
and their centers are shifted by finite momentum, proportional to $\Delta_{x}$.

\begin{figure}[t]
\includegraphics[width=.48\textwidth]{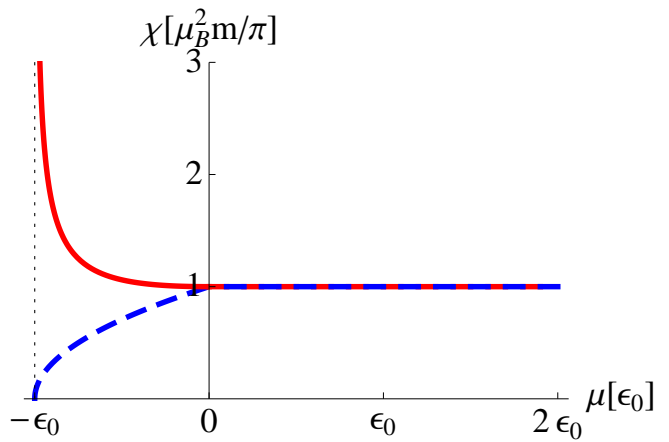}
\caption{(COLOR ONLINE) Spin susceptibility of free Rashba fermions
(in units of $\chi_0=\mu_B^2m/\pi$) as a function of the chemical
potential $\mu$ (in units of $\epsilon_0=m\alpha^2/2$). Solid (red):
$\chi^{\{0\}}_{xx}$; dashed (blue): $\chi^{\{0\}}_{zz}$. Note that $\chi^{\{0\}}_{xx}\neq \chi^{\{0\}}_{zz}$
if only the lowest Rashba branch is occupied  ($-\epsilon_0\leq\mu<0$)
but $\chi^{\{0\}}_{xx}=\chi^{\{0\}}_{zz}=\chi_0$ if both branches are occupied ($\mu>0$)
at $T=0$.}
\label{fig:ChiNoSOT0}
\end{figure}

We now give a brief summary of results for the susceptibility in the absence
of electron-electron interaction, which sets the zeroth order of the
perturbation theory (for more details, see Appendix \ref{app:0thOrder}). The in-plane
rotational symmetry of the Rashba Hamiltonian guarantees that $\chi^{\{0\}}
_{yy}=\chi^{\{0\}} _{xx}$. The static uniform susceptibility (defined in the limit
of zero frequency and vanishingly small wavenumber) is still diagonal even
in the presence of the SOI: $\chi^{\{0\}} _{ij}=\delta _{ij}\chi^{\{0\}} _{ii},$ although $%
\chi^{\{0\}} _{xx}=\chi^{\{0\}} _{yy}\neq \chi^{\{0\}} _{zz}$ in general.
The susceptibility depends strongly on whether both or only the
lower of the two Rashba branches are occupied, see Fig.~(\ref{fig:ChiNoSOT0}). In the latter case, the spin
response is strongly anisotropic. At $T=0$,
\begin{subequations}
\begin{eqnarray}
\chi _{zz}^{\{0\}} &=&\chi _{0}\sqrt{1+\mu /\epsilon _{0}}  \label{chizz0} \\
\chi _{xx} ^{\{0\}}&=&\chi _{0}\frac{1+\mu /2\epsilon _{0}}{\sqrt{1+\mu /\epsilon
_{0}}},  \label{chixx0}
\end{eqnarray}
\end{subequations}
where $\chi _{0}\equiv \mu _{B}^{2}m/\pi $ is the spin susceptibility of 2D electrons in the absence of the SOI, $\epsilon _{0}=m\alpha ^{2}/2$
is the depth of the energy minimum of the lower branch and the chemical
potential $\mu $ is within the range $-\epsilon _{0}\leq \mu \leq 0.$ The
in-plane susceptibility exhibits a 1D-like van Hove singularity at the
bottom of the lower branch, i.e., for $\mu \rightarrow -\epsilon _{0}$.
On the other hand, if both branches are occupied (which is the case for $\mu
>0)$, the spin susceptibility is isotropic and the same as in the absence of
the SOI
\begin{equation}
\chi^{\{0\}} _{zz}=\chi^{\{0\}} _{xx}=\chi _{0}.
\end{equation}
This isotropy can be related to a hidden symmetry of the Rashba Hamiltonian
manifested by conservation of  the square of the electron's velocity operator ${\hat v}$.\cite{RashbaJS} The eigenvalue of ${\hat v}^2$, given by $2\epsilon/m+2\alpha^2$ with $\epsilon$ being the energy, is the same
for both branches. The square of the group velocity
$v_g^2=\left(\nabla_{\bf k}\epsilon_{\bf k}^{\pm}\right)^2=2\epsilon/m+\alpha^2$ also does not depend on the branch index. Therefore, the total density of states
\begin{equation}
\nu(\epsilon)=\frac{1}{2\pi} \frac{k^{+}+k^{-}}
{|v_g|}=\frac{m}{\pi},
\end{equation}
where $k^{\pm}=\mp m\alpha+\sqrt{m^2\alpha^2+2m\epsilon}$ are the momenta of the $\pm$ branches
corresponding to energy $\epsilon$,
is the same as without the SOI, if both branches are occupied.
One can show also that isotropy of the spin susceptibility is not specific for the Rashba coupling but is there also in the presence of both Rashba and (linear) Dresselhaus interactions.\cite{ali}

\begin{figure}[t]
\includegraphics[width=.46\textwidth]{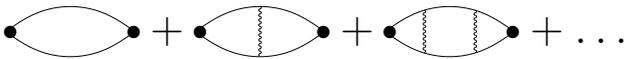}
\caption{The RPA diagrams for the spin susceptibility corresponding to Eq.~(\ref{eq:RPA}).}
\label{fig:RPA}
\end{figure}

As one step beyond the free-electron model, we consider a Stoner-like enhancement of the spin susceptibility by the electron-electron interaction.
In the absence of the SOI and for a point-like interaction $U$, the renormalized spin susceptibility is given by
\begin{equation}
\chi=\frac{\chi_0}{1-mU/\pi}.
\label{stoner1}
\end{equation}
In the presence of the SOI, the ladder series for the susceptibility, shown in Fig.~\ref{fig:RPA}
is given by
\begin{widetext}
\begin{eqnarray}
\chi_{ii}&=&\chi_{ii}^{\{0\}}+U\mu_B^2\sum_{K,P}\mathrm{Tr}\left[\hat\sigma_i\hat G(K+Q)\hat G(P+Q)\hat\sigma_i\hat G(P)\hat G(K)\right]
\notag\\&&
-U^2\mu_B^2\sum_{K,P,L}\mathrm{Tr}\left[\hat\sigma_i\hat G(K+Q)\hat G(P+Q)\hat G(L+Q)\hat\sigma_i\hat G(L)\hat G(P)\hat G(P)\right]+\dots\label{eq:RPA}
\end{eqnarray}
\end{widetext}
where $Q=(\Omega=0,\mathbf{q}\to 0)$ and $i=x,z$. Although the traces do look different for $\chi_{xx}$ and $\chi_{zz}$, these differences disappear after angular integrations, and the resulting series are the same. As we have shown above, the zero-order susceptibilities are also the same (and equal to $\chi_0$) if both Rashba branches are occupied; hence
\begin{equation}
\chi_{ii}=\mu_B^2\left(\frac{m}{\pi}+\frac{m^2U}{\pi^2}+\frac{m^3U^2}{\pi^3}+\dots\right)=\frac{\chi_0}{1-mU/\pi},\label{stoner2}
\end{equation}
which is the same result as in Eq.~(\ref{stoner1}). Therefore, at the mean-field level, the spin susceptibility remains isotropic and independent of the SOI. In the rest of the paper, we will show that none of these two features survives beyond the mean-field level: the actual spin susceptibility is anisotropic and both its components do depend on the SOI.

We now come back to the free-electron model and discuss the $T$ dependence of the spin susceptibility.
A special feature of a 2DEG in the absence of the SOI is a
breakdown of the Sommerfeld expansion at finite $T$: since the density of
states does not depend on the energy, all power-law terms of this expansion
vanish, and the resulting $T$ dependence of $\chi^{\{0\}} $ is only exponential. The
SOI leads to the energy dependence of the density of states for the
individual branches, and one would expect the Sommerfeld expansion to be
restored. This is what indeed happens if only the lower branch is occupied.
In this case,
\begin{subequations}
\begin{eqnarray}
\chi^{\{0\}} _{zz}(T) &=&\chi^{\{0\}} _{zz}(0)-\chi _{0}\frac{\pi ^{2}}{24}\left( \frac{T}{%
\epsilon _{0}}\right) ^{2}\frac{1}{\left( 1+\mu /\epsilon _{0}\right)
^{3/2}} \notag\\\label{eq:FreeChiZZmu<0} \\
\chi^{\{0\}} _{xx}(T) &=&\chi^{\{0\}} _{xx}(0)+\chi _{0}\frac{\pi ^{2}}{48}\left( \frac{T}{%
\epsilon _{0}}\right) ^{2}\frac{2-\mu /\epsilon _{0}}{\left( 1+\mu /\epsilon
_{0}\right) ^{5/2}},\notag\\ \label{eq:FreeChiXXmu<0}
\end{eqnarray}
\end{subequations}
provided that $T\ll \min \left\{ -\mu ,\epsilon _{0}+\mu \right\}$.
[Here, $\chi^{\{0\}} _{zz}(0)$ and $\chi^{\{0\}} _{xx}(0)$
are the zero temperature values given by Eqs. (\ref{chizz0}) and (\ref{chixx0})].
However, if both branches are occupied, the energy dependent terms in the branch
densities of states cancel out, and the resulting dependence is
exponential, similar to the case of no SOI, although with different
pre-exponential factors:
\begin{subequations}
\begin{eqnarray}
\chi^{\{0\}} _{zz}(T) &=&\chi _{0}\left( 1-\frac{T}{2\epsilon _{0}}e^{-\mu
/T}\right) \label{eq:FreeChiZZsmallT} \\
\chi^{\{0\}} _{xx}(T) &=&\chi _{0}\left( 1+\frac{T^{2}}{4\epsilon _{0}^{2}}\,e^{-\mu
/T}\right) \label{eq:FreeChiXXsmallT}
\end{eqnarray}
for $T\ll \epsilon _{0}\ll \mu $,
and
\end{subequations}
\begin{subequations}
\begin{eqnarray}
\chi^{\{0\}} _{zz}(T) &=&\chi _{0}\left[ 1-\left( 1-\frac{2\epsilon _{0}}{3T}%
\right) e^{-\mu /T}\right] \label{eq:FreeChiZZlargeT} \\
\chi^{\{0\}} _{xx}(T) &=&\chi _{0}\left[ 1-\left( 1-\frac{4\epsilon _{0}}{3T}%
\right) e^{-\mu /T}\right] \label{eq:FreeChiXXlargeT}
\end{eqnarray}
\end{subequations}
for $\epsilon _{0}\ll T\ll \mu$.
In a similar way, one can show that that there are no power-law terms in the dependence of $\chi^{\{0\}}_{xx}$ and $\chi^{\{0\}}_{zz}$ on the external magnetic field.

Notice that $\chi_{xx}\neq \chi_{zz}$ at finite temperature, even if both Rashba subbands are occupied. This suggests that the hidden symmetry of the Rashba Hamiltonian
is, in fact, a rotational symmetry in a $(2+1)$ space with imaginary time being an extra dimension.\cite{Rashba_private} Finite temperature  should then play a role of finite size along the time axis breaking the rotational symmetry.

In what follows, we assume that the SOI is weak, i.e., $\left| \alpha \right|
\ll v_{F},$ and thus the energy scales describing SOI are small:
$m\alpha ^{2}\ll |\alpha| k_{F}\ll \mu .$ This condition also means
that both Rashba branches are occupied. [Also, from now on we relabel $\mu
\rightarrow E_{F}$.] The main result of this section is that, for a weak SO
coupling, the $T$ and $\Delta$ dependences of $\chi $ in the free case are at least exponentially
weak and thus cannot mask the power-law dependences arising from the
electron-electron interaction, which are discussed in the rest of the paper.

\section{\label{sec:SS}SPIN SUSCEPTIBILITY to second order in the
electron-electron interaction}

\subsection{\label{sec:general}General strategy}

\begin{figure*}[t]
\setlength{\unitlength}{1cm}
\begin{picture}(16,2.5)(0,0)
	\put(-0.8,0){\includegraphics[width=.2\textwidth]{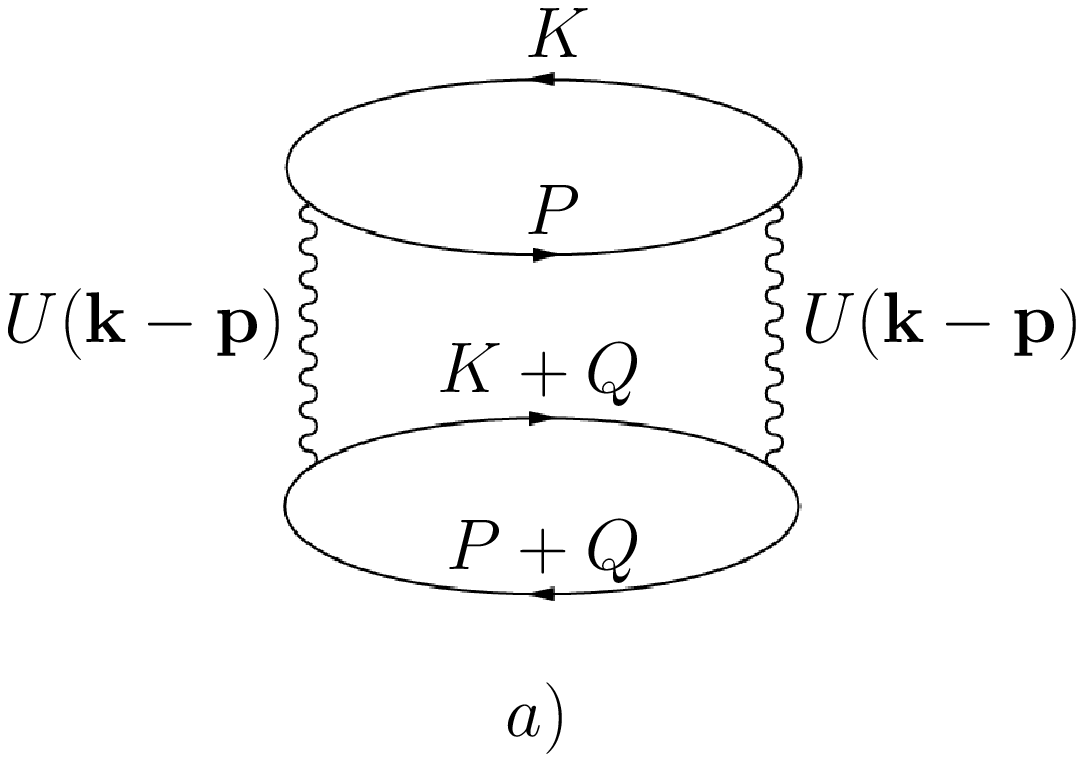}}
        \put(3.6,0){\includegraphics[width=.146\textwidth]{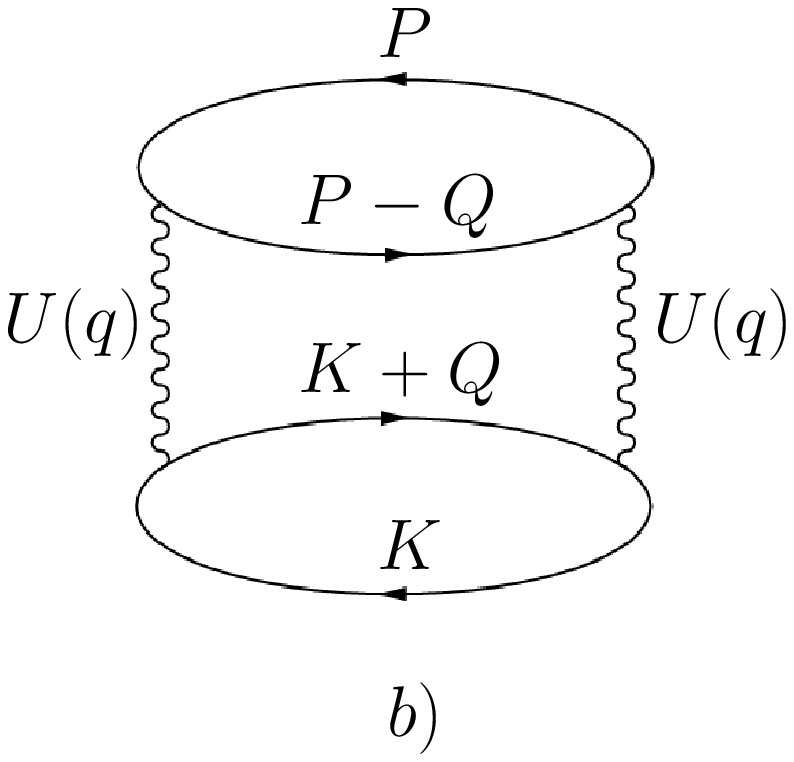}}
	\put(7,0){\includegraphics[width=.146\textwidth]{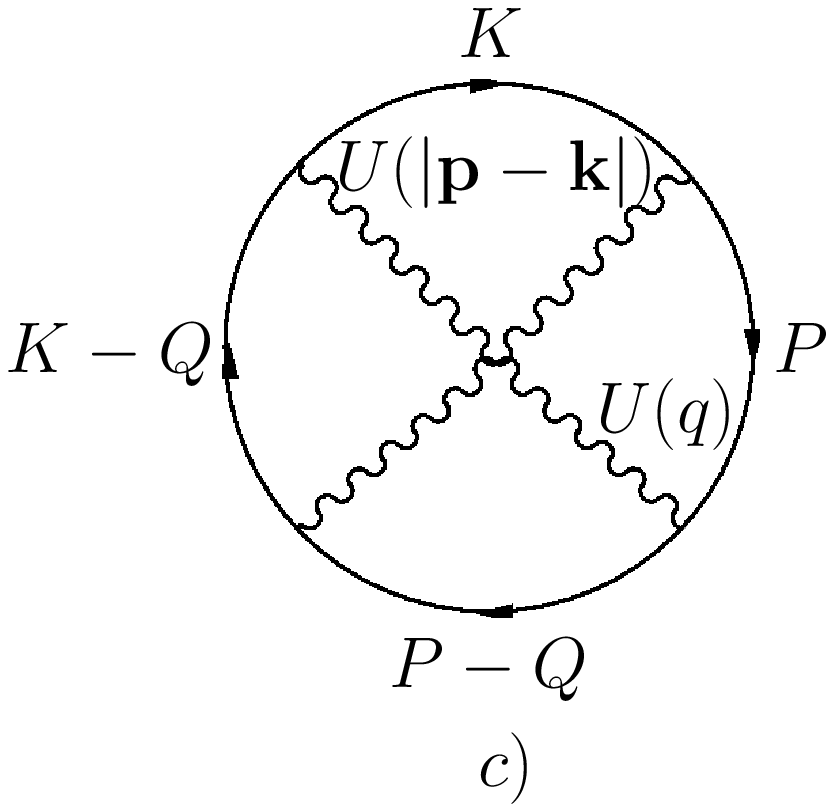}}
	\put(10.8,0){\includegraphics[width=.164\textwidth]{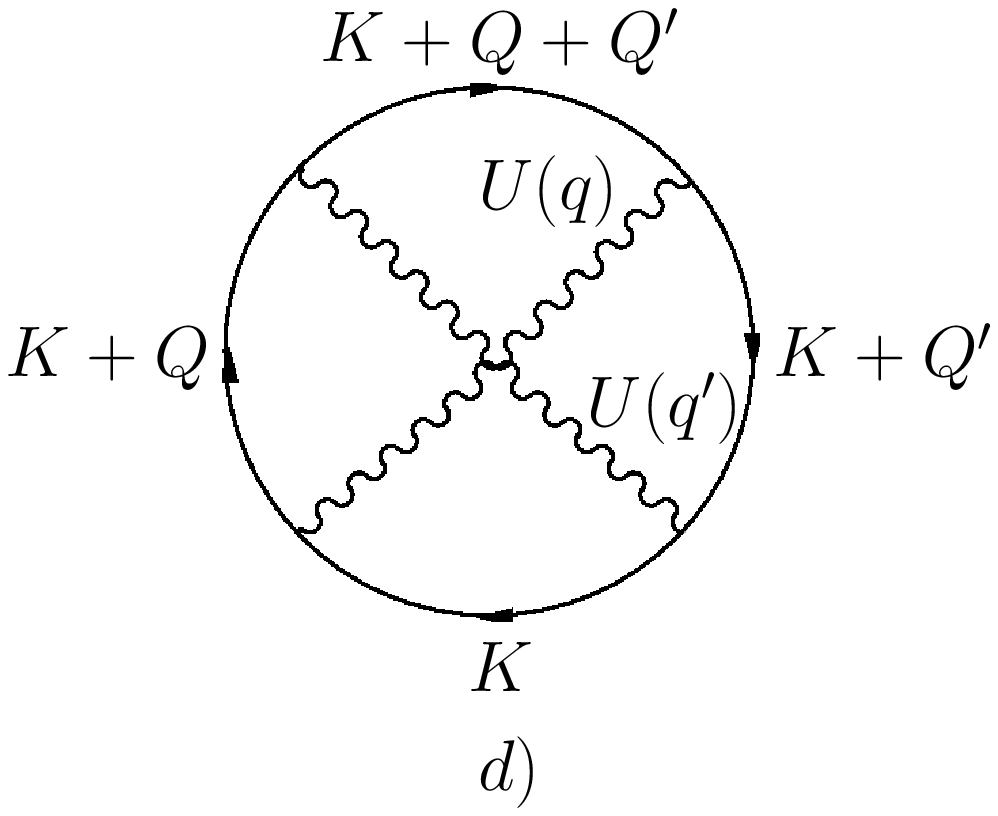}}
	\put(14.8,0){\includegraphics[width=.12\textwidth]{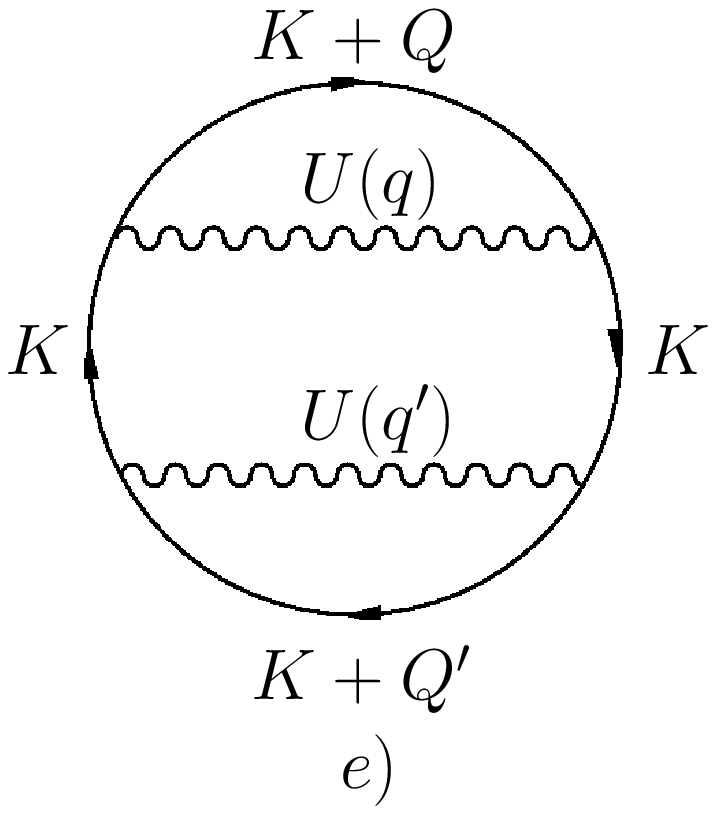}}
\end{picture}
\caption{Second order diagrams for the thermodynamic potential.
A nonanalytic contribution comes only from diagram $a$), where
the momenta are arranged in such a way that $\mathbf{k}\approx-\mathbf{p%
}$ while $Q$ is small; therefore, $k\approx p\approx k_{F}$,
and the momentum transfer in each scattering event is close to
$|\mathbf{k}-\mathbf{p}|\approx2k_{F}$. As is shown in Sec.~\ref{sec:Diags}, diagrams
$b$)-$e$) do not contribute to the nonanalytic behavior of
the spin susceptibility.}
\label{fig:Diagrams}
\end{figure*}

The spin susceptibility tensor $\chi_{ij}$ is related to the
thermodynamic (grand canonical) potential $\Xi(T,\alpha,\Delta)$ by the following identity
\begin{equation}
\chi_{ij}(T,\alpha)=-\frac{\partial^{2}\Xi}
{\partial B_i\partial B_j}\bigg|_{B=0}.
\end{equation}
To second order in the electron-electron interaction $U(q)$, there is only one diagram for the
thermodynamic potential that gives rise to a nonanalytic behavior: diagram $a$) in Fig.~\ref{fig:Diagrams}.
The rest of the diagrams in this figure can be shown to be irrelevant (cf. Sec.~\ref{sec:Diags}). Algebraically, diagram $a$) in Fig.~\ref{fig:Diagrams} reads
\begin{align}  \label{eq:Xi12kDef}
\notag \delta\Xi^{(2)} \equiv -\frac{1}{4}&\sum_Q\sum_K\sum_P U^{2}(|\mathbf{k}-%
\mathbf{p}|)\\
&\times \mathrm{Tr}(\hat{G}_K\hat{G}_P)\mathrm{Tr}(\hat{G}_{K+Q}\hat{G}%
_{P+Q}),
\end{align}
where $\sum_Q=T\sum_\Omega\int d^2q/(2\pi)^2$, $\sum_K=T\sum_\omega\int
d^2k/(2\pi)^2$, and we use \lq\lq relativistic\rq\rq\/ notation $K\equiv(\omega,%
\mathbf{k})$ with a fermionic frequency $\omega$ and $Q\equiv(\Omega,\mathbf{%
q})$ with a bosonic frequency $\Omega$.

Evaluation of the first spin trace in Eq.~(\ref{eq:Xi12kDef}) yields
\begin{equation}
\mathrm{Tr}(\hat{G}_{K}\hat{G}_{P})=\frac{1}{2}\sum_{ss^{\prime
}}B_{ss^{\prime }}(\mathbf{k},\mathbf{p})g_{s}(K)g_{s^{\prime }}(P),
\label{eq:Tr2}
\end{equation}
where
\begin{equation}
B_{ss^{\prime }}(\mathbf{k},\mathbf{p})\equiv 1+ss^{\prime }\frac{\alpha ^{2}%
\mathbf{k}\cdot \mathbf{p}+\alpha
\left[\boldsymbol\Delta\times(\mathbf{k}+\mathbf{p})\right]_z
+\Delta ^{2}}{\bar{%
\Delta}_{\mathbf{k}}\bar{\Delta}_{\mathbf{p}}}  \label{eq:Bxx}
\end{equation}
and ${\bar{\Delta}}_{\mathbf{k}}$ is given by Eq.~(\ref{zeeman}).
The second-order thermodynamic potential then becomes
\begin{widetext}
\begin{equation}  \label{eq:Xi12kDef_1}
\delta\Xi^{(2)} \equiv -\frac{1}{16}\sum_Q\sum_K\sum_P
U^{2}(|{\bf k}-{\bf p}|)B_{s_1s_3}({\bf k},{\bf p})B_{s_2s_4}({\bf k}+{\bf q},{\bf p}+{\bf q})
g_{s_1}(K)g_{s_2}(P)g_{s_3}(K+Q)g_{s_4}(P+Q).
\end{equation}
\end{widetext}
Equation~(\ref{eq:Xi12kDef_1}) describes the interaction among
electrons from all Rashba branches via an effective vertex $U(|\mathbf{k}-%
\mathbf{p}|)B_{ss^{\prime }}(\mathbf{k},\mathbf{p})$, which depends not only
on the momentum transfer $\mathbf{k}-\mathbf{p}$ but also on the initial momenta $\mathbf{k}$
and $\mathbf{p}$ themselves. This last dependence is due to
anisotropy of the Rashba spinors.

It has been shown in Refs.~\onlinecite{PhysRevB.68.155113,maslov06_09} that, at weak coupling, the main contribution to the nonanalytic part of the spin susceptibility comes from \lq\lq backscattering\rq\rq\/ processes, i.e., processes with $\mathbf{p}\approx-\mathbf{k}$ and small $q$ (compared to $k_F$). In particular, the second-order contribution is entirely of the backscattering type. The proof given in Ref.~\onlinecite{maslov06_09} applies to any kind of the angular-dependent vertex and, thus, also to vertices in Eq.~(\ref{eq:Xi12kDef_1}). Therefore, the calculation can be simplified dramatically by putting in $\mathbf{p}=-\mathbf{k}$ and neglecting $\mathbf{q }$ in the effective vertices. The last assumption is justified as long as the typical values of $q$ are determined by the smallest energy of the
problem, i.e., $q\sim\max\{T/v_F,m|\alpha|\}\ll k_F$. By the same argument, the
magnitudes of $\mathbf{k}$ and $\mathbf{p}$ in the vertices can be replaced by $k_F$. The bare
interaction is then evaluated at $|\mathbf{k}-\mathbf{p}|\approx 2k_F$, and
we introduce a coupling constant $U\equiv U(2k_F)$. Ignoring the angular dependence of
$B_{s_is_j}$ for a moment, Eq.~(\ref{eq:Xi12kDef_1})
is reduced to a convolution of two particle-hole bubbles, formed by
electrons belonging to either the same or different Rashba branches
\begin{equation}
\Pi _{s_is_j}(Q)=\sum_{K}g_{s_i}(K)g_{s_j}(K+Q).  \label{piph}
\end{equation}
By assumption, both components of $Q$ in $\Pi_{s_is_j}(Q)$, i.e., $\Omega$ and $q$, are
small (compared to $E_F$ and $k_F$, correspondingly). It is important to
realize that, despite a two-band nature of the Rashba spectrum, $\Pi
_{s_is_j}(Q)$ has no threshold-like singularities at the momentum $q_0=2
m\alpha$, separating the Rashba subbands\cite{pletyukhov06} (for a detailed
derivation of this result, see Appendix \ref{app:smallq}). Therefore, the
nonanalytic behavior of the spin susceptibility comes only from the
Landau-damping singularity of the dynamic bubble, as it is also the case in
the absence of the SOI.

After the simplifications described above, Eq.~(\ref{eq:Xi12kDef_1}) becomes
\begin{widetext}
\begin{equation}  \label{eq:Xi12kDef_3}
\delta\Xi^{(2)} \equiv -\frac{U^2}{16}\sum_Q\sum_K\sum_P
B_{s_1s_3}B_{s_2s_4}
g_{s_1}(K)g_{s_2}(P)g_{s_3}(K+Q)g_{s_4}(P+Q),
\end{equation}
\end{widetext}
where
\begin{equation}  \label{eq:Bpi}
B_{ss^{\prime}}\equiv B_{ss^{\prime}}(\mathbf{k}_F,-\mathbf{k}_F)=
1+ss^{\prime} \frac{\Delta^2-\alpha^2k_F^2}{\bar{\Delta}_{\mathbf{k}_F} \bar{%
\Delta}_{-\mathbf{k}_F}}
\end{equation}
and $\mathbf{k}_F\equiv (\mathbf{k}/k)k_F$. Finally, to obtain the leading $T$ dependence of $\chi_{ij}$, it suffices to replace the fermionic Matsubara sums in Eq.~(\ref{eq:Xi12kDef_3}) by integrals but keep the
bosonic Matsubara sum as it is. The rest of the calculations proceed somewhat differently for the cases of the transverse and in-plane magnetic fields.

\subsection{Transverse magnetic field}

\label{sec:SS2zz} First, we consider a simpler case of the magnetic field
transverse to the 2DEG plane: $\boldsymbol{\Delta}=g\mu _{B}B\mathbf{e%
}_{z}/2$. In this case, the effective Zeeman energy is isotropic in the
momentum space; therefore, ${\bar{\Delta}}_{\mathbf{k}_{F}}={\bar{\Delta}}_{-%
\mathbf{k}_{F}}$ and
\begin{equation}
	B_{ss^{\prime }}  = 1+ss^{\prime }\frac{\Delta ^{2}-\alpha ^{2}k_{F}^{2}}
	{\bar{\Delta}_{\mathbf{k}_{F}}^{2}} = 1+ss^{\prime }\frac{\Delta ^{2}-\alpha ^{2}k_{F}^{2}}
	{\Delta ^{2}+\alpha^{2}k_{F}^{2}}.\label{eq:BOutDef}
\end{equation}
Thereby the integrals over $d^{3}K$ and $d^{3}P$ separate, and one obtains
\begin{equation}
\delta \Xi _{zz}^{(2)}=-\frac{U^{2}}{16}\sum_{s_{1}\dots
s_{4}=\pm 1}B_{s_{1}s_{3}}B_{s_{2}s_{4}}\sum_Q\Pi _{s_{1}s_{2}}(Q)\Pi
_{s_{3}s_{4}}(Q),  \label{sat1}
\end{equation}
where $\Pi _{s_{i}s_{j}}(Q)$ is  given by Eq.~(\ref{piph}).
The single-particle spectrum in the second Green's function in Eq.~(\ref{piph})
can be linearized with respect
to $q$ as $\xi _{\mathbf{k}+\mathbf{q}}\approx \xi _{%
\mathbf{k}}+v_{F}q\cos \phi _{kq}$ with $\phi _{\mathbf{kq}}\equiv \angle (\mathbf{k},%
\mathbf{q})$. Since, by assumption, the SOI is small, the effective
Zeeman energy in the Green's functions can be replaced by its value at $%
k=k_{F}$. Integration over $dkk$ can be then replaced by that
over $d\xi _{\mathbf{k}}$.
The integrals over $\omega $, $\xi_{\mathbf{k}}$ and $\phi
_{kq}$ are performed in the same way as in the absence of the SOI,
and we arrive at the following expression for the dynamic part of the
polarization bubble
\begin{equation}
\Pi _{ss^{\prime }}=\frac{m}{2\pi }\frac{|\Omega |}{\sqrt{\left[ \Omega
+i(s^{\prime }-s)\bar{\Delta}_{\mathbf{k}_{F}}\right] ^{2}+\left(
v_{F}q\right) ^{2}}}.  \label{jss}
\end{equation}

We pause here for a comment. The polarization bubble in Eq.~(\ref{jss}) is
very similar to the dynamic part of the polarization bubble for spin-up and
-down electrons in the presence of the magnetic field but in the absence of
the SOI \cite{maslov06_09}
\begin{equation}
\Pi _{\uparrow \downarrow }(\Omega ,q)=\frac{m}{2\pi }\frac{|\Omega |}{\sqrt{%
(\Omega -ig\mu _{B}B)^{2}+(v_{F}q)^{2}}}.
\end{equation}
As we already mentioned in Sec.~\ref{sec:Int}, the nonanalytic behavior of the spin susceptibility
is due to an effective interaction of Fermi-liquid
quasiparticles via particle-hole pairs with
small energies and momenta near $2k_F$. Our calculation is arranged in such a way that, on a technical level,
we deal with pairs with small momenta $q$. The spectral weight of these pairs is
proportional to the polarization bubble, which is singular for small $\Omega
$ and $q$. Since finite magnetic field cuts off the singularity in $\Pi
_{\uparrow \downarrow }(\Omega ,q)$, the nonanalytic dependence of $\chi $
on, e.g., temperature, saturates when $T$
becomes comparable to the Zeeman splitting $g\mu _{B}B$. At lower
energies, $\chi $ exhibits a nonanalytic dependence on $\Delta $: $\chi
\propto |\Delta|$. Likewise, the singularity in Eq.~(\ref{jss}) is cut at
the effective Zeeman energy ${\bar{\Delta}}_{\mathbf{k}_{F}}$, which reduces
to the SOI energy scale $|\alpha|k_{F}$, when the real magnetic field goes to
zero. Therefore, one should expect the nonanalytic $T$ dependence of $%
\chi _{zz}$ to be cut by the SOI. Although soft particle-hole pairs can
be still generated within a~given branch, i.e., for $s=s^{\prime }$, the
entire dependence on the Zeeman energy in this case is eliminated and
processes of this type do not affect the spin susceptibility. In the rest of
this section, we are going to demonstrate that the SOI indeed plays a role
of the magnetic field for $\chi _{zz}$.

For later convenience, we define a new quantity
\begin{equation}
\mathcal{P}_{s}\equiv \mathcal{P}_{s}(\Omega ,q)=\frac{1}{\sqrt{\left(
\Omega -2is\bar{\Delta}_{\bm{k}_{F}}\right) ^{2}+v_{F}^{2}q^{2}}}
\label{eq:JsXXDef}
\end{equation}
and sum over the Rashba branches in Eq.~(\ref{sat1}). The
contribution of the set $\{s_{1}=s_{2}\equiv s,s_{3}=s_{4}\equiv s^{\prime
}\}$ does not depend on ${\bar{\Delta}}_{\mathbf{k}_{F}}$:
\begin{equation}
\sum_{s,s^{\prime }}B_{ss^{\prime }}^{2}\Pi_{ss}^{2}=\left( \frac{m\Omega }{%
2\pi }\right) ^{2}\sum_{s,s^{\prime }}B_{ss^{\prime }}^{2}\mathcal{P}%
_{0}^{2}.
\end{equation}
The set $\{s_{1}=-s_{2}\equiv s,s_{3}=-s_{4}\equiv s^{\prime }\}$ gives
\begin{equation}
\sum_{s,s^{\prime }}B_{ss^{\prime }}B_{-s,-s^{\prime
}}\Pi_{s,-s}\Pi_{s^{\prime },-s^{\prime }}=\left( \frac{m\Omega }{2\pi }%
\right) ^{2}\sum_{s,s^{\prime }}B_{ss^{\prime }}^{2}\mathcal{P}_{s}\mathcal{P%
}_{s^{\prime }},
\end{equation}
where we used that $B_{ss}=B_{-s,-s}$. Finally, the sets $\{s_{1}=s_{2}\equiv
s,s_{3}=-s_{4}\equiv s^{\prime }\}$ and $\{s_{1}=-s_{2}\equiv
s,s_{3}=s_{4}\equiv s^{\prime }\}$ contribute
\begin{align}
\sum_{s,s^{\prime }}(B_{ss^{\prime }}B_{s,-s^{\prime }}&
\Pi_{ss}\Pi_{s^{\prime },-s^{\prime }}+B_{ss^{\prime }}B_{-s,s^{\prime
}}\Pi_{s,-s}\Pi_{s^{\prime }s^{\prime }})  \notag \\
& =2\left( \frac{m\Omega }{2\pi }\right) ^{2}\sum_{ss^{\prime
}}B_{ss^{\prime }}B_{s,-s^{\prime }}\mathcal{P}_{0}\mathcal{P}_{s^{\prime }},
\end{align}
where we relabeled the indices in the second sum ($s\rightarrow s^{\prime }$%
, $s^{\prime }\rightarrow s$) and used the symmetry property $B_{ss^{\prime
}}=B_{s^{\prime }s}$.

The angular integral contributes a unity, so that
\begin{align}  \label{eq:XiZZProof}
\delta\Xi_{zz}^{(2)} = -&\left(\frac{mU}{8\pi}\right)^2T\sum_{\Omega}%
\Omega^2 \int\frac{dqq}{2\pi}  \notag \\
&\times\sum_{ss^{\prime}} [B_{ss^{\prime}}^2(\mathcal{P}_0^2+\mathcal{P}_{s}%
\mathcal{P}_{s^{\prime}})+2B_{ss^{\prime}}B_{s,-s^{\prime}}\mathcal{P}_{0}%
\mathcal{P}_{s^{\prime}}].
\end{align}
Now we sum over $ss^{\prime}$, add and subtract a combination $%
2(\alpha^4k_F^4+4\alpha^2k_F^2\Delta^2+\Delta^4)\mathcal{P}_0^2$ inside the
square brackets, and obtain, after some algebra,
\begin{align}  \label{eq:XiZZMid}
& & \delta\Xi_{zz}^{(2)} = -\left(\frac{mU}{8\pi}\right)^2T\sum_{%
\Omega}\Omega^2 \int^{\infty}_0\frac{dqq}{2\pi}\frac{4}{\bar{\Delta}_{%
\mathbf{k}_F}^4}[4\bar{\Delta}_{\mathbf{k}_F}^4\mathcal{P}_0^2  \notag \\
& & + \Delta^4(\mathcal{P}_+^2 + \mathcal{P}_-^2 - 2\mathcal{P}_0^2) +
4\alpha^2k_F^2\Delta^2\mathcal{P}_0(\mathcal{P}_+ + \mathcal{P}_- - 2%
\mathcal{P}_0)],
\end{align}
where we used that $\int dqq\left(\mathcal{P}_+\mathcal{P}_- - \mathcal{P}%
_0^2\right)=0$. The first term in the square brackets in Eq.~(\ref{eq:XiZZMid})
 does not depend on the
effective field $\bar{ \Delta}_{\mathbf{k}_F}$ and, therefore, can be dropped.
Integration over $q$ in the remaining terms is performed as
\begin{equation}
	\int dqq\mathcal{P}_0(\mathcal{P}_+ + \mathcal{P}_- - 2\mathcal{P}_0) =
	\frac{1}{v_F^2}\ln\frac{\Omega^2}{\Omega^2+\bar{\Delta}_{\mathbf{k}_F}^2},
\end{equation}
\begin{equation}
	\int dqq(\mathcal{P}_+^2 + \mathcal{P}_-^2 - 2\mathcal{P}_0^2)
	= \frac{1}{v_F^2}\ln\frac{\Omega^2}{\Omega^2+4\bar{\Delta}_{\mathbf{k}_F}^2}.
\end{equation}
Collecting all terms together, we obtain
\begin{align}
\delta\Xi_{zz}^{(2)} = -\frac{2}{\pi}&\left(\frac{mU}{4\pi v_F}%
\right)^2\bigg[ \frac{\Delta^4}{4\bar{\Delta}_{\bm{k}_F}^4}T\sum_{\Omega}\Omega^2
\ln\frac{\Omega^2}{\Omega^2+4\bar{\Delta}_{\mathbf{k}_F}^2}  \notag \\
&+\frac{\alpha^2k_F^2\Delta^2}{\bar{\Delta}_{\bm{k}_F}^4}T\sum_{\Omega}\Omega^2
\ln\frac{\Omega^2}{\Omega^2+\bar{\Delta}_{\mathbf{k}_F}^2}\bigg].
\end{align}
The bosonic sum is evaluated by replacing the sum by an integral as
follows
\begin{equation}
T\sum_{\Omega}F(\Omega)=\int_{-\infty}^{\infty}\frac{d\Omega}{2\pi}\coth%
\frac{\Omega}{2T} \mathrm{Im}\left[\lim_{\delta\rightarrow0}F(-i\Omega+%
\delta)\right]
\end{equation}
and using the identity
\begin{align}
\mathrm{Im}\bigg[\lim_{\delta\rightarrow0}(-i\Omega+\delta)^2 \ln&\frac{%
(-i\Omega+\delta)^{2}}{(-i\Omega+\delta)^{2}+x^2}\bigg]  \notag \\
&= \pi\Omega^2\mathrm{sign}\Omega\Theta(x^2-\Omega^{2}),
\end{align}
where $\Theta(x)$ stands for the step function.

The thermodynamic potential then becomes
\begin{align}
\delta \Xi _{zz}^{(2)}=&-\frac{2}{\pi }\left( \frac{mU}{4\pi v_{F}}%
\right) ^{2}T^{3}\bigg[ \frac{\Delta ^{4}}{4\bar{\Delta}_{\mathbf{k}%
_{F}}^{4}}\mathcal{F}\left( \frac{2\bar{\Delta}_{\mathbf{k}_{F}}}{T}\right)
\notag\\
& +\frac{\alpha ^{2}k_{F}^{2}\Delta ^{2}}{\bar{\Delta}_{\mathbf{k}_{F}}^{4}}%
\mathcal{F}\left( \frac{\bar{\Delta}_{\mathbf{k}_{F}}}{T}\right) \bigg]
\label{XiF}
\end{align}
with
\begin{widetext}
\begin{eqnarray}  \label{eq:FDef}
\mathcal{F}(y) = \int_0^{y}dxx^2\coth\left(x/2\right)
=-(1/3)y^2\left[y+6\mathrm{Li}_1\left(e^{y}\right)\right]+4[\zeta(3)+y\mathrm{Li}
_2\left(e^{y}\right)-\mathrm {Li}_3\left(e^{y}\right)],
\end{eqnarray}
\end{widetext}
where $\mathrm{Li}_{n}(z)\equiv \sum_{k=1}^{\infty }z^{k}/k^{n}$ is the
polylogarithm function and $\zeta (z)$ is the Riemann zeta function. In
practice, the integral form of $\mathcal{F}(y)$ is more convenient as it can
be easily expanded in the limits of small and large argument. Indeed, for $%
y\ll 1,$ one expands $\coth (x/2)$ as $\coth (x/2)=2/x+x/6$ and, upon
integrating over $x$ in Eq. (\ref{eq:FDef}), obtains
\begin{equation}
\mathcal{F}\left( y\right) =y^{2}+\frac{y^{3}}{24}+\mathcal{O}\left(
y^{4}\right),\; \text{ for }0<y\ll 1.  \label{Fsmall}
\end{equation}
For $y\gg 1,$ one subtracts unity from the integrand and replaces the upper
limit in the remaining integral by infinity:
\begin{widetext}
\begin{eqnarray}
\mathcal{F}\left( y\right)  &=&\frac{y^{3}}{3}+\int_{0}^{y}dxx^{2}\left(
\coth \frac{x}{2}-1\right) =\frac{y^{3}}{3}+\int_{0}^{\infty }dxx^{2}\left(
\coth \frac{x}{2}-1\right) -\int_{y}^{\infty }dxx^{2}\left( \coth \frac{x}{2}%
-1\right)  \notag\\
&=&\frac{y^{3}}{3}+4\zeta \left( 3\right) +\mathcal{O}\left( e^{-y}\right),\;
\text{ for }y\gg 1.  \label{Flarge}
\end{eqnarray}
\end{widetext}

\subsubsection{Temperature dependence of $\protect\chi _{zz}$}

The (linear) spin susceptibility is given by $\chi _{zz}=-\partial^2\Xi/\partial B_z^2 |_{B=0}$, which means that only terms proportional to $\Delta
^{2} $ in the thermodynamic potential matter. Therefore, for
finite $\alpha $, the spin susceptibility comes entirely from the second
term in the square brackets of \ Eq. (\ref{XiF}), which is proportional to $%
\alpha ^{2}k_{F}^{2}\Delta ^{2}$. [On the other hand, for $\alpha =0$ the
second term vanishes, while $\Delta ^{4}/\bar{\Delta}_{{\bf k}_{F}}^{4}=1$, and the spin susceptibility comes exclusively from the first term: $\delta\Xi_{zz}^{(2)}$ still depends on the magnetic field through $\mathcal{F}(\bar\Delta_{{\bf k}_F}/T)$, where the $\Delta$-dependence must be retained; in this case, $\mathcal{F}(\bar\Delta_{{\bf k}_F}/T)$ is evaluated as shown in Ref.~\onlinecite{PhysRevB.68.155113}.]
Neglecting the first term and differentiating the second one, we
obtain the interaction correction to $\chi _{zz}$ for $\Delta\to 0$ as
\begin{equation}
\delta \chi _{zz}^{(2)}=2\chi _{0}\left( \frac{mU}{4\pi }\right)
^{2}\frac{T^{3}}{\alpha ^{2}k_{F}^{2}E_{F}}\mathcal{F}\left( \frac{\left|
\alpha \right| k_{F}}{T}\right) .  \label{eq:Chi2ZZRes}
\end{equation}
For $T\gg |\alpha |k_{F},$ the asymptotic expansion of $\mathcal{F}$
in Eq. (\ref{Fsmall}) gives
\begin{equation}
\delta \chi _{zz}^{(2)}\approx 2\chi _{0}\left( \frac{mU}{4\pi }%
\right) ^{2}\left[ \frac{T}{E_{F}}+\frac{1}{24}\frac{\alpha ^{2}k_{F}^{2}}{%
TE_{F}}+\dots \right] .  \label{eq:Delta<<T}
\end{equation}
The first term in Eq.~(\ref{eq:Delta<<T}) coincides with the result of Refs.~%
\onlinecite{PhysRevB.68.155113,maslov06_09,PhysRevB.74.205122,schwiete06}
obtained in the absence of the SOI, while the second term is a
correction due to the finite SOI. In the opposite limit, i.e., for $T\ll
|\alpha |k_{F},$ the asymptotic expansion of $\mathcal{F}$ in
Eq. (\ref{Flarge}) gives
\begin{equation}
\delta \chi _{zz}^{(2)}\approx 2\chi _{0}\left( \frac{mU}{4\pi }%
\right) ^{2}\left[ \frac{|\alpha |k_{F}}{3E_{F}}+4\zeta (3)\frac{T^{3}}{%
\alpha ^{2}k_{F}^{2}E_{F}}+\dots \right] .  \label{eq:Delta>>T}
\end{equation}
As it was anticipated, the SOI cuts off the nonanalytic $T$
dependence for $T\lesssim |\alpha |k_{F}.$ However, the $T$ dependence is
replaced by a nonanalytic $\left| \alpha \right| $ dependence on the SO
coupling.

\begin{figure}[t]
\includegraphics[width=.48\textwidth]{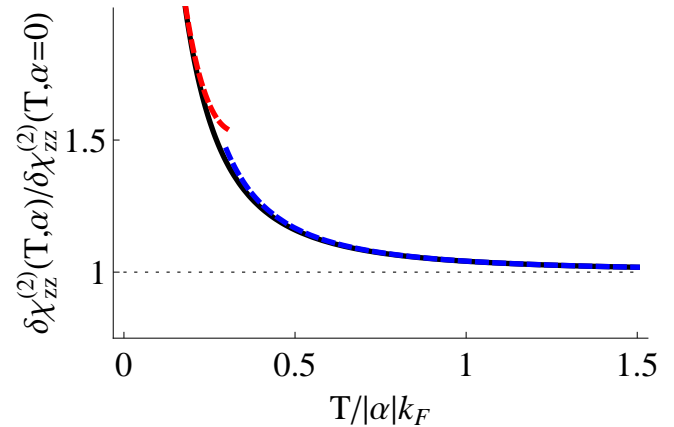}
\caption{(COLOR ONLINE)
The second-order nonanalytic correction to $\chi_{zz}$, normalized to its value in the absence of the SOI, as
a function of $T/|\alpha|k_F$ [cf. Eq.~(\ref{scaling})].
The asymptotic forms, given by Eqs.~(\ref{eq:Delta<<T}) and (\ref{eq:Delta<<T}), are shown by
dashed (red and blue) lines.}\label{fig:ChiZZofT}
\end{figure}

Normalizing Eq.~(\ref{eq:Chi2ZZRes}) to the leading $T$ dependent term for $%
\alpha =0$, i.e., by $\delta \chi _{zz}^{(2)}\left( T,\alpha =0\right) =$ $%
\chi _{0}\left( mU/4\pi \right) ^{2}\left( T/E_{F}\right) ,$ we express
$\delta \chi _{zz}^{(2)}$ via a scaling function of the
variable $T/|\alpha |k_{F}$
\begin{equation}
\frac{\delta \chi _{zz}^{(2)}\left( T,\alpha \right) }{\delta \chi
_{zz}^{(2)}\left( T,\alpha =0\right) }=\left( \frac{T}{|\alpha |k_{F}}%
\right) ^{2}\mathcal{F}\left( \frac{\left| \alpha \right| k_{F}}{T}\right) .
\label{scaling}
\end{equation}
The left-hand side of Eq. (\ref{scaling}) is plotted in Fig. \ref{fig:ChiZZofT}
along with its high and low $T$ asymptotic forms.

\begin{figure}[t]
\includegraphics[width=.45\textwidth]{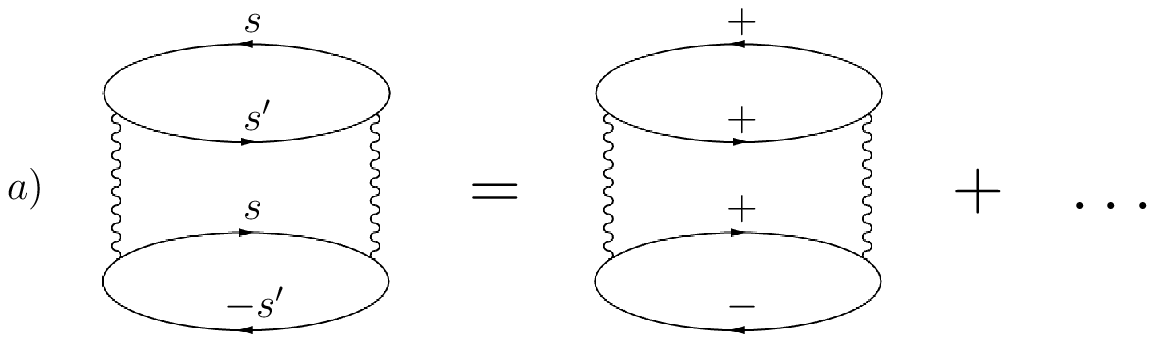}\\
\vspace*{2em}
\includegraphics[width=.45\textwidth]{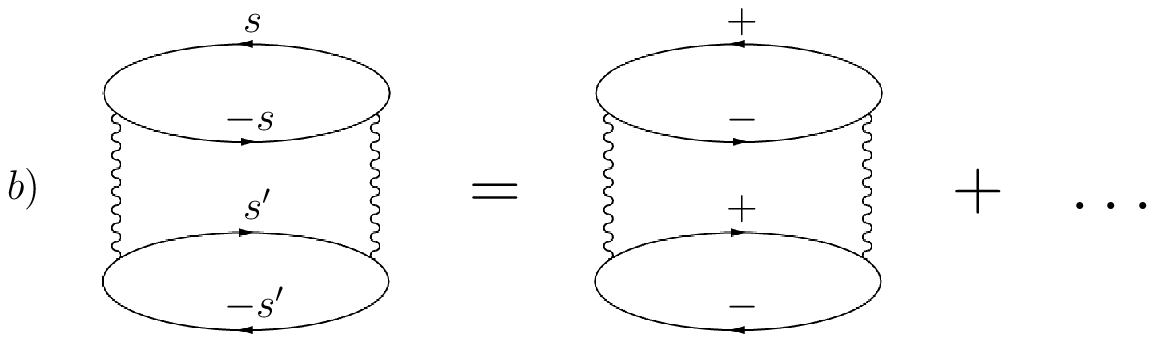}
\caption{Top: Diagrams contributing to nonanalytic behavior of $\chi_{zz}$.
There are eight such diagrams with the following choice of Rashba indices:
$+++-$, $++-+$, $+-++$, $-+++$, $---+$, $--+-$, $-+--$, and $+---$;
one of the them is shown on the right.
Bottom: Diagrams contributing to nonanalytic behavior of $\chi_{xx}$.
There are four such diagrams with the following choice of Rashba indices:
$+-+-$, $+--+$, $-++-$, and $-+-+$; one of them is shown on the right.}
\label{fig:Rashba}
\end{figure}

Now we can give a physical interpretation of the above results. Although the electron-electron interaction  mixes Rashba branches, the final result in Eq.~(\ref{eq:Chi2ZZRes})
comes only from a special combination of electron states. Namely, three out of four
electron states involved in the scattering process (two for the incoming and
two for the outgoing electrons) must belong to the same Rashba branch, while the last one
must belong to the opposite branch, as shown in Fig.~\ref{fig:Rashba}a. This can
be seen from Eq.~(\ref{eq:XiZZMid}) by considering four terms in the square
brackets. The first term, proportional to $\bar{\Delta}_{\bm{k}_{F}}^{4}$, does
not depend on the field upon the cancellation with an overall factor of $%
\bar{\Delta}_{\bm{k}_{F}}^{4}$ in the denominator; the second term, proportional
to $\alpha ^{4}k_{F}^{4}$, vanishes; the third term is already proportional
to $\Delta ^{4}$ and, thus, cannot affect the spin susceptibility for finite
$\alpha $. Therefore, the effect comes exclusively from the last term,
proportional to $\Delta ^{2}$, \ because the product $B_{ss^{\prime
}}B_{s,-s^{\prime }}$ is equal to $4\Delta ^{2}\left( \alpha k_{F}\right)
^{2}/(\alpha^2k_F^2+\Delta^2)^2\approx4\Delta^2/(\alpha k_F)^2$ for any choice of
$s$ and $s^{\prime}$. This term corresponds to the structure $\sum_{ss^{\prime }}B_{ss^{\prime
}}B_{s,-s^{\prime }}\Pi_{s,-s}\Pi_{s^{\prime }s^{\prime }}$ in Eq.~(\ref
{eq:XiZZProof}). The diagrams corresponding to this structure are shown in Fig.~%
\ref{fig:Rashba}a. Pairing electron Green's functions from different bubbles,
we always obtain a combination $\sum_{K}g_{\pm }\left( K\right) g_{\pm
}\left( K+Q\right) ,$ which depends neither on the SOI nor on the
magnetic field, and a combination $\sum_{K}g_{\pm }\left( K\right) g_{\mp
}\left( K+Q\right) ,$ which depends on both via the effective Zeeman
energy $\bar{\Delta}_{\mathbf{k}_{F}}=\sqrt{\alpha^2 k_{F}^2+\Delta^2}$.
In the weak-field limit, one needs to keep $\Delta ^{2}$
only in the prefactor. The singularity in the combination $\sum_{K}g_{\pm
}\left( K\right) g_{\mp }\left( K+Q\right) $ is then regularized by finite
SOI, which is the reason why the nonanalytic $T$ dependence is cut
off by the SOI.

\subsubsection{Magnetic-field dependence of $\protect\chi _{zz}$}

Now we consider the case of $T\ll \max\{\Delta ,|\alpha| k_{F}\}$ when, to first
approximation, one can set $T=0.$ In this case, one can define a non-linear
susceptibility $\chi_{zz}\left( B_z,\alpha \right) =-\partial ^{2}\Xi_{zz}
/\partial B_z^{2}$ evaluated at finite rather than zero magnetic field. For $%
T\rightarrow 0,$ we replace the scaling function $\mathcal{F}$ in Eq. (\ref
{XiF}) by the first term in its large-argument asymptotic form (\ref{Flarge}%
) to obtain
\begin{equation}
\delta \Xi _{zz}^{(2)}=-\frac{2}{3\pi }\left( \frac{mU}{4\pi v_{F}}%
\right) ^{2}\frac{2\Delta ^{4}+\alpha ^{2}k_{F}^{2}\Delta^2}{\sqrt{\Delta
^{2}+\alpha ^{2}k_{F}^{2}}} .
\end{equation}
Differentiating twice with respect to the field, we find
\begin{equation}
\delta\tilde{\chi}^{(2)}_{zz}\left( B_z,\alpha \right) =\chi _{0}\left( \frac{%
mU}{4\pi }\right) ^{2}\frac{\left| \Delta \right| }{E_{F}}\mathcal{G}%
\left( \frac{|\alpha| k_{F}}{\Delta }\right) ,
\end{equation}
where
\begin{equation}
\mathcal{G}\left( x\right) =\frac{2x^{6}+23x^{4}+30x^{2}+12}{3\left(
1+x^{2}\right) ^{5/2}}
\end{equation}
has the following asymptotics
\begin{eqnarray}
\mathcal{G}\left( x\ll 1\right) &=&4+x^{4}/6+\dots  \notag \\
\mathcal{G}\left( x\gg 1\right) &=&\left( 2/3\right) |x|+6/|x|+\dots
\end{eqnarray}
For $\left| \Delta \right| \gg \left| \alpha \right| k_{F},$ the nonanalytic correction $\tilde{\chi}%
_{zz}\left( B_z,\alpha \right) $ reduces to the result of Ref. \onlinecite{maslov06_09}, obtained in the absence of the SOI, plus a correction term
\begin{equation}
\delta\chi^{(2)}_{zz}\left( B_z,\alpha \right) =4\chi _{0}\left( \frac{%
mU}{4\pi }\right) ^{2}\left[ \frac{\left| \Delta \right| }{E_{F}}+%
\frac{1}{24}\frac{\alpha ^{4}k_{F}^{4}}{\left| \Delta \right| ^{3}E_{F}}%
+\dots \right].
\label{chizzB>}
\end{equation}
In the opposite limit of $\left| \Delta \right| \ll \left| \alpha \right|
k_{F}$, the nonanalytic field dependence is cut off by the SOI
\begin{align}
\delta\tilde{\chi}_{zz}^{(2)}\left( B_z,\alpha \right) =\frac{2}{3}\chi
_{0}&\left( \frac{mU}{4\pi }\right) ^{2}\bigg[ \frac{\left| \alpha
\right| k_{F}}{E_{F}}\notag\\
&+9\frac{\Delta ^{2}}{|\alpha |k_{F}E_{F}}+\dots \bigg].
\label{eq:chizzB}
\end{align}

\subsection{\label{sec:SS2xx}In-plane magnetic field}
If the magnetic field is along the $x$-axis, $\boldsymbol{\Delta}%
=\mu _{B}B\mathbf{e}_{x}/2$, the effective Zeeman energies, $\bar{\Delta}%
_{\pm \mathbf{k}_{F}}\equiv \sqrt{\alpha ^{2}k_{F}^{2}\pm 2\alpha k_{F}\sin
\theta _{\mathbf{k}}\Delta +\Delta ^{2}}$, depends on the angle $\theta _{%
\mathbf{k}}$ between $\mathbf{k}$ and the direction of the field, chosen as
the $x$ axis. Coming back to Eq.~(\ref{eq:Xi12kDef_3}), we integrate first
over the fermionic frequencies, then over the magnitudes of the fermionic
momenta, then over the angle between $\mathbf{p}$ and $\mathbf{q,}$ and
finally over the angle between $\mathbf{q}$ and $\mathbf{k}$ (at fixed $%
\mathbf{k).}$ This yields
\begin{align}\label{eq:Xi2XX}
\delta \Xi _{xx}^{(2)}=-\frac{U^{2}}{16}T\sum_{\Omega }& \int \frac{%
d\theta _{\mathbf{k}}}{2\pi }\int \frac{dqq}{2\pi }  \notag \\
& \times \sum_{\{s_{i}\}}B_{s_{1}s_{3}}B_{s_{2}s_{4}}\Pi _{s_{1}s_{2}}^{+%
\mathbf{k}_{F}}\Pi _{s_{3}s_{4}}^{-\mathbf{k}_{F}},
\end{align}
where
\begin{align}\label{eq:PiAngleDep}
	\Pi _{ss^{\prime }}^{\pm \mathbf{k}_{F}}
	& \equiv \Pi _{ss^{\prime }}^{\pm\mathbf{k}_{F}}(\Omega ,q;\theta _{\mathbf{k}})
	={\sum_{K}}^\prime g_{s}^{\pm \mathbf{k}_{F}}(K)g_{s^{\prime }}^{\pm \mathbf{k}_{F}}(K+Q)
	\notag \\ & =\frac{m}{2\pi }\frac{|\Omega |}{\sqrt{\left[ \Omega +i(s^{\prime }-s)
	\bar{\Delta}_{\pm \mathbf{k}_{F}}\right] ^{2}+v_{F}^{2}q^{2}}},
\end{align}
with $g_{s}^{\pm \mathbf{k}_{F}}(K)=1/(i\omega -\xi_{\mathbf{k}}-s\bar{\Delta}_{\pm
\mathbf{k}_{F}})$, and $\sum'_{K}$ indicates that the integration
over $\theta _{\mathbf{k}}$ is excluded. The remaining integration over $%
\theta _{\mathbf{k}}$ is performed last, after integration over $q$ and
summation over $\Omega .$

\begin{widetext}
As in Sec.~\ref{sec:SS2zz}, it is convenient to define a new quantity
\begin{equation}
\mathcal{P}_{s}^{\pm \mathbf{k}_{F}}\equiv \mathcal{P}_{s}^{\pm \mathbf{k}%
_{F}}(\Omega ,q;\theta _{\mathbf{k}})=\frac{1}{\sqrt{\left( \Omega -2is\bar{%
\Delta}_{\pm \mathbf{k}_{F}}\right) ^{2}+v_{F}^{2}q^{2}}},
\end{equation}
and to re-write the thermodynamic potential as
\begin{equation}
\delta \Xi _{xx}^{(2)}=-\left( \frac{mU}{8\pi }\right)
^{2}T\sum_{\Omega }\Omega ^{2}\int \frac{d\theta _{\mathbf{k}}}{2\pi }\int
\frac{dqq}{2\pi }\sum_{ss^{\prime }}\left [B_{ss^{\prime }}B_{s,-s^{\prime }}\mathcal{P}_{0}(%
\mathcal{P}_{s^{\prime }}^{+\mathbf{k}_{F}}+\mathcal{P}_{s^{\prime }}^{-%
\mathbf{k}_{F}})+B_{ss^{\prime }}^{2}(\mathcal{P}_{0}^{2}+\mathcal{P}_{s}^{+%
\mathbf{k}_{F}}\mathcal{P}_{s^{\prime }}^{-\mathbf{k}_{F}})\right].\label{eq:Chi2XXProof}
\end{equation}
Subsequently, we sum over $s$ and $s^{\prime }$, add and subtract $4(\bar{%
\Delta}_{\mathbf{k}_{F}}^{2}\bar{\Delta}_{-k_{F}}^{2}-2\alpha
^{2}k_{F}^{2}\sin 2\theta _{\mathbf{k}})\mathcal{P}_{0}^{2}$ inside the
square brackets, and, after some algebraic manipulations, obtain
\begin{align}
\delta \Xi_{xx}^{(2)}=-2 \left( \frac{mU}{8\pi }\right)
^{2}&T\sum_{\Omega }\Omega ^{2}\int \frac{d\theta _{\mathbf{k}}}{2\pi }\int
\frac{dqq}{2\pi }\Big[8\mathcal{P}_{0}^{2} +a_{0}\mathcal{P}_{0}\left(\mathcal{P}_{+}^{-\mathbf{k}_{F}}+\mathcal{P}_{+}^{+%
\mathbf{k}_{F}}+\mathcal{P}_{-}^{-\mathbf{k}_{F}}+\mathcal{P}_{-}^{+\mathbf{k%
}_{F}}-4\mathcal{P}_{0}\right) \notag\\
 &+a_{+}\left(\mathcal{P}_{+}^{+\mathbf{k}_{F}}\mathcal{P}_{-}^{-\mathbf{k}_{F}}+%
\mathcal{P}_{+}^{-\mathbf{k}_{F}}\mathcal{P}_{-}^{+\mathbf{k}_{F}}-2\mathcal{%
P}_{0}^{2}\right)
+a_{-}\left(\mathcal{P}_{-}^{-\mathbf{k}_{F}}\mathcal{P}_{-}^{+\mathbf{k}_{F}}+%
\mathcal{P}_{+}^{-\mathbf{k}_{F}}\mathcal{P}_{+}^{+\mathbf{k}_{F}}-2\mathcal{%
P}_{0}^{2}\right)\Big],\label{eq:XiXXBrackets}
\end{align}
where
\begin{eqnarray}
a_{0}=\frac{4\alpha ^{2}k_{F}^{2}\Delta ^{2}\cos ^{2}\theta_{\mathbf{k}}}{%
\bar{\Delta}_{\mathbf{k}_{F}}^{2}\bar{\Delta}_{-\mathbf{k}_{F}}^{2}};\;
a_{\pm }=\frac{\alpha ^{4}k_{F}^{4}+\Delta ^{4}-2\alpha
^{2}k_{F}^{2}\Delta ^{2}\sin ^{2}\theta_{\mathbf{k}}\pm \bar{\Delta}_{\mathbf{k}_{F}}%
\bar{\Delta}_{-\mathbf{k}_{F}}(\alpha ^{2}k_{F}^{2}-\Delta ^{2})}{\bar{\Delta%
}_{\mathbf{k}_{F}}^{2}\bar{\Delta}_{-\mathbf{k}_{F}}^{2}}.
\end{eqnarray}
The first term in the square brackets in Eq.~(\ref{eq:XiXXBrackets}) does
not depend on the effective field $\bar{\Delta}_{\pm \mathbf{k}_{F}}$ and
can be dropped. The remaining integrals over $q$ are equal to
\begin{align}
\int dqq& \mathcal{P}_{0}[\mathcal{P}_{+}^{-\mathbf{k}_{F}}+\mathcal{P}%
_{+}^{+_{\mathbf{k}_{F}}}+\mathcal{P}_{-}^{-\mathbf{k}_{F}}+\mathcal{P}%
_{-}^{+\mathbf{k}_{F}}-4\mathcal{P}_{0}]
 =\frac{1}{v_{F}^{2}}\ln \frac{\Omega ^{2}}{\Omega ^{2}+\bar{\Delta}_{%
\mathbf{k}_{F}}^{2}}+\frac{1}{v_{F}^{2}}\ln \frac{\Omega ^{2}}{\Omega ^{2}+%
\bar{\Delta}_{-\mathbf{k}_{F}}^{2}},
\end{align}
\begin{align}
\int dqq&[\mathcal{P}_{+}^{+\mathbf{k}_{F}}\mathcal{P}_{-}^{-\mathbf{k}_{F}}+%
\mathcal{P}_{+}^{-\mathbf{k}_{F}} \mathcal{P}_{-}^{+\mathbf{k}_{F}}-2%
\mathcal{P}_{0}^{2}]
 =\frac{1}{v_{F}^{2}}\ln \frac{\Omega ^{2}}{\Omega ^{2}+(\bar{\Delta}_{%
\mathbf{k}_{F}}-\bar{\Delta}_{-\mathbf{k}_{F}})^{2}},
\end{align}
and
\begin{align}\label{eq:CombinationOfP}
\int dqq&[\mathcal{P}_{-}^{-\mathbf{k}_{F}}\mathcal{P}_{-}^{+\mathbf{k}_{F}}+%
\mathcal{P}_{+}^{-\mathbf{k}_{F}} \mathcal{P}_{+}^{+\mathbf{k}_{F}}-2%
\mathcal{P}_{0}^{2}]
 =\frac{1}{v_{F}^{2}}\ln \frac{\Omega ^{2}}{\Omega ^{2}+(\bar{\Delta}_{%
\mathbf{k}_{F}}+\bar{\Delta}_{-\mathbf{k}_{F}})^{2}}.
\end{align}
Evaluating the Matsubara sum in the same way as in Sec.~\ref{sec:SS2zz}, we
obtain
\begin{align}
\delta \Xi _{xx}^{(2)}= -2\left( \frac{mU}{8\pi v_{F}}\right) ^{2}T^{3}\int
\frac{d\theta_{\mathbf k}}{(2\pi )^{2}}\Bigg\{&a_{0}\left[\mathcal{F}\left(\frac{\bar{\Delta}_{\mathbf{k}%
_{F}}}{T}\right)+\mathcal{F}\left(\frac{\bar{\Delta}_{-\mathbf{k}_{F}}}{T}\right)\right] \notag \\
 &+a_{+}\mathcal{F}\left(\frac{|\bar{\Delta}_{\mathbf{k}_{F}}-\bar{\Delta}_{-\mathbf{k}%
_{F}}|}{T}\right)+a_{-}\mathcal{F}\left(\frac{\bar{\Delta}_{\mathbf{k}_{F}}+\bar{\Delta}_{-%
\mathbf{k}_{F}}}{T}\right)\Bigg\},  \label{eq:Xixx}
\end{align}
\end{widetext}
where the function $\mathcal{F}\left( x\right) $ and its asymptotic limits
are given by Eqs. (\ref{eq:FDef}-\ref{Flarge}).

The angular integral cannot be performed analytically because the function $%
\mathcal{F}(y)$ depends in a complicated way on the angle $\theta_{\mathbf{k}}$
through $\bar{\Delta}_{\pm \mathbf{k}_{F}}$. Therefore, we consider two limiting cases
below.

\subsubsection{Temperature dependence of $\protect\chi _{xx}$}

First, we consider the limit of a weak magnetic field: $|\Delta |\ll \max\{|\alpha
|k_{F},T$\}. The main difference between the cases of in- and transverse
orientations of the field is in the term proportional to $a_{+}$ in Eq. (\ref{eq:Xixx}). The
argument  $\mathcal{F}$ in this term vanishes in the limit of
$\Delta \rightarrow 0$, whereas the arguments of $\mathcal{F}$ in the rest of the terms reduce to a scaling variable $|\alpha |k_{F}/T$, as it was
also the case for the transverse field. Therefore, the SOI energy scale, $%
|\alpha |k_{F},$ and temperature are interchangeable in the rest of the
terms, which means that a nonanalytic $T$ dependence arising from these terms is
cut off by the SOI (and vice versa). However, the $a_{+}$ term does
not depend on $\alpha $ and produces a nonanalytic $T$ dependence which is
\emph{not }cut off by the SOI. To see this, we expand prefactors $%
a_{0}$ and $a_{\pm }$ to leading order in $\Delta $ as
\begin{eqnarray}
a_{0} &=&4\frac{\Delta ^{2}}{\alpha ^{2}k_{F}^{2}}\cos ^{2}\theta _{\mathbf{k%
}}+\mathcal{O}(\Delta ^{4}),  \notag \\
a_{+} &=&2-4\frac{\Delta ^{2}}{\alpha ^{2}k_{F}^{2}}\cos ^{2}\theta _{%
\mathbf{k}}+\mathcal{O}(\Delta ^{4}),  \notag \\
a_{-} &=&2\frac{\Delta ^{4}}{\alpha ^{4}k_{F}^{4}}\cos ^{4}\theta _{\mathbf{k%
}}+\mathcal{O}(\Delta ^{6}).
\end{eqnarray}
The last term in Eq. (\ref{eq:Xixx}), proportional to $a_{-}$, does not
contribute to order $\Delta ^{2}$, and we focus on the first two terms. In
the term $a_{0}$, we replace $\bar{\Delta}_{\pm \mathbf{k}_{F}}=\left| \alpha
\right| k_{F}$ in the argument of the $\mathcal{F}$ function. In the $a_{+}$
term, we replace $|\bar{\Delta}_{\bm{k}_{F}}-\bar{\Delta}_{-\bm{k}_{F}}|=2|\Delta \sin
\theta _{\mathbf{k}}|+\mathcal{O}(\Delta ^{2})$, and then expand $\mathcal{F}%
\left( 2|\Delta \sin \theta _{\mathbf{k}}|/T\right) =4\Delta ^{2}\sin
^{2}\theta _{\mathbf{k}}/T^{2}$ using Eq. (\ref{Fsmall}). Integrating over $%
\theta_{\mathbf{k}}$ and differentiating the result twice with respect to $B,$ we obtain
\begin{eqnarray}
\delta \chi _{xx}^{(2)}\left( T,\alpha \right) &=&\chi _{0}\left( \frac{%
mU}{4\pi }\right) ^{2}\left[ \frac{T^{3}}{\alpha ^{2}k_{F}^{2}}%
\mathcal{F}\left( \frac{\left| \alpha \right| k_{F}}{T}\right) +\frac{T}{%
E_{F}}\right]  \notag \\
&=&\frac{1}{2}\chi _{zz}^{(2)}\left( T,\alpha \right) +\frac{1}{2}\chi
_{0}\left( \frac{mU}{4\pi }\right) ^{2}\frac{T}{E_{F}},
\label{eq:Chi2xxres}
\end{eqnarray}
where $\chi _{zz}^{(2)}\left( T,\alpha \right) $ is the correction to $\chi
_{zz}$ given by Eq. (\ref{eq:Chi2ZZRes}). [Notice a remarkable similarity
between Eq. (\ref{eq:Chi2xxres}) and the relation between $\chi _{xx}$ and $%
\chi _{zz}$ in the non-interacting case, Eq. (\ref{eq:ChiXXvsZZ}).] Equation
(\ref{eq:Chi2xxres}) is one of the main results of this paper. It shows that
a nonanalytic $T$ dependence of $\chi _{xx}$, given by the stand-alone $%
T/E_{F}$ term survives in the presence of the SOI. Explicitly, the $%
T $ dependence is
\begin{equation}
\delta \chi _{xx}^{(2)}= 2\chi _{0}\left( \frac{mU}{4\pi }%
\right) ^{2}\left[ \frac{T}{E_{F}}+\frac{\alpha ^{2}k_{F}^{2}}{48TE_{F}}\right]   \label{eq:ChiXX2ndT}
\end{equation}
for $T\gg |\alpha |k_{F}$ and
\begin{equation}
\delta \chi _{xx}^{(2)}= 2\chi _{0}\left( \frac{mU}{4\pi }%
\right) ^{2}\left[ \frac{|\alpha |k_{F}}{6E_{F}}+\frac{T}{2E_{F}}+2\zeta (3)%
\frac{T^{3}}{\alpha ^{2}k_{F}^{2}E_{F}}\right]   \label{eq:ChiXX2ndAlpha}
\end{equation}
for $T\ll |\alpha |k_{F}$.

As it is also the case for the transverse magnetic field, the first term in the
second line of Eq.~(\ref{eq:ChiXX2ndT}) is due to particle-hole pairs formed
by electrons from three identical and one different Rashba branches. On the
other hand, the linear-in-$T$ term, absent in $\delta\chi _{zz}^{(2)}$, is
comes from processes involving electrons from the different Rashba branches in each
particle-hole bubble, see Fig.~\ref{fig:Rashba}b. Indeed, pairing electrons and holes,
which belong to different Rashba branches and move in the same direction, we
obtain particle-hole bubbles $\mathcal{P}_{\pm }^{\pm \mathbf{k}_{F}}$
[cf. Eq.~(\ref{eq:XiXXBrackets})].
The product of two
such bubbles, $\mathcal{P}_{+
}^{\pm \mathbf{k}_{F}}\mathcal{P}_{-}^{\mp \mathbf{k}_{F}}$, being integrated over $q$ and summed over $\Omega $,
depends on the
difference of the Zeeman energies $|{\bar{\Delta}}_{\mathbf{k}_{F}}-{\bar{\Delta}}_{-\mathbf{k}%
_{F}}|$. Since $\sin \theta _{\mathbf{k}}$ is odd upon $\mathbf{%
k}\rightarrow \mathbf{-k}$, this difference is finite and proportional to $%
|\Delta |$ for $\Delta \rightarrow 0$ but does not depend on $\alpha $. This
is a mechanism by which one gets an $\mathcal{O}(\Delta ^{2})$ contribution
to the thermodynamic potential and, therefore, a $T$ dependent contribution to $\chi _{xx}$,
which does not involve the SOI.

\subsubsection{Magnetic-field dependence of $\protect\chi_{xx}$}

Now we analyze the non-linear in-plane susceptibility ${\chi}%
_{xx}(B_x,\alpha )=-\partial^2\Xi_{xx}/\partial B_x^2$ at  $T=0$. Replacing  $\mathcal{F}$ in
Eq.~(\ref{eq:Xixx}) by its large-argument asymptotics from Eq.~(\ref{Flarge}), we obtain
\begin{widetext}
\begin{align}
\delta \Xi _{xx}^{(2)}= -\frac{2}{3}\left( \frac{mU}{8\pi v_{F}}\right) ^{2}\int
\frac{d\phi _{kx}}{(2\pi )^{2}}\left\{a_{0}\left[\bar{\Delta}_{\mathbf{k}%
_{F}}^3+\bar{\Delta}_{-\mathbf{k}_{F}}^3\right]
 +a_{+}|\bar{\Delta}_{\mathbf{k}_{F}}-\bar{\Delta}_{-\mathbf{k}%
_{F}}|^3+a_{-}\left[\bar{\Delta}_{\mathbf{k}_{F}}+\bar{\Delta}_{-%
\mathbf{k}_{F}}\right]^3\right\}. \label{eq:XixxT0}
\end{align}
\end{widetext}
The angular integral can now be solved explicitly in the limiting
cases of $|\Delta |\ll |\alpha |k_{F}$ and $|\Delta |\gg |\alpha |k_{F}$.
Since our primary interest is just to see whether a nonanalytic
field-dependence survives in the presence of the SOI, we will
consider only the weak-field case: $|\Delta |\ll |\alpha |k_{F}$. The
$a_{-}$ term in Eq.~(\ref{eq:XixxT0}) can then be dropped, while the $a_{0}$ and $a_{+}$ ones yield
\begin{equation}
\delta {\chi}_{xx}^{(2)}\left( B_x,\alpha \right) =\frac{1}{3}\chi
_{0}\left( \frac{mU}{4\pi }\right) ^{2}\left[ \frac{\left| \alpha
\right| k_{F}}{E_{F}}+\frac{16}{\pi }\frac{|\Delta |}{E_{F}}\right] .
\label{eq:chixxB}
\end{equation}
The first term in Eq.~(\ref{eq:chixxB}) is just half of the first term in $%
\delta {\chi}_{zz}^{(2)}$ [cf. Eq.~(\ref{eq:chizzB})]. However, the
second term represents a nonanalytic dependence on the field which is not
cut off by the SOI.

\subsection{\label{sec:Diags}Remaining second order diagrams}

Besides the diagrams considered so far, there are other second order
diagrams, which -- in principle -- could contribute to the spin
susceptibility. These diagrams are depicted in Fig.~\ref{fig:Diagrams}$b$-$e$. In the
absence of the SOI, these diagrams are irrelevant because the
electron-electron interaction conserves spin. This means that the spins of
electrons in each of the bubbles in, e.g,  diagram $b$) are the same and,
therefore, the Zeeman energies, entering the Green's functions, can be absorbed into the chemical potential.
The same argument also goes for the other two diagrams. In the presence of
the SOI, this argument does not work because spin is not a good quantum number
and the interaction mixes states from all Rashba
branches with different Zeeman energies. However, one can show that
the net result is the same as without the SOI: diagrams in Fig. \ref{fig:Diagrams}$b$-$e$
do not contribute to the nonanalytic behavior of the spin
susceptibility. This is what we are going to show in this section.

We begin with diagram $b$), which is a small momentum-transfer
counterpart of diagram in $a)$:
\begin{equation}
\delta \Xi _{b}^{(2)}= -U^{2}(0)T\sum_{Q}\left[\mathrm{Tr}\hat{\Pi}(Q)%
\right]^2,
\end{equation}
where
\begin{eqnarray}
\hat{\Pi}(Q)&=&\frac{1}{2}\sum_{K}\hat{G}(K)\hat{G}(K+Q)  \notag \\
&=&\frac{1}{2}\sum_{s,t=\pm 1}\hat{\Omega}_{s}\hat{\Omega}_{t}\sum_{K}g_{s}%
\left( K\right) g_{t}\left( K+Q\right)
\end{eqnarray}
is the full (summed over Rashba branches) polarization bubble, and both the
space- and time-like components of $Q\equiv(\Omega,\mathbf{q})$ are
small. As it is also the case in the absence of the SOI, the small-$Q$ bubble does not depend on the magnetic field. Indeed, noticing that the
matrix $\hat{\zeta}$ in Eq.~(\ref{eq:OmegaDef}) has the following properties
\begin{equation}
\hat{\zeta}^{2}=\hat{I}\; \mathrm{and}\; \mathrm{Tr}\hat{\zeta}=0,
\label{eq:proj}
\end{equation}
it is easy to show that
\begin{equation}
\hat{\Omega}_{s}\hat{\Omega}_{t}=\frac{1}{4}\left[ (1+st)\hat{I}+\left(
s+t\right) \hat{\zeta}\right].
\end{equation}
Consequently, $\hat{\Omega}_{+}\hat{\Omega}_{+}=$ $\hat{\Omega}_{-}\hat{%
\Omega}_{-}=(1/2)\hat{I}$ and $\hat{\Omega}_{+}\hat{\Omega}_{-}=\hat{\Omega}%
_{-}\hat{\Omega}_{+}=0.$ Therefore, electrons from different branches do not
contribute to $\hat{\Pi}(Q)$, while the Zeeman energies in the contributions
from the same branch can be absorbed into the chemical potential. As
result, the dynamic part of the bubble depends neither on the field nor on
the SOI (as long as a weak dependence of the Fermi velocity for a
given branch on $\alpha $ is neglected):
\begin{equation}
\hat{\Pi}(Q)=\hat{I}\frac{m}{2\pi }\frac{\left| \Omega \right| }{\sqrt{%
\Omega ^{2}+v_{F}^{2}q^{2}}}.  \label{ld}
\end{equation}
Therefore diagram $b$) does not contribute to the spin susceptibility. We
remind the reader that, since there are no threshold-like singularities in
the static polarization bubble (see discussion in Sec.~\ref{sec:SS}), the
Landau-damping singularity in Eq. ~(\ref{ld}) is the only singularity which
may have contributed to a nonanalytic behavior of the spin susceptibility.
However, as we have just demonstrated, Landau damping is not effective in
diagrams with small momentum transfers.

Similarly, diagram $c$) with two crossed interaction lines, one of which carries a small momentum and the other one carries a momentum near $2k_F$, is expressed via a small $Q$ bubble as
\begin{equation}
\delta \Xi _{c}^{(2)}= -U(0)U(2k_F)\sum_{Q}\mathrm{Tr}\left[\hat{\Pi}%
^{2}(Q)\right]
\end{equation}
and, therefore, does not depend on the magnetic field. Diagram $d$), with both interaction lines carrying small momenta,
contains a trace of four Green's functions
\begin{widetext}
\begin{eqnarray}
\delta\Xi _{d}^{\left( 2\right) } &=&-\frac{U^{2}(0)}{4}\!\!\!\sum_{Q,Q^{\prime },K}\!\!\!
\text{Tr}\left[ \hat{G}\left( K\right) \hat{G}(K+Q)\hat{G}(K+Q+Q^{\prime })%
\hat{G}(K+Q^{\prime })\right]  \\
&=&-\frac{U^{2}(0)}{4}\sum_{Q,Q^{\prime },K}\sum_{p,r,s,t=\pm}\text{Tr}\left[
\hat{\Omega}_{p}\hat{\Omega}_{r}\hat{\Omega}_{s}\hat{\Omega}_{t}\right]
g_{p}\left( K\right) g_{r}\left( K+Q\right) g_{s}\left( K+Q+Q^{\prime }\right)
g_{t}\left( K+Q^{\prime }\right),
\end{eqnarray}
\end{widetext}
where $Q$ and $Q^{\prime}$ are small, so that dependence of  $\hat\Omega_{l}$ on either of the bosonic momenta can be neglected. Using again the properties of the projection operator from Eq.~(\ref{eq:proj}), we find that
\begin{widetext}
\begin{eqnarray}
\mathrm{Tr}\left[\hat{\Omega}_{p}\hat{\Omega}_{r}\hat{\Omega}_{s}\hat{\Omega}_{t}\right]&=&\frac{1}{16}\mathrm{Tr}%
\left[ \left\{ \left(1+pr\right) \left( 1+st\right)+(p+r)(s+t)\right\}{\hat I} +\left\{ \left( 1+pr\right)
\left( s+t\right) +\left( 1+st\right) \left( p+r\right)\right\} \hat{\zeta}\right]\notag\\
&=&\frac{1}{8}\left[\left( 1+pr\right) \left( 1+st\right)+(p+r)(s+t)\right].
\end{eqnarray}
\end{widetext}
This expression vanishes if at least one of the indices from the set $%
\left\{ p,r,s,t\right\} $ is different from the others. Therefore, only
electrons from the same branch contribute to $\delta\Xi _{d}^{\left( 2\right) }$,
the Zeeman energy can again be absorbed into the chemical potential, and $%
\delta\Xi _{d}^{\left( 2\right) }$ does not depend on the magnetic field.

Finally, the last diagram, $\delta\Xi _{e}^{(2)}$ corresponds to the
first-order self-energy inserted twice into the zeroth order thermodynamic
potential. Such an insertion only shifts the chemical potential and, for a $q$ dependent $U$,
gives a regular correction to the electron effective mass but
does not produce any nonanalytic behavior.

\begin{figure}[t]
\includegraphics[width=.4\textwidth]{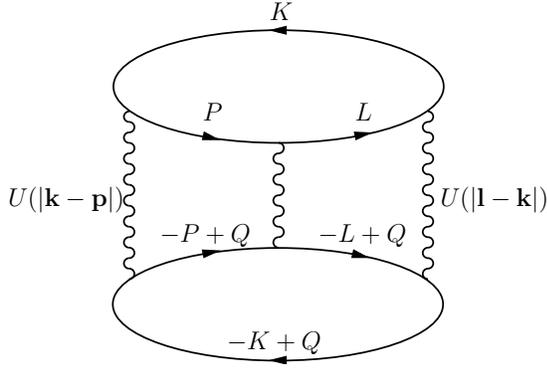}
\caption{The third-order Cooper-channel diagram for the thermodynamic
potential.}\label{fig:Chi3rd}
\end{figure}

\section{\label{sec:Inf}
Cooper-channel renormalization}

\subsection{General remarks}

\label{sec:remarks}

The second-order nonanalytic contribution to the spin susceptibility comes
from ``backscattering''\/ processes in which two fermions moving in almost
opposite directions experience almost complete backscattering. Because the
total momentum of two fermions is small, the backscattering process is a
special case of the interaction in the Cooper (particle-particle) channel.
Higher order processes in this channel lead to logarithmic renormalization
of the second-order result.~\cite{chubukov05b,chubukov06,maslov06_09,chubukov07,aleiner06,PhysRevB.74.205122,ProcNatlAcadSci.103.15765,schwiete06,PhysRevB.77.045108,PhysRevLett.98.156401,PhysRevB.79.115445}
For a weak interaction, considered throughout this paper, this is the
leading higher-order effect. In the absence of the SOI, resummation
of all orders in the Cooper channel leads to the following scaling form of
the spin susceptibility
\begin{equation}
\delta\chi \propto \frac{E}{\ln ^{2}\left(E/\Lambda\right)},  \label{eq:cooperinfnoso}
\end{equation}
where $\Lambda $ is the ultraviolet cutoff and $E\equiv \max \{T,|\Delta
|\}$ is small enough so that $mU|\ln \left( E/\Lambda \right) |\gg 1$. As we
see, both the linear-in-$E$ term, which occurs already at second order,
and its logarithmic renormalization contain the same energy scale. The
reason for this symmetry is very simple: an arbitrary order Cooper diagram
for the thermodynamic potential contains two bubbles joined by a ladder. The
spin susceptibility is determined only by diagrams with opposite fermion
spins in each of the bubbles. Therefore, all Cooper bubbles in such
diagrams are formed by fermions with opposite spins, so that the logarithmic
singularity of the bubble is cut off at the largest of the two energy
scales, i.e, temperature or Zeeman energy. At zero incoming momentum and
frequency, the Cooper bubble is
\begin{equation}
\Pi _{C}^{\uparrow \downarrow }=\frac{m}{2\pi }\ln \frac{\Lambda }{\max
\left\{ T,\Delta \right\} },  \label{picupdown}
\end{equation}
hence the symmetry of the result with respect to interchanging $T$ and $%
\Delta $ follows immediately. It will be shown in this section that this
symmetry does not hold in the presence of the SOI. The reason is that the
Rashba branches are not the states with definite spins, and diagrams with
Cooper bubbles formed by electrons from the same branches also contribute to
the spin susceptibility. Although a Cooper bubble formed by electrons from
branches $s$ and $s^{\prime }$
\begin{eqnarray}
\Pi _{C}^{s,s^{\prime }} &=&T\sum_{\omega }\frac{m}{2\pi }\int \frac{d\theta
_{\mathbf{p}}}{2\pi }\int d\xi _{p}g_{s}(\omega ,\mathbf{p})g_{s^{\prime
}}(-\omega ,-\mathbf{p})  \notag \\
&=&\frac{m}{2\pi }\ln \frac{\Lambda }{\max \left\{ T,\left| s-s^{\prime
}\right| \bar{\Delta}_{\mathbf{k}_{F}}\right\} }  \label{eq:TLn}
\end{eqnarray}
looks similar to that in the absence of the SOI [Eq. (\ref{picupdown}%
)], its diagonal element $\Pi _{C}^{ss}$ depends only on $T$ even if $T\ll
\bar{\Delta}_{\mathbf{k}_{F}}.$ Therefore, for $\Delta =0,$ the Cooper
logarithm in Eq. (\ref{eq:cooperinfnoso}) will depend only on $T$ in the
limit of $T\rightarrow 0$ while the energy $E$ in the numerator may be
given either by $T$ or by $|\alpha|k_{F}$.

\subsection{Third-order Cooper channel contribution to $\protect\chi _{zz}$}
\label{sec:cooper3}

Is this section, we obtain the third-order Cooper channel contribution to $
\chi _{zz}$. This calculation will help to understand the general strategy
employed later, in Secs.~\ref{sec:InfZZ} and \ref{sec:InfXX}, in resumming Cooper diagrams to all orders.

The third-order Cooper diagram for the thermodynamic potential, depicted in
Fig.~\ref{fig:Chi3rd}, is given by
\begin{align}
\delta \Xi ^{(3)}_{zz}= \frac{U^{3}}{6}\sum_{K,P,L,Q}&\mathrm{Tr}(\hat{G}_{K}%
\hat{G}_{P}\hat{G}_{L})\notag \\
&\times\mathrm{Tr}(\hat{G}_{-K+Q}\hat{G}_{-P+Q}\hat{G}_{-L+Q}).
\end{align}
First, we evaluate the traces $\mathrm{Tr}(\hat{G}_{K}\hat{G}_{P}\hat{G}%
_{L})=\sum_{rst}\mathcal{B}_{rst}g_{r}(K)g_{s}(P)g_{t}(L)$ with the
coefficients
\begin{widetext}
\begin{align}
\mathcal{B}_{rst}\equiv\mathrm{Tr}[\hat{\Omega}_{r}(\mathbf{k})\hat{\Omega}%
_{s}(\mathbf{p})\hat{\Omega}_{t}(\mathbf{l})]=& \frac{1}{4\bar{\Delta}_{%
\mathbf{k}}\bar{\Delta}_{\mathbf{p}}\bar{\Delta}_{\mathbf{l}}}[\bar{\Delta}_{%
\mathbf{k}}\bar{\Delta}_{\mathbf{p}}\bar{\Delta}_{\mathbf{l}}+irst\alpha
^{2}\Delta (\mathbf{k}\times \mathbf{p}+\mathbf{p}\times \mathbf{l}+\mathbf{l%
}\times \mathbf{k})_{z}  \notag \\
& +rs(\alpha ^{2}\mathbf{k}\cdot \mathbf{p}+\Delta ^{2})\bar{\Delta}_{%
\mathbf{l}}+rt(\alpha ^{2}\mathbf{k}\cdot \mathbf{l}+\Delta ^{2})\bar{\Delta}%
_{\mathbf{p}}+st(\alpha ^{2}\mathbf{l}\cdot \mathbf{p}+\Delta ^{2})\bar{%
\Delta}_{\mathbf{k}}].
\end{align}
\end{widetext}
Since $\mathbf{q}$ is small and $\mathcal{B}_{rst}$ is an even function of the fermionic momenta,
\begin{eqnarray}
&&\mathrm{Tr}\left[\hat\Omega_{r'}(-\mathbf{k}+\mathbf{q})\hat\Omega_{s'}(-\mathbf{p}+\mathbf{q})\hat\Omega_{t'}(-\mathbf{l}+\mathbf{q})\right]\notag\\
&&\approx \mathrm{Tr}\left[\hat\Omega_{r'}(-\mathbf{k})\hat\Omega_{s'}(-\mathbf{p})\hat\Omega_{t'}(-\mathbf{l})\right]=\mathcal{B}_{r's't'},
\end{eqnarray}
and the thermodynamic potential becomes
\begin{eqnarray}
\delta \Xi _{zz}^{(3)}&=& \frac{U^{3}}{6}\sum_{K,P,L,Q}\sum_{rst}%
\sum_{r^{\prime }s^{\prime }t^{\prime }}\mathcal{B}_{rst}\mathcal{B}%
_{r^{\prime }s^{\prime }t^{\prime }}g_{r}(K)g_{r^{\prime
}}(-K+Q)\notag\\
&&\times g_{s}(P)g_{s^{\prime }}(-P+Q)g_{t}(l)g_{t^{\prime }}(-L+Q).
\end{eqnarray}
Each pair of the Green's functions with opposite momenta forms a Cooper
bubble, which depends logarithmically on the largest of
the two energy scales--temperature or the effective Zeeman energy, see Eq. (%
\ref{eq:TLn}). The third-order contribution contains one such logarithmic
factor which can be extracted from any of the three Cooper bubbles; this
gives an overall factor of three:
\begin{eqnarray}
\delta \Xi _{zz}^{(3)}&=& \frac{U^{3}}{2}\sum_{K,P,Q}\sum_{rst}\sum_{r^{%
\prime }s^{\prime }t^{\prime }}\int \frac{d\theta _{\mathbf{kl}}}{2\pi }%
\mathcal{B}_{rst}\mathcal{B}_{r^{\prime }s^{\prime }t^{\prime
}}\Pi
_{C}^{t,t^{\prime }}\notag\\&&\times
g_{r}(K)g_{r^{\prime }}(-K+Q)g_{s}(P)g_{s^{\prime }}(-P+Q),
\end{eqnarray}
where $\theta _{\mathbf{kl}}\equiv \angle (\mathbf{k},\mathbf{l}).$ With
this procedure, the third-order diagram reduces effectively to the second-order
one, but with a new set of coefficients. Since we already know that
the nonanalytic part of the second-order diagram comes from processes with $%
\mathbf{k}\approx -\mathbf{p}$, the coefficient $\mathcal{B}_{rst}$ (and
its primed counterpart) simplify significantly because $\mathbf{k}\times
\mathbf{p}=\mathbf{0}$ and $\mathbf{p}\times \mathbf{l}=\mathbf{l}\times
\mathbf{k}$. Integrating over $\theta _{\mathbf{kl}},$ we obtain
\begin{widetext}
\begin{align}
\mathcal{A}_{rstr^{\prime }s^{\prime }t^{\prime }}\equiv & \int \frac{%
d\theta _{\mathbf{kl}}}{2\pi }\mathcal{B}_{rst}\mathcal{B}_{r^{\prime
}s^{\prime }t^{\prime }}=\frac{1}{16\bar{\Delta}_{\mathbf{k}_{F}}^{6}}\Big\{%
-2rstr^{\prime }s^{\prime }t^{\prime }\alpha ^{4}k_{F}^{4}\Delta ^{2}+\frac{1%
}{2}(rtr^{\prime }t^{\prime }+sts^{\prime }t^{\prime }-rts^{\prime
}t^{\prime }-r^{\prime }t^{\prime }st)\alpha ^{4}k_{F}^{4}\bar{\Delta}_{%
\mathbf{k}_{F}}^{2}  \notag \\
& +\bar{\Delta}_{\mathbf{k}_{F}}^{2}[\bar{\Delta}_{\mathbf{k}%
_{F}}^{2}+(rs+rt+st)\Delta ^{2}-rs\alpha ^{2}k_{F}^{2}][\bar{\Delta}_{%
\mathbf{k}_{F}}^{2}+(r^{\prime }s^{\prime }+r^{\prime }t^{\prime }+s^{\prime
}t^{\prime })\Delta ^{2}-r^{\prime }s^{\prime }\alpha ^{2}k_{F}^{2}]\Big\},
\end{align}
and
\begin{equation}
\delta \Xi _{zz}^{(3)} = \frac{U^{3}}{2}\sum_{rst}\sum_{r^{\prime
}s^{\prime }t^{\prime }}\mathcal{A}_{rstr^{\prime }s^{\prime }t^{\prime
}}\Pi _{C}^{t,t^{\prime }} \sum_{K,P,Q}g_{r}(K)g_{r^{\prime}}(-K+Q)g_{s}(P)g_{s^{\prime }}(-P+Q).
\label{mon1}
\end{equation}
The convolution of two Cooper bubbles in Eq. (\ref{mon1}) can be re-written via a convolution of two particle-hole bubbles by relabeling $Q\rightarrow Q+K+P$:
\begin{eqnarray}
\sum_{K,P,Q}g_{r}(K)g_{r^{\prime }}(-K+Q)g_{s}(P)g_{s^{\prime }}(-P+Q)
=\sum_{Q}\sum_{K}g_{r}(K)g_{s^{\prime }}(K+Q)\sum_{P}g_{s}(P)g_{r^{\prime}}\left( P+Q\right)
\sum_{Q}\Pi _{rs^{\prime }}(Q)\Pi _{sr^{\prime }}(Q),
\end{eqnarray}
where $\Pi _{s,s^{\prime }}(Q)$ is a particle-hole bubble defined in Eq.~(\ref{piph}). Summing over the Rashba indices, integrating over the momentum, and
assuming that the Zeeman energy is the smallest energy in the problem, i.e., that
$\Delta \ll \max\{T,\left| \alpha \right| k_{F}\}$, we find
\begin{align}
\delta \Xi _{zz}^{(3)}=-\frac{1}{2\pi v_{F}^{2}}\left( \frac{mU}{4\pi }%
\right) ^{3}\frac{\Delta ^{2}T^{3}}{\alpha ^{2}k_{F}^{2}}\Bigg\{&
\left[ 12\mathcal{F}\left( \frac{|\alpha |k_{F}}{T}\right) -\mathcal{F}%
\left( \frac{2|\alpha |k_{F}}{T}\right) \right] \ln \frac{T}{\Lambda } \notag\\
&+\left[ 4\mathcal{F}\left( \frac{|\alpha |k_{F}}{T}\right) +\mathcal{F}%
\left( \frac{2|\alpha |k_{F}}{T}\right) \right] \ln \frac{\max \{T,|\alpha |k_F\}}{\Lambda }\Bigg\}
\label{eq:ChiXX3rdRes}
\end{align}
with $\mathcal{F}(y)$ given by Eq.~(\ref{eq:FDef}).
\end{widetext}

The asymptotic behavior of $\chi_{zz}$ for $T\gg |\alpha |k_{F}$ is computed from Eq.~(\ref{eq:ChiXX3rdRes})
\begin{equation}
\delta \chi _{zz}^{(3)}=
8\chi _{0}\bigg(\frac{mU}{4\pi }\bigg)^{3}\frac{T}{%
E_{F}}\ln \frac{T}{\Lambda}
\end{equation}
and, as to be expected,  $\delta \chi _{zz}^{(3)}$ scales as $T\ln T$.

In the opposite limit of $T\ll |\alpha|k_F$,
\begin{align}
\delta\chi_{zz}^{(3)} &=
\frac{2}{3}\chi_0\bigg(\frac{mU}{4\pi}\bigg)^3
\frac{|\alpha|k_F}{E_F}\bigg(\ln\frac{T}{\Lambda}+3\ln\frac{|\alpha|k_F}{\Lambda}%
\bigg)  \notag \\
&\approx
\frac{2}{3}\chi_0\bigg(\frac{mU}{4\pi}\bigg)^3\frac{|\alpha|k_F}{%
E_F}\ln\frac{T}{\Lambda},
\end{align}
since $|\ln(T/\Lambda)|\gg|\ln(|\alpha|k_F/\Lambda)|$. As it was advertised in Sec.~\ref{sec:remarks},
the $T\ln T$ scaling at high temperatures is replaced by the $|\alpha|\ln T$ scaling at low temperatures
which implies that the energy scales $|\alpha|k_F$ and $T$ are not interchangeable.

\subsection{\label{sec:Resum} Re-summation of all diagrams in the Cooper channel}
\subsubsection{\label{sec:Resum_1}Scattering amplitude in the chiral basis}

\begin{figure}[t]
\includegraphics[width=.45\textwidth]{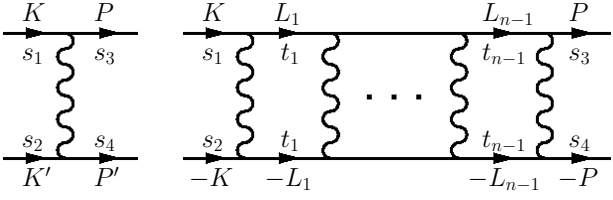}
\caption{Left: the effective scattering amplitude $\Gamma
_{s_{1}s_{2};s_{3}s_{4}}^{(1)}(\mathbf{k,k}^{\prime };\mathbf{p,p}^{\prime
}) $ in the chiral basis. Right: a generic $n$-th order ladder diagram in
the Cooper channel, $\Gamma _{s_{1}s_{2};s_{3}s_{4}}^{(n)}(\mathbf{k},-%
\mathbf{k}; \mathbf{p},-\mathbf{p}) $.}
\label{fig:Ladder}
\end{figure}

It is more convenient to resum the Cooper ladder diagrams in the chiral basis,
in which the Green's functions are
diagonal. Introducing Rashba spinors $\vert\mathbf{k},s\rangle$, we
re-write the number-density operator as
\begin{equation}
	\hat{\rho}_{\mathbf{q}}=\sum_{\mathbf{k}}\sum_{s_{1},s_{2}}\langle
	\mathbf{k}+\mathbf{q},s_{2}\vert\mathbf{k},s_{1}\rangle
	 \hat{c}_{\mathbf{k+q},s_{2}}^{\dagger }\hat{c}_{\mathbf{k},s_{1}},
\end{equation}
so that the Hamiltonian of the four-fermion interaction becomes
\begin{widetext}
\begin{align}
\hat{H}_{\text{int}}=\frac{1}{2}\sum_{\mathbf{q}}U\left( \mathbf{q}\right) \hat{\rho}%
_{\mathbf{q}}\hat{\rho}_{-\mathbf{q}}
 =\frac{1}{2}\sum_{\mathbf{p,p}^{\prime },\mathbf{k,k}^{\prime
}}\sum_{\{s_{i}\}}\Gamma _{s_{1}s_{2};s_{3}s_{4}}^{(1)}(\mathbf{k,k}^{\prime
};\mathbf{p,p}^{\prime })\hat{c}_{\mathbf{p}^{\prime },s_{4}}^{\dagger }\hat{%
c}_{\mathbf{p},s_{3}}^{\dagger }\hat{c}_{\mathbf{k},s_{1}}\hat{c}_{%
\mathbf{k}',s_{2}},
\end{align}
\end{widetext}
where the effective scattering amplitude is defined by (cf. Fig. \ref{fig:Ladder})
\begin{eqnarray}\label{eq:GammaDef}
\Gamma _{s_{1}s_{2};s_{3}s_{4}}^{(1)}(\mathbf{k,k}^{\prime };\mathbf{p,p}%
^{\prime }) &=&U\left( \mathbf{k-p}\right) \langle\mathbf{p},s_{3}\vert\mathbf{k}
,s_{1}\rangle \langle\mathbf{p}^{\prime },s_{4}|\mathbf{k}',s_{2}\rangle.  \notag \\
&&
\end{eqnarray}
In the absence of the magnetic field,
\begin{equation}
\vert\mathbf{k,}s\rangle=\frac{1}{\sqrt{2}}\left(
\begin{array}{c}
-ise^{-i\theta _{\mathbf{k}}} \\
1
\end{array}
\right) ,
\end{equation}
where $\theta _{\mathbf{k}}\equiv \angle \left( \mathbf{k,}\hat{x}\right) ,$
and Eq.~(\ref{eq:GammaDef}) gives~\cite{Gorkov01}
\begin{eqnarray}\label{eq:Gamma1}
\Gamma _{s_{1}s_{2};s_{3}s_{4}}^{(1)}(\mathbf{k,k'};\mathbf{p,p}%
^{\prime }) &=&\frac{1}{4}U\left( \mathbf{k-p}\right) \left[
1+s_{1}s_{3}e^{i\left( \theta _{\mathbf{p}}-\theta _{\mathbf{k}}\right) }\right]  \notag \\
&&\times \left[ 1+s_{2}s_{4}e^{i\left( \theta _{\mathbf{p}'}-\theta _{\mathbf{%
k}^{\prime }}\right) }\right].
\end{eqnarray}

\begin{figure}[t]
\includegraphics[width=.4\textwidth]{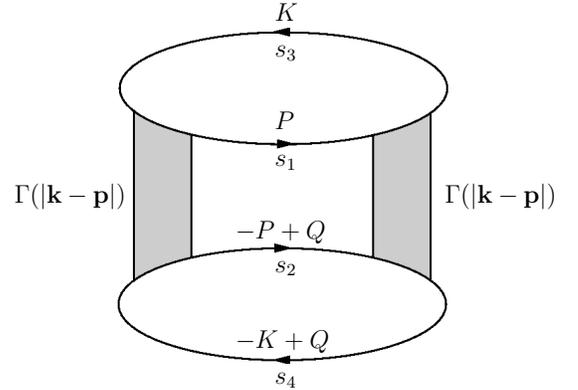}
\caption{A skeleton diagram for the thermodynamic potential~$\Xi$.}
\label{fig:DiagDressed}
\end{figure}

In order to resum the ladder diagrams for the thermodynamic potential to infinite order, we consider a
skeleton diagram depicted in Fig.~\ref{fig:DiagDressed}, which is obtained from the
second-order diagram --shown in Fig.~\ref{fig:Diagrams}a-- by replacing
the bare interaction $U(\mathbf{q})$ with the dressed
scattering amplitudes: $\Gamma _{s_{1}s_{2};s_{3}s_{4}}(\mathbf{k},\mathbf{%
-k+q;p,-p+q})$ and its time reversed counterpart. The dressed amplitudes contain infinite sums
of the Cooper ladder diagrams shown in Fig.~\ref{fig:Ladder}. We will be
interested in the limit of vanishingly small magnetic fields and
temperatures smaller than the SOI energy scale: $\Delta \ll T\ll |\alpha|k_{F}$.
In this limit, the largest contribution to the ladder
diagrams comes from the internal Cooper bubbles formed by electrons from the
same Rashba subbands. Each ''rung'' of this ladder contributes a large
Cooper logarithm $L\equiv\left( m/2\pi \right) \ln \left( \Lambda /T\right)$,
which depends only on the temperature, and one has to select the diagrams with
a~maximum number of $L$ factors.

\subsubsection{\label{sec:Renormalization}
Renormalization group for scattering amplitudes}
Resummation of Cooper diagrams is performed most conveniently via the Renormalization Group (RG) procedure \cite{PhysRevB.79.115445}.
In the Cooper channel, the
bare amplitude is given by Eq. (\ref{eq:Gamma1}) with $\bm{p}'=-\bm{p}$ and $\bm{k}'=-\bm{k}$ or, equivalently, $\theta_{\bm{k}'}=\theta_{\bm{k}}+\pi$ and $\theta_{\bm{p}'}=\theta_{\bm{p}}+\pi$:
\begin{align}\label{eq:GammaCooperByAngle}
	\notag &\Gamma_{s_{1}s_{2};s_{3}s_{4}}^{(1)}(\bm{k,-k};\bm{p,-p})
	= U_{s_{1}s_{2};s_{3}s_{4}}^{(1)}\\
	&\,\,\,\,\,\,+ V_{s_{1}s_{2};s_{3}s_{4}}^{(1)}e^{i(\theta_{\bm{p}}-\theta _{\bm{k}})}
	+ W_{s_{1}s_{2};s_{3}s_{4}}^{(1)}e^{2i(\theta_{\bm{p}}-\theta _{\bm{k}})},
\end{align}
where the three terms correspond to orbital momenta $\ell=0,1,2$, respectively.
The bare values of partial amplitudes are given by
\begin{subequations}
\begin{align}
	U^{(1)}_{s_1s_2;s_3s_4} &= U/4,\label{eq:uvwzerob1}\\
	V^{(1)}_{s_1s_2;s_3s_4} &= (U/4)(s_1s_3+s_2s_4),\\
	W^{(1)}_{s_1s_2;s_3s_4} &= (U/4)s_1s_2s_3s_4.\label{eq:uvwzerob3}
\end{align}
\end{subequations}

Consider now a ladder diagram consisting of $n$ interaction lines and $2(n-1)$ internal fermionic lines, as shown in Fig.~\ref{fig:Ladder}. As we have already pointed out, in the limit $T\ll |\alpha |k_{F}$, the dominant logarithmic-in-$T$ contribution originates from those Cooper bubbles which
are formed by electrons from the same Rashba branch. Therefore, the $n$-th
order Cooper ladder can be written iteratively as
\begin{align}\label{eq:GammaNDef}
	\notag &\Gamma _{s_{1}s_{2};s_{4}s_{3}}^{(n)}(\mathbf{k,-k};\mathbf{p,-p})=\\
	&= -L\int_{\theta _{\mathbf{l}}}\sum_{s}
	\Gamma _{s_{1}s_{2};ss}^{(n-1)}(\mathbf{k,-k};\mathbf{l},-\mathbf{l})
	\Gamma_{ss;s_{3}s_{4}}^{(1)}(\mathbf{l},-\mathbf{l};\mathbf{p,-p})
\end{align}
where $n\geq2$.  Since only
\lq\lq charge-neutral\rq\rq\/ terms of the type $e^{i(\theta_{\bm{p}}-\theta_{\bm{l}})}e^{i(\theta_{\bm{l}}-\theta_{\bm{k}})}$ survive upon
averaging over $\theta_\bm{l}$,
different partial harmonics are renormalized independently of each other, i.e.,
we have the following group property:
\begin{align}
	\notag &\Gamma_{s_{1}s_{2};s_{3}s_{4}}^{(n)}(\bm{k,-k};\bm{p,-p})
	= (-L)^{n-1}[U_{s_{1}s_{2};s_{3}s_{4}}^{(n)}\\
	&\,\,\,\,\,\,+ V_{s_{1}s_{2};s_{3}s_{4}}^{(n)}e^{i(\theta_{\bm{p}}-\theta _{\bm{k}})}
	+ W_{s_{1}s_{2};s_{3}s_{4}}^{(n)}e^{2i(\theta_{\bm{p}}-\theta _{\bm{k}})}].
\end{align}
Differentiating Eq.~(\ref{eq:GammaNDef}) for $n=2$ with respect to $L$ we obtain three decoupled
one-loop RG flow equations
\begin{subequations}
\begin{align}
	\label{eq:RGzz1}
	-\frac{d}{dL}U_{s_1s_2;s_3s_4}(L)
	&= \sum_s U_{s_1s_2;ss}(L)U_{ss;s_3s_4}(L),\\
	-\frac{d}{dL}V_{s_1s_2;s_3s_4}(L)
	&= \sum_s V_{s_1s_2;ss}(L)V_{ss;s_3s_4}(L),\\
	\label{eq:RGzz3}
	-\frac{d}{dL}W_{s_1s_2;s_3s_4}(L)
	&= \sum_s W_{s_1s_2;ss}(L)W_{ss;s_3s_4}(L)
\end{align}
\end{subequations}
with initial conditions specified by $U_{s_1s_2;s_3s_4}(0)=U^{(1)}_{s_1s_2;s_3s_4}$, $V_{s_1s_2;s_3s_4}(0)=V^{(1)}_{s_1s_2;s_3s_4}$, and $V_{s_1s_2;s_3s_4}(0)=V^{(1)}_{s_1s_2;s_3s_4}$.

Solving this system of RG equations and substituting the results into the backscattering amplitude,
\begin{align}
	\notag \Gamma_{s_{1}s_{2};s_{3}s_{4}}&(\bm{k,-k};\bm{-k,k})
	= U_{s_{1}s_{2};s_{3}s_{4}}(L)\\
	&- V_{s_{1}s_{2};s_{3}s_{4}}(L) + W_{s_{1}s_{2};s_{3}s_{4}}(L)
\end{align}
which is a special case of the Cooper amplitude for $\bm{p}=-\bm{k}$,
we obtain
\begin{equation}\label{eq:ZeroAmplitude_1}
	\Gamma_{ss;\pm s\pm s}(\mathbf{k},\mathbf{-k};\mathbf{-k},\mathbf{k})
	=\frac{U}{2+UL}\mp\frac{U}{2(1+UL)},
\end{equation}

\begin{equation}\label{eq:NonZeroAmplitude}
	\Gamma_{s-s;\mp s\pm s}(\mathbf{k},\mathbf{-k};\mathbf{-k},\mathbf{k})
	= \frac{U}{2+UL}\pm\frac{U}{2},
\end{equation}
\begin{equation}\label{eq:ZeroAmplitude_2}
	\Gamma_{\sigma(\pm\mp;\mp\mp)}(\mathbf{k},\mathbf{-k};\mathbf{-k},\mathbf{k})=0,
\end{equation}
where $\sigma(s_1s_2;s_3s_4) \equiv$ $\{(s_1s_2;s_3s_4)$, $(s_2,s_3;s_4s_1)$,
$(s_3,s_4;s_1s_2)$, $(s_4,s_1;s_2,s_3)\}$ stands for all cyclic permutations of indices.

We see that the RG flow described by Eqs.~(\ref{eq:RGzz1}-\ref{eq:RGzz3}) has a non-trivial solution: whereas the amplitudes in Eqs.~(\ref{eq:ZeroAmplitude_1}) and (\ref{eq:ZeroAmplitude_2}) flow to zero in the limit of $L\to\infty$, the amplitudes in Eq.~(\ref{eq:NonZeroAmplitude}) approach RG-invariant values of $\pm U/2$. This behavior is in a striking contrast to what one finds in the absence of the SOI, when the repulsive interaction is renormalized to zero in the Cooper channel. Notice that
we consider only the energy scales below the SOI energy, while the conventional behavior is recovered at energies above the SOI scale.

The scattering amplitudes can be also derived by iterating Eq.~(\ref{eq:GammaNDef}) directly.
Examining a few first orders, one recognizes the pattern for the $n$-th order partial amplitudes to be
\begin{equation}
	\Gamma_{ss;\pm s\pm s}^{(n)}(\mathbf{k},\mathbf{-k};\mathbf{-k},\mathbf{k})
	=U^n(-L)^{n-1}\frac{1\mp2^{n-1}}{2^n},
\end{equation}
\begin{equation}
	\Gamma_{s-s;\mp s\mp s}^{(n)}(\mathbf{k},\mathbf{-k};\mathbf{-k},\mathbf{k})
	=U^n(-L)^{n}\left(\frac{1}{2^n}\pm\frac12\delta_{n,1}\right),
\end{equation}
\begin{equation}
	\Gamma_{\sigma(\pm\mp;\mp\mp)}^{(n)}(\mathbf{k},\mathbf{-k};\mathbf{-k},\mathbf{k})=0.
\end{equation}
Summing these amplitudes over $n$, one reproduces the RG result.

\subsubsection{\label{sec:InfZZ}Renormalization of $\protect\chi_{zz}$}

The infinite-order result for the thermodynamic potential is obtained by
replacing the bare contact interaction $U$ by its \lq\lq dressed\rq\rq\/ counterpart $\Gamma$
in the second-order skeleton diagram Fig.~\ref{fig:DiagDressed}
\begin{align}
	\notag \delta\Xi _{zz}=-\frac{1}{4}&\sum_{Q}{\sum_{\{s_{i}\}}}
	\Gamma_{s_1s_4;s_3s_2}(\mathbf{k},\mathbf{-k};\mathbf{-k},\mathbf{k})\\
	&\times \Gamma_{s_3s_2;s_1s_4}(\mathbf{-k},\mathbf{k};\mathbf{k},\mathbf{-k})
	\Pi _{s_1s_2}\Pi _{s_3s_4}  \label{wed1}
\end{align}
with $\Gamma_{s_3s_4;s_1s_2}(\mathbf{-k},\mathbf{k};\mathbf{k},\mathbf{-k})
=\Gamma_{s_1s_2;s_3s_4}(\mathbf{k},\mathbf{-k};\mathbf{-k},\mathbf{k})$.

Now, we derive the asymptotic form of $\delta\chi_{zz}$ valid in the limit of strong Cooper renormalizaton, i.e., for $UL\gg 1$.
In this limit, the only non-vanishing
scattering amplitude is given by Eq.~(\ref{eq:NonZeroAmplitude}). Replacing the full $\Gamma$ by its RG-invariant asymptotic limit $\Gamma_{s-s;\mp s\pm s}(\mathbf{k},\mathbf{-k};\mathbf{-k},\mathbf{k}) = \pm U/2$, we obtain
\begin{equation}\label{eq:Xi_dressed}
	\delta\Xi _{zz} = - \frac{U^2}{16}T\sum_{\Omega}\int\frac{qdq}{2\pi}\left[
	(\Pi_{+-}^2+\Pi_{-+}^2-2\Pi_0^2)+4\Pi_0^2\right]
\end{equation}
In contrast to the perturbation theory, where the magnetic-field dependence of the thermodynamic potential was provided by the vertices while the polarization bubbles supplied the dependence on the temperature and on the SOI, the vertices in the non-perturbative result (\ref{eq:Xi_dressed}) depend neither on the field nor on the SOI. Therefore, the dependences of $\Xi$ on all three parameters ($B$, $T$, and $\alpha$) must come from the polarization bubbles. The integral over $q$ along with the sum over the Matsubara frequency $\Omega$ have already been performed in Sec. \ref{sec:SS}. Note that the last term in square brackets
(proportional to $\Pi_0^2$) does not depend on the magnetic field
and thus can be dropped. The final result reads
\begin{equation}
	\delta\Xi _{zz} = -\frac{T^3}{8\pi v_F^2}\left(\frac{mU}{2\pi}\right)^2
	\mc{F}\left(\frac{2\bar\Delta_{\bm{k}_F}}{T}\right)
\end{equation}
so that
\begin{equation}\label{eq:ChiZZRen}
	\delta\chi_{zz} = \frac{\chi_0}{2}\left(\frac{mU}{4\pi}\right)^2\frac{|\alpha|k_F}{E_F},
\end{equation}
where use was made of the expansion
\begin{equation}
\frac{\partial ^2}{\partial \Delta^2}\mc{F}\left(\frac{2\bar\Delta_{\bm{k}_F}}{T}\right)\approx
\frac{2}{|\alpha|k_F T} \mc{F}'\left(\frac{2|\alpha|k_F}{T}\right)
\end{equation} and made use of the asymptotic form (\ref{Flarge}) of the function $\mc{F}$ to find that $\mc{F}'(x)\approx x^2$ for $x\gg1$.
Comparing the non-perturbative and second-order results for $\chi_{zz}$ [given by Eqs.~(\ref{eq:ChiZZRen}) and (\ref{eq:Delta>>T}), respectively], we see that the only effect of Cooper renormalization
is a change in the numerical coefficient of the nonanalytic part of $\chi_{zz}$. This is a consequence of a non-trivial fixed point in the Cooper channel
which corresponds to finite rather than vanishing Coulomb repulsion.

\begin{figure}[t]
\includegraphics[width=0.45\textwidth]{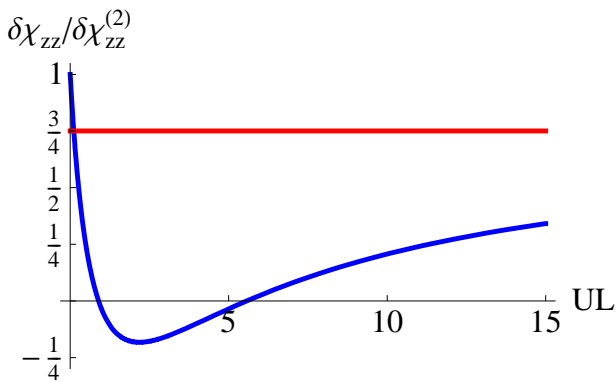}
\caption{Nonanalytic part of $\chi_{zz}$, normalized by the second-order result (\ref{eq:Delta>>T}), as a function of the Cooper-channel renormalization parameter $UL=(Um/2\pi)\ln(\Lambda/T)$. The horizontal line marks the low-temperature limit.}
\label{fig:chizzcooper}
\end{figure}

The temperature dependence of $\chi_{zz}$ can be also found for an arbitrary value of the Cooper renormalization parameter $UL$.
Deferring the details to Appendix \ref{app:RGzz}, we present here only the final result
\begin{widetext}
\begin{eqnarray}
\notag \delta\chi_{zz}
	&=&\chi_0\frac{2|\alpha|k_F}{E_F}\left(\frac{mU}{4\pi}\right)^2\bigg[
	\left(\frac{1}{2+UL}-\frac{1}{2}\right)^2+
	\frac{1}{3}\left(\frac{1}{2(1+UL)}+\frac{1}{2+UL}\right)^2\notag\\
&&+\frac{4}{3}\left(\frac{1}{2(1+UL)}-\frac{2}{(2+UL)^2}+\frac{2}{2+UL}\right)
	\left(\frac{1}{2+UL}-\frac{1}{2}\right)\bigg].\label{eq:chizzfinal}
\end{eqnarray}
\end{widetext}
In the limit of strong renormalization in the Cooper channel, i.e., for $UL\gg 1$, only the first term survives the logarithmic suppression, and Eq.~(\ref{eq:chizzfinal})  reduces to Eq.~(\ref{eq:ChiZZRen}).
In the absence of Cooper renormalization, i.e., for $L=0$, Eq.~(\ref{eq:chizzfinal}) reduces to the second-order result (\ref{eq:Delta>>T}).
In between these two limits, $\delta\chi_{zz}$ is a non-monotonic function of $UL$: as shown in Fig.~\ref{fig:chizzcooper}, $\delta\chi_{zz}$ exhibits a (shallow) minimum at $UL\approx 2.1$. In a wide interval of $UL$ ($0.9 \leq UL\leq 5.6$), the sign of $\delta\chi_{zz}$ is opposite (negative) to that in either of the high- and low-temperature limits. It is also seen from this plot that the low-$T$ asymptotic value (marked by a straight line) is reached only at very large ($\gtrsim 100$)  values of $UL$.

\subsubsection{\label{sec:InfXX}Renormalization of $\protect\chi_{xx}$}

The second-order result for the in-plane magnetic field is renormalized in a similar way with two exceptions.
First, because the Zeeman energy is anisotropic in this case -- the effective magnetic field $\bar{\Delta}_{\pm \mathbf{k}_{F}}(\theta _{\mathbf{l}})\equiv \sqrt{\alpha ^{2}k_{F}^{2}\pm 2\alpha k_{F}\Delta \sin \theta _{\mathbf{l}}+\Delta^2}$ depends on the direction of the electron
momentum $\mathbf{l}$ with respect to the field
$\theta _{\mathbf{l}}\equiv \angle (\mathbf{l},{\bf B})$-- integration in the ''rungs'' of the Cooper ladder can be performed only over
the fermionic frequency and the magnitude of the electron momenta (or,
equivalently, the variable $\xi _{\mathbf{l}}$).
Consequently, the elementary building block of the ladder
\begin{align}
L(\theta _{\mathbf{l}})& =T\sum_{\omega }\frac{m}{2\pi }%
\int d\xi _{\mathbf{l}}g_{t}(\omega ,\mathbf{l})g_{t^{\prime }}(-\omega ,-%
\mathbf{l})  \notag \\
& =\frac{m}{2\pi }\ln \frac{\Lambda }{\max \left\{ T,\left| t\bar{\Delta}_{%
\mathbf{k}_{F}}(\theta _{\mathbf{l}})-t^{\prime }\bar{\Delta}_{-\mathbf{k}%
_{F}}(\theta _{\mathbf{l}})\right| \right\} }
\end{align}
depends on $\theta _{\mathbf{l}}$.

In principle, the dependence of $L$ on the angle $\theta _{%
\mathbf{l}}$ should be taken into account when averaging over $\theta _{\bm{l}}$.
However, in the limit of $\Delta \ll T\ll |\alpha |k_{F}$, the
angle-dependent term under the logarithm can be approximated as $\left| t%
\bar{\Delta}_{k_{F}}(\theta _{\bm{l}})-t^{\prime }\bar{\Delta}_{-k_{F}}(\theta
_{\bm{l}})\right| \approx \left| t-t^{\prime }\right| |\alpha |k_{F}$ and, as it
was also the case for $\chi_{zz}$, $L=(m/2\pi )\ln (\Lambda /T)$ provided that $t=t^{\prime }$.

Second, the particle-hole bubbles also depend on the direction of the electron momentum, hence, the infinite-order result for the thermodynamic potential is found again by replacing the bare interaction $U$ in the second-order diagram by $\Gamma_{s_1s_2;s_3s_4}(\theta _{\mathbf{k}})$ and retaining the angular dependence of the bubbles.

The RG equations for the in-plane magnetic field are considerably more complicated. The main difference is that even the RG-invariant terms
depend on the magnetic field.
A detailed discussion of Cooper renormalization of the scattering amplitudes and spin susceptibility for this case is given in Appendices \ref{app:RGxx} and \ref{app:ChiXXRen}, respectively.
Below, we only show the final result for the renormalized spin susceptibility
\begin{equation}\label{eq:chixxfullmain}
	\delta\chi_{xx} = \frac{\chi_0}{3}\left(\frac{mU}{4\pi}\right)^2
	\frac{|\alpha|k_F}{E_F}
	+\mc{O}\left(\frac{T}{\ln T}\right).
\end{equation}
The $T$-independent term
is the same as without the Cooper renormalization [cf. (\ref{eq:ChiXX2ndAlpha})].
The linear-in-$T$ term however is suppressed by at least a factor of $1/\ln T$,
similar to the case of no SOI, where it is suppressed by a $\ln^2T$ factor.

\section{\label{sec:Sum}SUMMARY and discussion}

We have considered a two-dimensional electron liquid in the
presence of the Rashba spin-orbit interaction (SOI). The main result of this paper is that
the combined effect of the electro-electron and spin-orbit interactions breaks isotropy
of the spin response, whereas either of these two mechanisms does not.
Namely, nonanalytic behavior of the spin susceptibility, as manifested by its temperature--
and magnetic-field dependences, studied in this paper, is different for different components
of the susceptibility tensor: whereas the nonanalytic behavior of $\chi_{zz}$
is cut off at the energy scale associated with the SOI (given by $|\alpha|k_F$ for the Rashba SOI), that
of $\chi_{xx}$ (and $\chi_{yy}=\chi_{xx}$)  continues through the SOI energy scale. The reason for this difference
is the dependence of the SOI-induced magnetic field on the electron momentum. If the external magnetic field
is perpendicular to the plane of motion, its effect is simply dual to that of the SOI-field: the $T$ dependence of $\chi_{zz}$
is cut by whichever of the two fields is larger. If the external field is in the plane of motion, it is always possible
to form a virtual particle-hole pair, which mediates the long-range interaction between quasiparticles, from the states belonging to the same Rashba branch.
The energy of such a pair depends on the external
but not on the effective field, so that the SOI effectively drops out of the result. We have also studied a non-perturbative renormalization of the spin susceptibility in the Cooper channel of the electron-electron interaction.
It turns out the RG flow of scattering amplitudes is highly non-trivial. As a result, the spin susceptibility exhibits a non-monotonic dependence on the Cooper-channel renormalization parameter ($\ln T$) and eventually saturates as a temperature-independent value, proportional to the SOI coupling $|\alpha|$.

Notably, all the results of the paper are readily applicable to the systems with large Dresselhaus SOI and negligible Rashba SOI. In this case the Rashba spin-orbit coupling should be simply replaced by the Dresselhaus spin-orbit coupling.

Now we would like to discuss possible implications of these results for (in)stability of a second-order ferromagnetic quantum critical point (QCP). This phenomenon depends crucially on the sign of the nonanalytic correction. In this regard, we should point out that we limited our analysis to the simplest possible model, which does not involve the Kohn-Luttinger superconducting instability and higher-order processes in the particle-hole channel. Therefore, the sign of our nonanalytic correction is \lq\lq anomalous\rq\rq\/, i.e., the spin susceptibility increases with the corresponding energy scale. As in the absence of the SOI, however, either of these two effects (Kohn-Luttinger and particle-hole) can reverse the sign of nonanalyticity. Therefore, it is instructive to consider consequences of both signs.

In the absence of the SOI, a nonanalyticty of the anomalous sign renders a second-order ferromagnetic QCP unstable with respect to either a first-order phase transition or a transition into a spiral state. \cite{maslov06_09,green09} This result was previously believed to be relevant only to systems with a $SU(2)$ symmetry of electron spins; in particular it was shown in Ref.~\onlinecite{pepin06}b that there is no nonanalyticity in $\chi$ for a model case of the Ising-like exchange interaction between electrons.
We have shown here that broken (by the SOI) $SU(2)$ symmetry is not sufficient for eliminating a nonanalyticity in the in-plane component of the spin susceptibility ($\chi_{xx}$). Based on our results for the magnetic-field dependence of $\chi_{zz}$ and $\chi_{xx}$ [cf. Eqs.~(\ref{eq:chizzB}) and (\ref{eq:chixxB})], we can construct a model form for the free energy as a function of the magnetization $M$. The most interesting case for us is the one in which the Zeeman energy due to spontaneous magnetization, $\sim M/m \mu_B$ is larger than the SOI energy scale $|\alpha|k_F$, so that $\chi_{xx}\neq \chi_{zz}$. Ignoring Cooper-channel renormalization, we  can write the free energy as
\begin{equation}
F=aM^2-b\left(|M_x|^3+|M_y|^3\right)+M^4,
\label{fe}
\end{equation}
where $M=\left(M_x^2+M_y^2+M_z^2\right)^{1/2}$ and the coefficient of the quartic term was absorbed into the overall scale of $F$, which is irrelevant for our discussion.
An important difference of this free energy, compared to the case of no SOI, is easy-plane anisotropy of the nonanalytic, cubic term. In the absence of the cubic term ($b=0$), a second-order quantum phase transitions occurs when $a=0$; in the paramagnetic phase, $a$ is positive but small near the QCP.
Since $\chi$ is isotropic at the mean-field level [cf. Eq.~(\ref{stoner2})], the regular, $M^2$ and $M^4$ terms in Eq.~(\ref{fe}) are isotropic as well.
If $b>0$ (which corresponds to the anomalous sign of nonanalyticity), the cubic term leads to a minimum of $F$ at finite $M$; when the minimum value of $F$ reaches zero, the states with zero and finite magnetization become degenerate,
and a first-order phase transition occurs. The first-order critical point is specified by the following equations
\begin{equation}
\frac{\partial F}{\partial M_z}=0,\;\frac{\partial F}{\partial M_x}=0,\; \frac{\partial F}{\partial M_y}=0,\;\mathrm{and}\;F=0.
\end{equation}
For $a>0$, the only root of the first equation is $M_z=0$, i.e.,
there is no net magnetization in the $z$ direction. Substituting $M_z=0$ into the remaining equations and employing in-plane symmetry ($M_x=M_y$),
we find that the first-order phase transition occurs at $a=b^2/8$. The broken-symmetry state is an $XY$ ferromagnet with spontaneous in-plane magnetization
$M^{\{c\}}_x=M^{\{c\}}_y=b/4$.
The first-order transition to an $XY$ ferromagnet occurs if the Zeeman energy, corresponding to a jump of the magnetization at the critical point, is larger than the SOI energy, i.e., $M_{x}^{\{c\}}/m\mu_B\gg |\alpha|k_F$.
In the opposite case, the SOI is irrelevant, and the first-order transition is to a Heisenberg ferromagnet.

If the nonanalyticity is of the \lq\lq normal\rq\rq\/ sign ($b<0$), the transition remains second order and occurs at $a=0$. However, the critical indices are different for the in-plane and transverse magnetization: in the broken-symmetry phase $(a<0$), $M_x=M_y\propto (-a)$ while $M_z\propto (-a)^{1/2}$. Since $|a|\ll 1$, the resulting state is an Ising-like ferromagnet with $M_{z}\gg M_{x}=M_y$.

A detailed study of the $|\mathbf{q}|$ dependence of $\chi$ in the presence of the SOI will be presented elsewhere.\cite{unpub:zak}

\acknowledgements
We thank A. Ashrafi, S. M. Badalyan, S. Chesi, A. Chubukov, P. Kumar, E. Mishchenko, M. Raikh, and E. I. Rashba
for stimulating discussions. R.A.\.Z. and D.L. acknowledge financial support from the
Swiss NF and the NCCR Nanoscience Basel. D.L.M. acknowledges support from the Basel QC2
visitor program and from NSF via grant DMR-0908029.
\appendix

\begin{widetext}

\section{\label{app:0thOrder}Temperature dependence of the spin susceptibility of free Rashba fermions}
In this Appendix, we consider the temperature dependence of the spin
susceptibility of free 2D electrons in the presence of the Rashba
SOI. The transverse and parallel spin susceptibilities, $\chi^{\{0\}} _{zz}$ and $%
\chi^{\{0\}} _{xx},$ are given by
\begin{align}
\chi^{\{0\}} _{zz}& =-\sum_{K}\mathrm{Tr}\left[ \hat{G}\left( K\right) \hat{\sigma}%
_{z}\hat{G}\left( K^{\prime }\right) \hat{\sigma}_{z}\right] \\
\chi^{\{0\}} _{xx}& =-\sum_{K}\mathrm{Tr}\left[ \hat{G}\left( K\right) \hat{\sigma}%
_{x}\hat{G}\left( K^{\prime }\right) \hat{\sigma}_{x}\right] ,
\label{eq:Chi0XXDef}
\end{align}
where $K=\left( \omega ,\mathbf{k}\right) $ and $K^{\prime }=\left( \omega ,%
\mathbf{k+q}\right) $ with $q\rightarrow 0.$ Evaluating the traces, we
obtain for $\chi^{\{0\}} _{zz}$
\begin{equation}
\chi^{\{0\}} _{zz}=-T\sum_{\omega }\int \frac{d^{2}k}{\left( 2\pi \right) ^{2}}%
\sum_{s,t}\frac{1}{2}(1-st)g_{s}(\omega ,\mathbf{k})g_{t}(\omega ,\mathbf{k}%
)=-2T\sum_{\omega }\int \frac{dkk}{2\pi}g_{+}(\omega ,\mathbf{k})g_{-}(\omega ,\mathbf{k}),
\end{equation}
where we took advantage of the isotropy of $g_{\pm }(\omega ,\mathbf{k})$ and
put  $q=0$, because the poles in the Green's
functions of different branches reside on
opposite sides of the real axis. We see that $\chi^{\{0\}} _{zz}$ is determined only
by inter-subband transitions. Similarly, we obtain for $\chi^{\{0\}} _{xx}$%
\begin{equation}
\chi^{\{0\}} _{xx}=-T\sum_{\omega }\int \frac{d^{2}k}{\left( 2\pi \right) ^{2}}%
\sum_{s,t}\frac{1}{2}\left[1-st\cos(2\theta _{\mathbf{k}}\right) ]g_{s}(\omega ,%
\mathbf{k})g_{t}(\omega ,\mathbf{k+q})|_{q\rightarrow 0}.
\end{equation}
Since $\cos \theta _{\mathbf{k}}$ averages to zero, $\chi^{\{0\}} _{xx}$ can be written as
\begin{equation}\label{eq:ChiXXvsZZ}
\chi^{\{0\}} _{xx}=\frac{1}{2}\chi^{\{0\}} _{zz}+\frac{1}{2}\delta \chi^{\{0\}} _{xx},
\end{equation}
where
\begin{equation}
\delta \chi^{\{0\}} _{xx}\equiv -T\sum_{\omega }\int \frac{d^{2}k}{\left( 2\pi
\right) ^{2}}[g_{+}(\omega ,\mathbf{k})g_{+}(\omega ,\mathbf{k+q}%
)+g_{-}(\omega ,\mathbf{k})g_{-}(\omega ,\mathbf{k+q})]|_{q\rightarrow 0}
\label{eq:DeltaChixx}
\end{equation}
is the contribution from intra-subband transitions, absent in $\chi^{\{0\}} _{zz}.$

Let us evaluate $\chi^{\{0\}} _{zz}$ first. Performing the fermionic Matsubara sum,
we obtain
\begin{equation}
\chi^{\{0\}} _{zz}=-2\int_{0}^{\infty }\frac{kdk}{2\pi }\,\,T\sum_{\omega }\frac{1}{%
2\alpha k}\left( \frac{1}{i\omega -\xi_{\mathbf{k}}-\alpha k}-\frac{1}{i\omega -\xi
_{k}+\alpha k}\right) =\frac{1}{2\pi \alpha }\int_{0}^{\infty
}dk[n_{F}(\xi_{\mathbf{k}}-\alpha k)-n_{F}(\xi_{\mathbf{k}}+\alpha k)]
\end{equation}
with a Fermi function $n_{F}(\epsilon )=\left[ e^{(\epsilon -\mu )/T}+1%
\right] ^{-1}$ and $\xi_{\mathbf{k}}=k^{2}/2m-\mu $. Changing variables in the first
integral to
$\xi_{\mathbf{k}}-\alpha k=\epsilon_{-}-\mu$, we find two roots: $%
k_{-}^{(1)}=m\alpha -\sqrt{(m\alpha )^{2}+2m\epsilon_{-}}$, valid for $%
-\epsilon _{0}<\epsilon_{-}<0$ with $dk_{-}^{(1)}/d\epsilon_{-}<0,$ and $%
k_{-}^{(2)}=m\alpha +\sqrt{(m\alpha )^{2}+2m\epsilon_{-}},$ valid for $%
\epsilon_{-}>0$ with $dk_{-}^{(2)}/d\epsilon_{-}>0$, where $\epsilon
_{0}\equiv m\alpha ^{2}/2$. Similarly, we change variables in the second
integral to
$\xi_{\mathbf{k}}+\alpha k=\epsilon_{+}-\mu$ and obtain only one
positive root $k_{+}=-m\alpha +\sqrt{(m\alpha )^{2}+2m\epsilon_{+}},$
valid for $\epsilon_{+}>0$ with $dk_{+}/d\epsilon_{+}>0$. Notice
that the absolute values of the (inverse) group velocities are the same for both branches: $|dk_{-}^{(1,2)}/d\epsilon _{-}|=|dk_{+}/d\epsilon _{+}|=m[(m\alpha )^{2}+2m\epsilon
_{\pm}]^{-1/2}$. Therefore,
\begin{equation}
\chi^{\{0\}} _{zz}=\frac{1}{2\pi \alpha }\left( \int_{-\epsilon _{0}}^{0}d\epsilon
\left| \frac{dk_{-}^{(1)}}{d\epsilon }\right| +\int_{0}^{\infty }d\epsilon
\left| \frac{dk_{-}^{(2)}}{d\epsilon }\right| -\int_{0}^{\infty }d\epsilon
\left| \frac{dk_{+}}{d\epsilon }\right| \right) n_{F}(\epsilon )=\frac{%
m}{2\pi \alpha }\int_{-\epsilon _{0}}^{0}d\epsilon \frac{n_{F}(\epsilon )}{%
\sqrt{(m\alpha )^{2}+2m\epsilon }},  \label{eq:Chi0ZZ}
\end{equation}
where we dropped the index on the integration variable $\epsilon $. Notably, the
high energy contributions from the two Rashba branches cancel each other for
any value of the chemical potential and the spin susceptibility is
determined exclusively by the bottom part of the lower Rashba branch.
Integration by parts yields
\begin{equation}
\chi^{\{0\}} _{zz}=\chi _{0}\left( n_{F}(0)-\int_{-\epsilon _{0}}^{0}d\epsilon \sqrt{%
1+\frac{\epsilon }{\epsilon _{0}}}\frac{\partial n_{F}(\epsilon )}{\partial
\epsilon }\right) ,  \label{sun1}
\end{equation}
where $n_{F}(0)=[e^{-\mu /T}+1]^{-1}$. In order to evaluate this integral,
it is convenient to consider three limiting cases.

For $T\ll \epsilon _{0}\ll \mu $, i.e., when both Rashba subbands are
occupied and the temperature is lower than the minimum of the lower subband,
we approximate $n_{F}(0)\approx 1-e^{-\mu /T}$, $-T\partial n_{F}(\epsilon
)/\partial \epsilon \approx e^{(\epsilon -\mu )/T}$, and $\sqrt{1+\epsilon
/\epsilon _{0}}\approx 1+\epsilon /2\epsilon _{0},$ so that
\begin{equation}
\chi^{\{0\}} _{zz}=\chi _{0}\left[ (1-e^{-\mu /T})+\frac{1}{T}\,e^{-\mu
/T}\int_{0}^{\infty }d\epsilon \left( 1-\frac{\epsilon }{2\epsilon _{0}}%
\right) e^{-\epsilon /T}\right] =\chi _{0}\left( 1-\frac{T}{2\epsilon _{0}}%
\,e^{-\mu /T}\right) .
\end{equation}
At $T=0,$ $\chi^{\{0\}} _{zz}=\chi _{0}.$

For $\epsilon _{0}\ll T\ll \mu $, i.e., when again both Rashba subbands are
occupied but the temperature is higher than the minimum of the lower
subband, we keep the same approximations for $n_{F}(0)$ and $-T\partial
n_{F}(\epsilon )/\partial \epsilon $ \ but neglect the $\epsilon $
dependence of the Fermi function in the integrand:
\begin{equation}
-\int_{-\epsilon _{0}}^{\{0\}}d\epsilon \sqrt{1+\frac{\epsilon }{\epsilon _{0}}}%
\frac{\partial n_{F}(\epsilon )}{\partial \epsilon }\approx \frac{1}{T}%
\,e^{-\mu /T}\int_{0}^{\epsilon _{0}}d\epsilon \sqrt{1+\frac{\epsilon }{%
\epsilon _{0}}}\,e^{-\epsilon /T}\approx \frac{1}{T}\,e^{-\mu
/T}\int_{0}^{\epsilon _{0}}d\epsilon \sqrt{1+\frac{\epsilon }{\epsilon _{0}}}%
=\frac{2\epsilon _{0}}{3T}e^{-\mu /T}.
\end{equation}
Thus,
\begin{equation}
\chi^{\{0\}} _{zz}=\chi _{0}\left[ 1-\left( 1-\frac{2\epsilon _{0}}{3T}\right)
e^{-\mu /T}\right] .
\end{equation}

Finally, for $\mu <0$, i.e., when only the lower Rashba subband is occupied,
the first term in Eq.~(\ref{sun1}) gives only an exponentially weak
temperature dependence, while the Sommerfeld expansion of the second term generates a $T^2$
contribution because the density of states depends on $\epsilon$:
\begin{equation}
\chi^{\{0\}} _{zz}=\chi _{0}\left[\sqrt{1-|\mu |/\epsilon _{0}}-\frac{\pi ^{2}}{24}%
\left(\frac{T}{\epsilon _{0}}\right)^2\frac{1}{(1-|\mu |/\epsilon _{0})^{3/2}%
}\right]
\end{equation}
At zero temperature, $\chi^{\{0\}} _{zz}=\chi _{0}\sqrt{1-|\mu |/\epsilon _{0}}$
vanishes at the bottom of the lower subband.

We now calculate $\delta \chi^{\{0\}} _{xx}$ given by Eq.~(\ref{eq:DeltaChixx}),
which can be written as
\begin{equation}
\delta \chi^{\{0\}} _{xx}=-\int_{0}^{\infty }\frac{dkk}{2\pi }\left( \frac{\partial
n_{F}(\epsilon_{-})}{\partial \epsilon_{-}}+\frac{\partial n_{F}(\epsilon
_{+})}{\partial \epsilon_{+}}\right) .
\end{equation}
The change of variables is straightforward in the second integral, since
the equation
$\epsilon_{+}-\mu=\xi_{\mathbf{k}}+\alpha k$ has only one positive root $%
k_{+}$
and the density
of states is given by
\begin{equation}\label{eq:DOS2+}
\nu_{+}(\epsilon )=\frac{k_{+}}{2\pi }\frac{dk_{+}}{%
d\epsilon }=\frac{m}{2\pi }\frac{\sqrt{1+\epsilon /\epsilon _{0}}-1}{\sqrt{%
1+\epsilon /\epsilon _{0}}}
\end{equation}
with $\epsilon _{0}\equiv m\alpha ^{2}/2$. In the first integral more care
must be taken: for $\epsilon>0$, we have $k_{-}^{(2)}=m\alpha +\sqrt{%
(m\alpha )^{2}+2m\epsilon}$ and the density of states is given by
\begin{equation}\label{eq:DOS1+}
\nu_{-}^{>}(\epsilon )=\frac{k_{-}^{(2)}}{2\pi }\frac{dk_{-}^{(2)}}{%
d\epsilon }=\frac{m}{2\pi }\frac{\sqrt{1+\epsilon /\epsilon _{0}}+1}{\sqrt{%
1+\epsilon /\epsilon _{0}}};
\end{equation}
however, for $-\epsilon _{0}<\epsilon<0$ both roots enter the density
of states
\begin{equation}\label{eq:DOS1}
\nu^{<}_{-}(\epsilon )=\int_{0}^{\infty }\frac{kdk}{2\pi }\delta (\epsilon
-\epsilon_{-})=\frac{1}{2\pi }\left( k_{-}^{(1)}\left| \frac{dk_{-}^{(1)}}{%
d\epsilon_{-}}\right| +k_{-}^{(2)}\left| \frac{dk_{-}^{(2)}}{d\epsilon_{-}}%
\right| \right) =\frac{m}{\pi }\frac{1}{\sqrt{1+\epsilon /\epsilon _{0}}}.
\end{equation}
Summing up the contributions from all energies, we find
\begin{equation}
\delta \chi^{\{0\}} _{xx}=\left( \int_{-\epsilon _{0}}^{0}d\epsilon \nu
^{<}_{-}(\epsilon )+\int_{0}^{\infty }d\epsilon \nu_{-}^{>}(\epsilon
)+\int_{0}^{\infty }d\epsilon \nu _{+}(\epsilon )\right) \left(-\frac{%
\partial n_{F}(\epsilon )}{\partial \epsilon }\right)=\int_{-\epsilon
_{0}}^{0}d\epsilon \nu ^{<}_{-}(\epsilon )\left(-\frac{\partial n_{F}(\epsilon )}{%
\partial \epsilon }\right)+\chi _{0}n_{F}(0),
\end{equation}
where the second and third integrals are easily evaluated because $\epsilon $
drops out of the sum $\nu_{+}+\nu _{-}^{>}=m/\pi $. Combining the
above result with Eqs.~(\ref{eq:Chi0XXDef}) and (\ref{eq:Chi0ZZ}), we get
\begin{equation}
\chi^{\{0\}} _{xx}=\chi _{0}\left( n_{F}(0)-\int_{-\epsilon _{0}}^{0}d\epsilon \frac{%
1+ \epsilon /2\epsilon _{0}}{\sqrt{1+\epsilon /\epsilon _{0}}}\frac{\partial
n_{F}(\epsilon )}{\partial \epsilon }\right) .  \label{eq:Chi0XXInt}
\end{equation}
For $T\ll \epsilon _{0}\ll \mu $, we approximate $n_{F}(0)\approx 1-e^{-\mu
/T}$ and $-T\partial n_{F}(\epsilon )/\partial \epsilon =e^{(\epsilon -\mu
)/T}(e^{(\epsilon -\mu )/T}+1)^{-2}\approx e^{(\epsilon -\mu )/T}$ as
before, and expand $(1+\epsilon /2\epsilon _{0})/
\sqrt{1+\epsilon /\epsilon _{0}}\approx 1
+\epsilon^2 /8\epsilon _{0}^2$, so that
\begin{equation}
\chi^{\{0\}} _{xx}=\chi _{0}\left[ n_{F}(0)+\frac{1}{T}\,e^{-\mu /T}\int_{0}^{\infty
}d\epsilon \left( 1 +\frac{ \epsilon^2 }{ 8\epsilon _{0}^2}\right)
e^{-\epsilon /T}\right] =\chi _{0}\left( 1+\frac{T^2}{4\epsilon _{0}^2}%
\,e^{-\mu /T}\right) .
\end{equation}
For $\epsilon _{0}\ll T\ll \mu $, we keep the same approximations for the
Fermi function but, as it was also the case for $\chi^{\{0\}} _{zz},$ neglect the $%
\epsilon $ dependence of the Fermi function in the integrand
\begin{equation}
-\int_{-\epsilon _{0}}^{0}d\epsilon \frac{1+ \epsilon /2\epsilon _{0}}{\sqrt{%
1+\epsilon /\epsilon _{0}}}\frac{\partial n_{F}(\epsilon )}{\partial
\epsilon }\approx \frac{1}{T}\,e^{-\mu /T}\int_{0}^{\epsilon _{0}}d\epsilon
\frac{1- \epsilon /2\epsilon _{0}}{\sqrt{1-\epsilon /\epsilon _{0}}}%
\,e^{-\epsilon /T}\approx \frac{1}{T}\,e^{-\mu /T} \int_{0}^{\epsilon
_{0}}d\epsilon \frac{1- \epsilon /2\epsilon _{0}}{\sqrt{1-\epsilon /\epsilon
_{0}}}= \frac{ 4\epsilon _{0}}{3T}e^{-\mu /T}.
\end{equation}
Thus,
\begin{equation}
\chi^{\{0\}} _{xx}=\chi _{0}\left[ 1-\left( 1-\frac{ 4\epsilon _{0}}{3T}\right)
e^{-\mu /T}\right] .
\end{equation}
Finally, for $\mu <0$, the Sommerfeld expansion of the second term in Eq.~(%
\ref{eq:Chi0XXInt}) yields
\begin{equation}
\chi^{\{0\}} _{xx}=\chi _{0}\left[\frac{1-|\mu|/2\epsilon_0}{\left(1-|\mu |/\epsilon
_{0}\right)^{1/2}}+\frac{\pi ^{2}}{48}\left(\frac{T}{\epsilon _{0}}\right)^2%
\frac{2+|\mu|/\epsilon_0}{\left(1-|\mu |/\epsilon _{0}\right)^{5/2}}\right]
\end{equation}
for $T\ll \mathrm{min}\{|\mu |,\epsilon _{0}-|\mu |\}$. At zero temperature,
$\chi^{\{0\}} _{xx}=\chi _{0}\left(1-|\mu|/2\epsilon_0\right)/\left(1-|\mu
|/\epsilon _{0}\right)^{1/2}$ diverges at the bottom of the lower subband.

\section{\label{app:smallq}Absence of a small-$q$ singularity in the static polarization bubble with spin-orbit interaction}

In Secs. \ref{sec:SS} and \ref{sec:Diags}, we argued that there is no contribution to the
nonanalytic behavior of the spin susceptibility from the region of small
bosonic momenta, $q\ll k_{F}.$  This statement contradicts Ref.~\onlinecite{chen99}
where it was argued that, in the presence of the SOI, a static particle-hole bubble has a square-root singularity
at $q=q_0\equiv 2m\alpha$ (in addition to the Kohn anomaly  which is also modified
by the SOI).  For a weak
SOI, $q_0$ is much smaller than $k_{F}$ and thus the region of small $q$ may also contribute to the nonanalytic behavior.
Later on, however, Refs.~\onlinecite{pletyukhov06,pletyukhov2} showed that there is no singularity at $q=q_0$. According to Ref.~\onlinecite{pletyukhov06},
the reason is related to a subtlety in approaching the static limit of a
dynamic bubble. While we agree with the authors of Refs. \onlinecite{pletyukhov06} in that
there are no small-$q$ singularities in the bubble, we find that a cancellation of
singular terms occurs in the calculation of a purely static bubble. The same result
was obtained in an unpublished work.~\cite{mishchenko} For the sake of
completeness, we present our derivation in this Appendix.

Evaluating the spin trace, we obtain for the static polarization bubble
\begin{equation}
\Pi (q)\equiv \sum_{K}\mathrm{Tr}[\hat{G}_{\omega,\mathbf{k+q} }\hat{G}_{%
\omega,\mathbf{k}}]=\frac{1}{2}\sum_{K}\sum_{s,t}\left[ 1+st\cos (\varphi
_{\mathbf{k}+\mathbf{q}}-\varphi _{\mathbf{k}})\right] g_{s}(\omega,\mathbf{k+q})
g_{t}(\omega,\mathbf{k}),
\end{equation}
where, as before, \ $K=\left(\omega,\mathbf{k}\right)$, $\varphi _{%
\mathbf{k}+\mathbf{q}}\equiv \angle (\mathbf{k}+\mathbf{q},\mathbf{e}_{x})$
and $\varphi _{\mathbf{k}}\equiv \angle (\mathbf{k},\mathbf{e}_{x})$. We
divide $\Pi (q)$ into intra- and intersubband contributions as
\begin{equation}
\Pi (q)=\Pi _{++}(q)+\Pi _{--}(q)+\Pi _{\pm }(q),
\label{eq:total}\end{equation}
where
\begin{subequations}
\begin{eqnarray}
\Pi _{\pm \pm }(q) &\equiv &\frac{1}{2}\sum_{K}[1+\cos (\varphi _{\mathbf{k}%
+\mathbf{q}}-\varphi _{\mathbf{k}})]g_{\pm }(\omega,\mathbf{k+q})g_{\pm }(%
\omega,\mathbf{k}),  \label{eq:Pi_pmpm} \\
\Pi _{\pm }(q) &\equiv &\frac{1}{2}\sum_{K}[1-\cos (\varphi _{\mathbf{k}+%
\mathbf{q}}-\varphi _{\mathbf{k}})][g_{+}(\omega,\mathbf{k+q})g_{-}(\omega,\mathbf{k})
+g_{-}(\omega,\mathbf{k+q})g_{+}(\omega,\mathbf{k})]\notag \\
&=&\sum_{K}[1-\cos (\varphi _{\mathbf{k}+\mathbf{q}}-\varphi _{\mathbf{k}}%
)]g_{+}(\omega,\mathbf{k+q})g_{-}(\omega,\mathbf{k})\label{eq:Pi_pm}.
\end{eqnarray}
\end{subequations}
In the last line, we employed obvious symmetries of the Green's function.

First, we focus on the intersubband part, $\Pi _{\pm }(q)$. Summation over
the Matsubara frequency yields
\begin{equation}
T\sum_{\omega }g_{+}(\mathbf{p,}\omega )g_{-}(\mathbf{k,}\omega )=\frac{%
n_{F}(\xi _{\mathbf{p}}^{+})-n_{F}(\xi _{\mathbf{k}}^{-})}{\xi^{+}_{\mathbf{p}}-\xi^{-}_{\mathbf{k}} },
\end{equation}
where $\mathbf{p=k+q}$, $\xi^{\pm} _{\mathbf{k}}=\xi_{\mathbf{k}}\pm \alpha
{k}$ and, as before, $\xi _{\mathbf{k}}=k^{2}/2m-E_{F}.$ Introducing additional integration over the momentum $\mathbf{p}$, as it was done in Ref.~\onlinecite{chen99},  Eq.\ (\ref
{eq:Pi_pm}) can be re-written as
\begin{equation}
\Pi _{\pm }(q)=\frac{2}{(2\pi )^{2}}\int_{0}^{2\pi
}d\theta \int_{0}^{\infty }dkk\int_{0}^{\infty }dpp\,\delta (p^{2}-|\mathbf{k%
}+\mathbf{q}|^{2})\left( 1-\frac{\mathbf{k}\cdot(\mathbf{k}+\mathbf{q})}{kp}%
\right) \frac{n_{F}(\xi^{+}_{\mathbf{p}})-n_{F}(\xi^{-}_{\mathbf{k}})}
{\xi^{+}_{\mathbf{p}}-\xi^{-}_{\mathbf{k}}},
\end{equation}
where $\theta=\angle(\mathbf{k},\mathbf{q})$.
Integration over $\theta $ yields
\begin{equation}
\int_{0}^{2\pi }d\theta \delta (p^{2}-k^{2}-q^{2}-2kq\cos \theta )\left( 1-%
\frac{k^{2}+kq\cos \theta }{kp}\right) =\frac{1}{kp}\frac{q^{2}-(k-p)^{2}}{%
\sqrt{(k+q)^{2}-p^{2}}\sqrt{p^{2}-(k-q)^{2}}},  \label{eq:DeltaInt}
\end{equation}
which imposes a constraint on the  range of integration over $p$, i.e., $%
|k-q|<p<k+q$. Since we assume that $q\ll k\approx p\approx k_{F}$, Eq.~(\ref
{eq:DeltaInt}) can be simplified to
\begin{equation}
\frac{\sqrt{q^{2}-(k-p)^{2}}}{2k_{F}^{3}},
\end{equation}
and $\Pi _{\pm }(q)$ becomes
\begin{equation}
\Pi _{\pm }(q)=\frac{1}{4\pi ^{2}k_{F}^{3}}\int_{0}^{\infty
}dkk\int_{|k-q|}^{k+q}dpp\,\sqrt{q^{2}-(k-p)^{2}}\,\frac{n_{F}(
\xi^+_{\mathbf{p}})-n_{F}(\xi^{-}_{\mathbf{k}})}
{\xi^{+}_{\mathbf{p}}-\xi^{-}_{\mathbf{k}}}.
\end{equation}
For a weak SOI ($m|\alpha| \ll k_{F}$) , $\xi _{\mathbf{k}}^{\pm }\approx
\xi _{\mathbf{k}}\pm \alpha k_{F}.$ Switching from integration over $k$ and $%
p$ to  integration over $\xi _{\mathbf{k}}$ and $\xi _{\mathbf{p}}$, we find
\begin{equation}
\Pi _{\pm }(q)=\frac{1}{4\pi ^{2}v_{F}^{3}k_{F}}\int_{-\infty }^{\infty
}d\xi_{\mathbf{k}}\int_{\xi_{\mathbf{k}}-v_{F}q}^{\xi _{\mathbf{k}}+v_{F}q}d\xi_{\mathbf{%
p}}\,\sqrt{(v_{F}q)^{2}-(\xi_{\mathbf{k}}-\xi_{\mathbf{p}})^{2}}\,\frac{n_{F}
(\xi _{\mathbf{p}}+\alpha k_{F})-n_{F}(\xi _{\mathbf{k}}-\alpha k_{F})}
{\xi _{\mathbf{p}}-\xi _{\mathbf{k}}+2\alpha k_{F}}.
\end{equation}
Shifting the integration variables as $\xi_{\mathbf{p}}\rightarrow \xi _{\mathbf{p}}-\alpha
k_{F}$, $\xi _{\mathbf{k}}\rightarrow \xi _{\mathbf{k}}+\alpha k_{F}$, we
eliminate the dependence of the Fermi functions on $\alpha k_{F}$. Assuming
also that $T=0,$ we obtain
\begin{equation}
\Pi _{\pm }(q)=\frac{1}{4\pi ^{2}v_{F}^{3}k_{F}}\int_{-\infty }^{\infty
}d\xi _{\mathbf{k}}\int_{\xi _{\mathbf{k}}+2\alpha k_{F}-v_{F}q}^{\xi _{\mathbf{k}}%
+2\alpha k_{F}+v_{F}q}d\xi _{\mathbf{p}}\,\sqrt{(v_{F}q)^{2}-(\xi _{\mathbf{p}}-\xi
_{\mathbf{k}}-2\alpha k_{F})^{2}}\,\frac{\Theta (-\xi _{\mathbf{p}})-\Theta
(-\xi _{\mathbf{k}})}{\xi _{\mathbf{p}}-\xi _{\mathbf{k}}},\label{eq:pipm}
\end{equation}
where $\Theta \left( x\right) $ is the step function. Notice that the integrand is finite only if $\xi _{\mathbf{k}}\xi _{\mathbf{p}}<0$, which imposes further constraints on the integration range.

We will now prove that $\Pi _{\pm }(q)$ given by Eq.~(\ref{eq:pipm}) is continuous at $q=q_{0}$.
To this end, it is convenient to consider the cases of $q<q_{0}$ and $q>q_{0}
$.  Combining all the constraints together,  we find that $\Pi
_{\pm }$ for  $q<q_{0}$ can be written as
\begin{align}
\Pi _{\pm }^{<}(q)\equiv \Pi _{\pm }(q<q_{0})=& -\frac{1}{4\pi
^{2}v_{F}^{3}k_{F}}\bigg(\int_{-v_{F}q-2\alpha k_{F}}^{v_{F}q-2\alpha
k_{F}}d\xi _{\mathbf{k}}\int_{0}^{\xi_{\mathbf{k}}+2\alpha k_{F}+v_{F}q}d\xi _{\mathbf{p}}
\notag  \label{eq:Pi_q<} \\
& +\int_{v_{F}q-2\alpha k_{F}}^{0}d\xi _{\mathbf{k}}\int_{\xi_{\mathbf{k}}+2\alpha
k_{F}-v_{F}q}^{\xi_{\mathbf{k}}+2\alpha k_{F}+v_{F}q}d\xi _{\mathbf{p}}\bigg)\frac{\sqrt{%
(v_{F}q)^{2}-(\xi _{\mathbf{p}}-\xi _{\mathbf{k}}-2\alpha k_{F})^{2}}}{\xi _{\mathbf{p}}%
-\xi _{\mathbf{k}}},
\end{align}
Reversing the sign of $\xi _{\mathbf{k}}$,  absorbing $\xi _{\mathbf{k}}$ into
$\xi _{\mathbf{p}}$, and defining the dimensionless variables $x=\xi _{\mathbf{k}}/v_{F}q$
and $y=\xi _{\mathbf{p}}/v_{F}q$, we obtain
\begin{equation}
\Pi _{\pm }^{<}(q)=-\frac{q^{2}}{4\pi ^{2}v_{F}k_{F}}\left( \int_{\beta
-1}^{\beta +1}dx\int_{x}^{\beta +1}dy+\int_{0}^{\beta -1}dx\int_{\beta
-1}^{\beta +1}dy\right) \frac{\sqrt{1-(y-\beta )^{2}}}{y},
\end{equation}
where $\beta \equiv q_{0}/q>1$. Next, we switch the order of integration in
the first term, so that the integrals over $x$ can be readily evaluated
\begin{align}
\Pi _{\pm }^{<}(q)& =-\frac{q^{2}}{4\pi ^{2}v_{F}k_{F}}\int_{\beta
-1}^{\beta +1}dy\left( \int_{\beta -1}^{y}dx+\int_{0}^{\beta -1}dx\right)
\frac{\sqrt{1-(y-\beta )^{2}}}{y}  \notag \\
& =-\frac{q^{2}}{4\pi ^{2}v_{F}k_{F}}\int_{\beta -1}^{\beta +1}dy\sqrt{%
1-(y-\beta )^{2}}=
-\frac{m}{2\pi }\left( \frac{q}{2k_{F}}\right) ^{2}.
\end{align}

For $q>q_{0},$ we have
\begin{eqnarray}
\Pi _{\pm }^{>}\left( q\right) &\equiv& \Pi _{\pm }(q>q_{0})= \frac{1}{4\pi
^{2}v_{F}^{3}k_{F}}\bigg(\int_{0}^{v_{F}q-2\alpha k_{F}}d\xi _{\mathbf{k}}%
\int_{\xi _{\mathbf{k}}+2\alpha k_{F}-v_{F}q}^{0}d\xi _{\mathbf{p}} \notag\\
&& -\int_{-v_{F}q-2\alpha k_{F}}^{0}d\xi _{\mathbf{k}}\int_{0}^{\xi_{\mathbf{k}}+2\alpha
k_{F}+v_{F}q}d\xi _{\mathbf{p}}\bigg)\frac{\sqrt{(v_{F}q)^{2}-(\xi _{\mathbf{p}}-\xi
_{\mathbf{k}}-2\alpha k_{F})^{2}}}{\xi _{\mathbf{p}}-\xi _{\mathbf{k}}}.
\label{eq:Pi_q>}\end{eqnarray}
Manipulations similar to those for the previous case yield
\begin{equation}
\Pi _{\pm }^{>}(q)=-\frac{q^{2}}{4\pi ^{2}v_{F}k_{F}}\left(
\int_{0}^{1-\beta }dx\int_{x}^{1-\beta }dy\frac{\sqrt{1-(y+\beta )^{2}}}{y}%
+\int_{0}^{1+\beta }dx\int_{x}^{1+\beta }dy\frac{\sqrt{1-(y-\beta )^{2}}}{y}%
\right)
\end{equation}
with $\beta =q_{0}/q<1$. Interchanging the order of integrations over $x$
and $y$, we find
\begin{equation}
\Pi _{\pm }(q>2m\alpha )=-\frac{q^{2}}{4\pi ^{2}v_{F}k_{F}}\left(
\int_{0}^{1-\beta }dy\sqrt{1-(y+\beta )^{2}}+\int_{0}^{1+\beta }dy\sqrt{%
1-(y-\beta )^{2}}\right) =-\frac{m}{2\pi }\left( \frac{q}{2k_{F}}\right)
^{2}.
\end{equation}
Since $\Pi _{\pm }^{<}(q=q_{0}-0^+)=\Pi _{\pm }^{>}(q=q_{0}+0^+)$, the function $%
\Pi _{\pm }(q)=-(m/2\pi )(q/2k_{F})^{2}$ is continuous at $q=q_{0}$ and,
thus, there is no singularity in the static particle-hole response function.
In addition, $\Pi _{\pm }(q)$ does not depend on the SOI. However,
since there is no $q^{2}$ term in the 2D bubble for  $q\leq 2k_{F}$ in the
absence of the SOI,  the $q^{2}$ term must be canceled out by
similar terms in $\Pi _{++}(q)$ and $\Pi _{--}(q),$ which is what we will
show below.

Having proven that $\Pi _{\pm }(q)$ is an analytic function of $q$, we can
re-derive its $q$ dependence simply by expanding the combination $\varphi _{\mathbf{k}+%
\mathbf{q}}-\varphi _{\mathbf{k}}$ in Eq.~(\ref{eq:Pi_pm}) for $q\ll k_F$ as $\varphi _{\mathbf{k}+\mathbf{q%
}}-\varphi _{\mathbf{k}}\approx (q/k_{F})\sin \varphi _{\mathbf{kq}},$ where
$\varphi _{\mathbf{kq}}\equiv \angle (\mathbf{k},\mathbf{q})$; then $1-\cos
(\varphi _{\mathbf{k}+\mathbf{q}}-\varphi _{\mathbf{k}})\approx
(q/k_{F})^{2}\sin ^{2}\left( \varphi _{\mathbf{kq}}\right) /2.$ Since we
already have a factor of $q^{2}$ up front, the Green's functions in Eq.~(\ref{eq:Pi_pm})
can be evaluated at $q=0.$ Accordingly, Eq. (\ref{eq:Pi_pm}) becomes
\begin{align}
\Pi _{\pm }(q)& =\frac{1}{2}\left( \frac{q}{k_{F}}\right) ^{2}\sum_{K}\sin
^{2}\varphi _{\mathbf{kq}}g_{+}(\mathbf{k},\omega )g_{-}(\mathbf{k,}\omega )=
\notag \\
& =\frac{m}{2\pi }\left( \frac{q}{2k_{F}}\right) ^{2}\int d\xi _{\mathbf{k}}%
\frac{n_{F}(\xi _{\mathbf{k}}^{+})-n_{F}(\xi _{\mathbf{k}}^{-})}{\xi _{\mathbf{k}}
^{+}-\xi _{\mathbf{k}}^{-}}=-\frac{m}{2\pi }\left( \frac{q}{2k_{F}}\right)
^{2}\int_{-\alpha k_{F}}^{\alpha k_{F}}d\xi _{\mathbf{k}}\frac{1}{2\alpha k_{F}}%
=-\frac{m}{2\pi }\left( \frac{q}{2k_{F}}\right) ^{2}.
\end{align}
Expanding Eq.~(\ref{eq:Pi_pmpm}) for the intraband contribution to
the bubble also to second order in $q$, we obtain
\begin{equation}
\Pi _{\pm \pm }(Q)=\sum_{K}g_{\pm }(\mathbf{k}+\mathbf{q,}\omega )g_{\pm }(%
\mathbf{k},\omega )|_{\mathbf{q}\rightarrow 0}-\frac{1}{4}\left( \frac{q}{%
k_{F}}\right)^{2}\sum_{K}\sin ^{2}\varphi _{\mathbf{kq}}g_{\pm }(\mathbf{k}+\mathbf{q},\omega
)g_{\pm }(\mathbf{k,}\omega )|_{q\rightarrow0}=-\frac{m}{2\pi }+\frac{m}{4\pi }\left( \frac{q%
}{2k_{F}}\right) ^{2},
\end{equation}
since
\begin{equation}
\sum_{K}g_{\pm }(\mathbf{k}+\mathbf{q,}\omega )g_{\pm }(\mathbf{k},\omega )|_{\mathbf{q}\rightarrow 0}
=-m\int \frac{d\xi _{\mathbf{k}}}{2\pi}\frac{\Theta(\xi _{\mathbf{k}+\mathbf{q}}^{\pm})
-\Theta(\xi _{\mathbf{k}}^{\pm})}{\xi _{\mathbf{k}+\mathbf{q}}^{\pm}-\xi _{\mathbf{k}}^{\pm}}
\bigg|_{q\rightarrow0}
=-m\int \frac{d\xi _{\mathbf{k}}}{2\pi}\delta(\xi_{\mathbf{k}}\pm|\alpha|k_F)=-\frac{m}{2\pi}.
\end{equation}
Thereby the total bubble (\ref{eq:total})
\begin{equation}
\Pi(q)=-\frac{m}{\pi }
\end{equation}
is independent of $q$ for $q\ll 2k_F$.

\section{\label{app:RG}Renormalizaton of the scattering amplitudes in the Cooper channel in the presence of the magnetic field}

In this Appendix, we present the derivation of the RG flow equations for the scattering amplitudes in the Cooper channel in the presence of the magnetic field.
These amplitudes are then used  to find the non-perturbative results for the spin susceptibility. Since the amplitudes are renormalized quite differently for the field applied perpendicularly and parallel to the 2DEG plane, we will treat these two cases separately.

\subsection{\label{app:RGzz}Transverse magnetic field}
\subsubsection{\label{RGzzGamma}RG flow of the scattering amplitudes}
If the  magnetic field is transverse to the 2DEG plane, $\bm{B}=B\bm{e}_z$, the eigenvectors of the Hamiltonian (\ref{eq:Ham}) read
\begin{equation}\label{eq:EigenvectorZZ}
	\vert {\bf k},s\rangle = \frac{1}{\sqrt{N_s(k)}}\left(
			\begin{array}{c}
				(\Delta -s\bar\Delta_{\bf k})ie^{-i\theta_{\bm{k}}}/\alpha k\\
				1
			\end{array}
			\right),
\end{equation}
where $N_s(k)=2+2\Delta (\Delta -s\bar\Delta_{\bf k})/(\alpha k)^{2}$ is the normalization factor and, as before, $\bar\Delta_{\bf k}\equiv(\Delta^2+\alpha^2k^2)^{-1/2}$ is the effective Zeeman energy. Since, by assumption, $|\alpha|k_F \ll E_F$,
we approximate $\bar\Delta_{{\bf k}}$ by $\bar\Delta_{{\bf k}_F}$. Substituting the above eigenvectors  into Eq.~(\ref{eq:GammaDef}), we find the scattering amplitude
 \begin{align}\label{eq:sazz}
	\notag \Gamma_{s_1s_2;s_4s_3}^{(1)}(\bm{k},\bm{k'};\bm{p},\bm{p'})
	=U\left(\prod_{i=1}^4\frac{1}
	{\sqrt{2+\frac{2\Delta(\Delta-s_i\bar\Delta_{{\bf k}_F})}{\alpha^2k_F^2}}}\right)
 	&\left[1+\frac{1}{\alpha^2k_F^2}(\Delta-s_1\bar{\Delta}_{{\bf k}_F})
	(\Delta-s_3\bar{\Delta}_{{\bf k}_F})e^{i(\theta_{\bm{p}}-\theta _{\bm{k}})}\right]\\
 	\times&\left[1+\frac{1}{\alpha^2k_F^2}(\Delta-s_2\bar{\Delta}_{{\bf k}_F})
	(\Delta-s_4\bar{\Delta}_{k_F})e^{i(\theta _{\bm{p'}}-\theta _{\bm{k'}})}\right].
\end{align}
To find the spin susceptibility, we need to know the scattering amplitude to second order in the magnetic field.
Expanding Eq.~(\ref{eq:sazz}) to second order in $\delta\equiv\Delta/|\alpha|k_F$  (note that $s_i^2=1$ and $s_i^3=s_i$) and projecting the amplitude onto the Cooper channel, where the  momenta are correlated in such a way that $\bm{k}'=-\bm{k}$ and $\bm{p}'=-\bm{p}$ or, equivalently, $\theta_{\bm{k}'}=\theta_\bm{k}+\pi$ and $\theta_{\bm{p}'}=\theta_\bm{p}+\pi$, we obtain
\begin{equation}
	\Gamma_{s_{1}s_{2};s_{3}s_{4}}^{(1)}(\bm{k,-k};\bm{p,-p})
	= U_{s_{1}s_{2};s_{3}s_{4}}
	+ V_{s_{1}s_{2};s_{3}s_{4}}e^{i(\theta_{\bm{p}}-\theta _{\bm{k}})}
	+ W_{s_{1}s_{2};s_{3}s_{4}}e^{2i(\theta_{\bm{p}}-\theta _{\bm{k}})},
\end{equation}
where we introduced partial amplitudes
\begin{equation}\label{eq:Uini}
	U_{s_{1}s_{2};s_{3}s_{4}} = \frac{U}{4} + \frac{U}{8}(s_1+s_2+s_3+s_4)\delta
	+ \frac{U}{16}(s_1s_2+s_1s_3+s_1s_4+s_2s_3+s_2s_4+s_3s_4-2)\delta^2 + \mc{O}(\delta^3),
\end{equation}
\begin{align}
	\notag V_{s_{1}s_{2};s_{3}s_{4}} =& \frac{U}{4}(s_1s_3+s_2s_4)
	+ \frac{U}{8}(s_1s_2s_3+s_1s_2s_4+s_1s_3s_4+s_2s_3s_4-s_1-s_2-s_3-s_4)\delta\\
	&+ \frac{U}{8}(1-s_1s_2-s_2s_3-s_3s_4-s_1s_4-s_1s_3-s_2s_4+s_1s_2s_3s_4)\delta^2 + \mc{O}(\delta^3),
\end{align}
\begin{align}\label{eq:Wini}
	\notag W_{s_{1}s_{2};s_{3}s_{4}} =& \frac{U}{4}s_1s_2s_3s_4
	- \frac{U}{8}(s_1s_2s_3+s_1s_2s_4+s_1s_3s_4+s_2s_3s_4)\delta\\
	&+ \frac{U}{16}(s_1s_2+s_1s_3+s_1s_4+s_2s_3+s_2s_4+s_3s_4-2s_1s_2s_3s_4)\delta^2 + \mc{O}(\delta^3).
\end{align}
For $\delta=0$, the partial amplitudes reduce back to Eqs.~(\ref{eq:uvwzerob1}-\ref{eq:uvwzerob3}).
The RG flow equations for the partial amplitudes are the same as in the absence of the magnetic field and are given by Eqs.~(\ref{eq:RGzz1})--(\ref{eq:RGzz3}) with initial conditions (\ref{eq:Uini})--(\ref{eq:Wini}).
Since the differential equations for $U$, $V$, and $W$ are identical, for the sake of argument we select the first one, copied below for the reader's convenience,
\begin{equation}
	-\frac{d}{dL}U_{s_1s_2;s_3s_4}(L) = \sum_s U_{s_1s_2;ss}(L)U_{ss;s_3s_4}(L)
\end{equation}
and introduce the following ansatz
\begin{align}
	U_{s_1s_2;s_3s_4}(L) = \frac{U_{s_1s_2;s_3s_4}(0)+a_{s_1s_2;s_3s_4}L}{1+bL},
\end{align}
which satisfies the initial condition. Substituting this formula into the differential equation for $U$ and multiplying the result by $(1+bL)^2$, we obtain an algebraic equation for $a_{s_1s_2;s_3s_4}$ and $b$
\begin{equation}
	bU_{s_1s_2;s_3s_4}(0)-a_{s_1s_2;s_3s_4}
	+\sum_s(U_{s_1s_2;ss}(0)+a_{s_1s_2;ss}L)(U_{ss;s_3s_4}(0)+a_{ss;s_3s_4}L)=0.
\end{equation}
Grouping coefficients of a polynomial in $L$, we obtain the following set of equations
\begin{align}
	\label{eq:AlgZZ1}
	&bU_{s_1s_2;s_3s_4}(0)-a_{s_1s_2;s_3s_4}+\sum_sU_{s_1s_2;ss}(0)U_{ss;s_3s_4}(0)=0,\\
	&\sum_s[U_{s_1s_2;ss}(0)a_{ss;s_3s_4}+a_{s_1s_2;ss}U_{ss;s_3s_4}(0)]=0,\\
	\label{eq:AlgZZ2}
	&\sum_s a_{s_1s_2;ss}a_{ss;s_3s_4}=0,
\end{align}
which are not independent. Thereby, we choose two out of three equations, namely, Eq.~(\ref{eq:AlgZZ1}) and Eq.~(\ref{eq:AlgZZ2}) with the $s_1=s_2=s_3=s_4=1$ combination of the Rashba indices,  so that there are 17 equations for 17 unknown variables: $a_{s_1s_2;s_3s_4}$ and $b$. The final solutions are listed below
\begin{align}
	&U_{\pm\pm;\pm\pm}(L) = \frac{U}{4}\frac{(1\pm\delta)^2}{1+(1+\delta^2)UL/2},\\
	&U_{ss;-s-s}(L) = U_{s-s;-ss}(L) = U_{s-s;s-s}(L) =
	\frac{U}{4}\frac{1-\delta^2}{1+(1+\delta^2)UL/2},\\
	&U_{\sigma(\pm\mp;\mp\mp)}(L) = \frac{U}{4}\frac{1\mp\delta-\delta^2/2}
	{1+(1+\delta^2)UL/2}.
\end{align}
The same procedure is repeated to derive the $V$- and $W$-amplitudes  listed below
\begin{align}
	&V_{ss;ss}(L) = \frac{U}{2}\frac{1-\delta^2}{1+(1-\delta^2)UL},\\
	&V_{ss;-s-s}(L) = -\frac{U}{2}\frac{1-\delta^2}{1+(1-\delta^2)UL},\\
	&V_{s-s;-ss}(L) = -\frac{U}{2}(1 - \delta^2)
	- \frac{U}{2}\frac{\delta^2 UL}{1+(1-\delta^2)UL},\\
	&V_{s-s;s-s}(L) = \frac{U}{2}(1 + \delta^2)
	-\frac{U}{2}\frac{\delta^2 UL}{1+(1-\delta^2)UL},\\
	&V_{\sigma(\pm\mp;\mp\mp)}(L) = \pm\frac{U}{2}\frac{\delta}{1+(1-\delta^2)UL}
\end{align}
and
\begin{align}
	&W_{\pm\pm;\pm\pm}(L) = \frac{U}{4}\frac{(1\mp\delta)^2}{1+(1+\delta^2)UL/2},\\
	&W_{ss;-s-s}(L) = W_{s-s;-ss}(L) = W_{s-s;s-s}(L) =
	\frac{U}{4}\frac{1-\delta^2}{1+(1+\delta^2)UL/2},\\
	&W_{\sigma(\pm\mp;\mp\mp)}(L) =-\frac{U}{4}\frac{1\pm\delta-\delta^2/2}
	{1+(1+\delta^2)UL/2}.
\end{align}
Summing up all the contributions to the backscattering amplitude, obtained from the Cooper amplitude for  a special choice of the momenta $\bm{p=-k}$, i.e., for $\theta_\bm{p}-\theta_\bm{k}=\pi$, we obtain
\begin{equation}
	\Gamma_{s_1s_2;s_3s_4}(\bm{k,-k};\bm{-k,k}) = U_{s_1s_2;s_3s_4}(L)
	- V_{s_1s_2;s_3s_4}(L) + W_{s_1s_2;s_3s_4}(L).
\end{equation}
To second order in the field, the backscattering amplitudes read
\begin{subequations}
\begin{align}
	&\Gamma_{ss;ss}(\bm{k,-k};\bm{-k,k}) = \left(\frac{U}{2+UL}-\frac{U}{2(1+UL)}\right)
	+ \left(\frac{U}{2(1+UL)^2}+\frac{2U}{(2+UL)^2}\right)\delta^2,\label{G1}\\
	&\Gamma_{ss;-s-s}(\bm{k,-k};\bm{-k,k}) = \left(\frac{U}{2+UL}+\frac{U}{2(1+UL)}\right)
	- \left(\frac{U}{2(1+UL)^2}-\frac{2U}{(2+UL)^2}+\frac{2U}{2+UL}\right)\delta^2,\label{G2}\\
	&\Gamma_{s-s;-ss}(\bm{k,-k};\bm{-k,k}) = \left(\frac{U}{2+UL}+\frac{U}{2}\right)
	- \left(\frac{U}{2(1+UL)}-\frac{2U}{(2+UL)^2}+\frac{2U}{2+UL}\right)\delta^2,\label{G3}\\
	&\Gamma_{s-s;s-s}(\bm{k,-k};\bm{-k,k}) = \left(\frac{U}{2+UL}-\frac{U}{2}\right)
	- \left(\frac{U}{2(1+UL)}-\frac{2U}{(2+UL)^2}+\frac{2U}{2+UL}\right)\delta^2,\label{G4}\\
	&\Gamma_{\sigma(\pm\mp;\mp\mp)}(\bm{k,-k};\bm{-k,k}) =
	\mp \left(\frac{U}{2(1+UL)}+\frac{U}{2+UL}\right)\delta.\label{G5}
\end{align}
\end{subequations}
For $\delta=0$, we reproduce Eqs.~(\ref{eq:ZeroAmplitude_1}-\ref{eq:ZeroAmplitude_2}). In the limit of strong renormalization in the Cooper channel, i.e., for $UL\to\infty$, the field-dependent terms in Eqs.~(\ref{G1}-\ref{G5}) vanish. {\em A posteriori}, this explains why we could obtain the renormalized result (\ref{eq:ChiZZRen}) for $\chi_{zz}$ in the main text using only the zero-field amplitudes.

\subsubsection{\label{app:ChiZZRen}Renormalization of $\chi_{zz}$ in the Cooper channel}

As in Sec.~\ref{sec:InfZZ}, the thermodynamic potential in the presence of Cooper renormalization is obtained by substituting the renormalized scattering amplitudes (\ref{G1}-\ref{G5}) into Eq.~(\ref{wed1})
\begin{subequations}
\begin{align}
	\delta\Xi _{zz}= - \frac12T\sum_{\Omega}\int\frac{qdq}{2\pi}\bigg[
	&\frac12\left(\frac{U}{2+UL}-\frac{U}{2}\right)^2(\Pi_{+-}^2+\Pi_{-+}^2-2\Pi_0^2)
	+\left(\frac{U}{2+UL}+\frac{U}{2(1+UL)}\right)^2(\Pi_{+-}\Pi_{-+}-\Pi_0^2)\label{Xi1}\\
	\notag -&\frac{U^2(16+32UL+22U^2L^2+6U^2L^3+U^4L^4)}{4(1+UL)^2(2+UL)^2}\Pi_0^2\bigg]\\
	\notag -\frac{\Delta^2}{\alpha^2k_F^2}T\sum_{\Omega}\int\frac{qdq}{2\pi}\bigg[
	&\left(\frac{U}{2(1+UL)}+\frac{U}{2+UL}\right)^2\Pi_0(\Pi_{+-}+\Pi_{-+}-2\Pi_0)\\
	+&\frac12\left(\frac{U}{2(1+UL)}-\frac{2U}{(2+UL)^2}+\frac{2U}{2+UL}\right)
	\left(\frac{U}{2+UL}
	-\frac{U}{2}\right)(\Pi_{+-}^2+\Pi_{-+}^2-2\Pi_0^2)\\
	\notag -&\left(\frac{U}{2(1+UL)^2}-\frac{2U}{(2+UL)^2}+\frac{2U}{2+UL}\right)
	\left(\frac{U}{2+UL}+\frac{U}{2(1+UL)}\right)(\Pi_{+-}\Pi_{-+}-\Pi_0^2)\\
	-&\frac{U^4L^2(12+18UL+7U^2L^2)}{2(1+UL)^2(2+UL)^3}\Pi_0^2\bigg].\label{Xi2}
\end{align}
\end{subequations}
The first part of $\delta\Xi_{zz}$ (\ref{Xi1}) came from the field-independent terms in the scattering amplitudes. This part depends on the magnetic field through the combinations of the polarization bubbles. The second part (\ref{Xi2}) already contains a field-dependent prefactor ($\Delta^2$) resulting from the field-dependent terms in the scattering amplitudes. Therefore, the polarization bubbles in this part can be evaluated in zero field.
The integrals over $q$ along with the summation over the Matsubara frequency $\Omega$
have already been performed in Sec. \ref{sec:SS}. Note that in each square bracket the last term
(proportional to $\Pi_0^2$) depends neither on the field nor on the SOI. In fact, it can be shown\cite{chubukov05a,chubukov05b} that this formally divergent contribution has a cubic dependence on temperature, $T\sum_{\Omega}\int qdq\Pi_0^2\propto T^3$, thus it adds a higher order correction and can be dropped. The final result reads
\begin{align}
	\notag \delta\Xi _{zz}
	=-&\frac{T^3}{8\pi v_F^2}\left(\frac{mU}{2\pi}\right)^2\bigg[
	\left(\frac{1}{2+UL}-\frac{1}{2}\right)^2
	\mc{F}\left(\frac{2\sqrt{\alpha^2k_F^2+\Delta^2}}{T}\right)\bigg]\\	
	\notag -&\frac{\Delta^2}{\alpha^2k_F^2}\frac{T^3}{2\pi v_F^2}\left(\frac{mU}{2\pi}\right)^2\bigg[
	\left(\frac{1}{2(1+UL)}+\frac{1}{2+UL}\right)^2\mc{F}\left(\frac{|\alpha|k_F}{T}\right)\\
	&+\frac12\left(\frac{1}{2(1+UL)}-\frac{2}{(2+UL)^2}+\frac{2}{2+UL}\right)
	\left(\frac{1}{2+UL}
	-\frac{1}{2}\right)\mc{F}\left(\frac{2|\alpha|k_F}{T}\right)\bigg].\label{Xi3}
\end{align}
The spin susceptibility is obtained by expanding Eq.~(\ref{Xi3}) further to order $\Delta^2$.
We also need to recall that our treatment of Cooper renormalization is only valid for $T\ll |\alpha|k_F$, because we kept  only the $T$-  but not $\alpha$-dependent Cooper logarithms (see the discussion at the end of Sec.~\ref{sec:Resum_1}). Therefore, the function $\mc{F}$ and its derivative should be replaced by their large-argument forms, Eq.~(\ref{Flarge}). Doing so, we obtain the final result for the nonanalytic part of the $\chi_{zz}$, presented in Eq.~(\ref{eq:chizzfinal}) in the main text.

\subsection{In-plane magnetic field}
\subsubsection{\label{app:RGxx} RG flow of the scattering amplitudes}

For the in-plane magnetic field, $\bm{B}=B\bm{e}_x$, the RG flow of the scattering amplitudes is more cumbersome because the eigenvectors of Hamiltonian (\ref{eq:Ham}) depend in a complicated way on the angle between the magnetic field and the electron momentum
\begin{equation}\label{eq:EigenvectorXX}
	\vert {\bf k},s\rangle = \frac{1}{\sqrt{2}}\left(
			\begin{array}{c}
				s\bar\Delta_{\bf k}/(\Delta-ie^{i\theta_\bm{k}}\alpha k)\\
				1
			\end{array}
			\right),
\end{equation}
where $\bar\Delta_{\bf k}\equiv(\Delta^2+2\alpha k\Delta\sin\theta_\bm{k}+\alpha^2k^2)^{-1/2}$ is the effective Zeeman energy. For that reason, the (double) Fourier series of the scattering amplitude (\ref{eq:GammaDef})  in the angles $\theta_\bm{k}$ and $\theta_\bm{p}$ contains infinitely many harmonics:
\begin{equation}
	\Gamma_{s_{1}s_{2};s_{3}s_{4}}^{(1)}(\bm{k,-k};\bm{p,-p})
	= \sum_{m,n=-\infty}^{\infty}\Gamma^{(1)\{m,n\}}_{s_{1}s_{2};s_{3}s_{4}}
	e^{i m\theta_{\bm{p}}}e^{i n\theta_{\bm{k}}},
\end{equation}
To second order in the field, however, the number of nonvanishing harmonics is limited to 15. Indeed, expanding the eigenvector (\ref{eq:EigenvectorXX}) to second order in $\delta\equiv\Delta/|\alpha|k_F$ as
\begin{equation}\label{eq:EigenvectorXX1}
	\vert {\bf k},s\rangle = \frac{1}{\sqrt{2}}\left(
			\begin{array}{c}
				sie^{-i\theta_{\bm{k}}}[1-i\cos\theta_\bm{k}\delta
				-(\cos^2\theta_\bm{k}-i\sin2\theta_\bm{k})\delta^2/2]\\
				1
			\end{array}
			\right) + \mc{O}(\delta^3)
\end{equation}
and substituting Eq.~(\ref{eq:EigenvectorXX1}) into the scattering amplitude  (\ref{eq:GammaDef}) with $\bm{k^\prime=-k}$ and $\bm{p^\prime=-p}$, we obtain
\begin{align}\label{eq:GammaXXIni}
	\notag \Gamma_{s_{1}s_{2};s_{3}s_{4}}^{(1)}&(\bm{k,-k};\bm{p,-p})
	= U_{s_{1}s_{2};s_{3}s_{4}}
	+ V_{s_{1}s_{2};s_{3}s_{4}}e^{i\theta_{\bm{p}}}e^{-i\theta_{\bm{k}}}
	+ W_{s_{1}s_{2};s_{3}s_{4}}e^{2i\theta_{\bm{p}}}e^{-2i\theta_{\bm{k}}}\\
	\notag &+ A_{s_{1}s_{2};s_{3}s_{4}}e^{i\theta_{\bm{p}}}e^{i\theta_{\bm{k}}}
	+ B_{s_{1}s_{2};s_{3}s_{4}}e^{-i\theta_{\bm{p}}}e^{-i\theta_{\bm{k}}}
	\notag + F_{s_{1}s_{2};s_{3}s_{4}}e^{2i\theta_{\bm{p}}}e^{-i\theta_{\bm{k}}}
	+ G_{s_{1}s_{2};s_{3}s_{4}}e^{i\theta_{\bm{p}}}e^{-2i\theta_{\bm{k}}}\\
	\notag &+ H_{s_{1}s_{2};s_{3}s_{4}}e^{i\theta_{\bm{p}}}
	+ J_{s_{1}s_{2};s_{3}s_{4}}e^{-i\theta_{\bm{k}}}
	\notag + L_{s_{1}s_{2};s_{3}s_{4}}e^{2i\theta_{\bm{p}}}
	+ M_{s_{1}s_{2};s_{3}s_{4}}e^{-2i\theta_{\bm{k}}}\\
	& + P_{s_{1}s_{2};s_{3}s_{4}}e^{3i\theta_{\bm{p}}}e^{-i\theta_{\bm{k}}}
	+ Q_{s_{1}s_{2};s_{3}s_{4}}e^{i\theta_{\bm{p}}}e^{-3i\theta_{\bm{k}}}
	+ R_{s_{1}s_{2};s_{3}s_{4}}e^{4i\theta_{\bm{p}}}e^{-2i\theta_{\bm{k}}}
	+ S_{s_{1}s_{2};s_{3}s_{4}}e^{2i\theta_{\bm{p}}}e^{-4i\theta_{\bm{k}}} + \mc{O}(\delta^3).
\end{align}
Here,
\begin{align}
	\label{eq:RGIniFirst}
	U_{s_{1}s_{2};s_{3}s_{4}} = \Gamma^{(1)\{0,0\}}_{s_{1}s_{2};s_{3}s_{4}}
	&= (U/16)[4+(s_1s_3+s_2s_4)\delta^2],\\
	V_{s_{1}s_{2};s_{3}s_{4}} = \Gamma^{(1)\{1,-1\}}_{s_{1}s_{2};s_{3}s_{4}}
	&= (U/8)(s_1s_3+s_2s_4)(2-\delta^2),\\
	W_{s_{1}s_{2};s_{3}s_{4}} = \Gamma^{(1)\{2,-2\}}_{s_{1}s_{2};s_{3}s_{4}}
	&= (U/16)[4s_1s_2s_3s_4+(s_1s_3+s_2s_4)\delta^2],\\
	A_{s_{1}s_{2};s_{3}s_{4}} = \Gamma^{(1)\{1,1\}}_{s_{1}s_{2};s_{3}s_{4}}
	&= (U/32)(s_1s_3+s_2s_4)\delta^2\\
	B_{s_{1}s_{2};s_{3}s_{4}} = \Gamma^{(1)\{-1,-1\}}_{s_{1}s_{2};s_{3}s_{4}}
	&= A_{s_{1}s_{2};s_{3}s_{4}},\\
	F_{s_{1}s_{2};s_{3}s_{4}} = \Gamma^{(1)\{2,-1\}}_{s_{1}s_{2};s_{3}s_{4}}
	&= (U/8)i(s_1s_3-s_2s_4)\delta,\\
	G_{s_{1}s_{2};s_{3}s_{4}} = \Gamma^{(1)\{1,-2\}}_{s_{1}s_{2};s_{3}s_{4}}
	&= -F_{s_{1}s_{2};s_{3}s_{4}},\\
	H_{s_{1}s_{2};s_{3}s_{4}} = \Gamma^{(1)\{1,0\}}_{s_{1}s_{2};s_{3}s_{4}}
	&= -F_{s_{1}s_{2};s_{3}s_{4}},\\
	J_{s_{1}s_{2};s_{3}s_{4}} = \Gamma^{(1)\{0,-1\}}_{s_{1}s_{2};s_{3}s_{4}}
	&= F_{s_{1}s_{2};s_{3}s_{4}},\\
	L_{s_{1}s_{2};s_{3}s_{4}} = \Gamma^{(1)\{2,0\}}_{s_{1}s_{2};s_{3}s_{4}}
	&= (U/16)(s_1s_3+s_2s_4+2s_1s_2s_3s_4)\delta^2,\\
	M_{s_{1}s_{2};s_{3}s_{4}} = \Gamma^{(1)\{0,-2\}}_{s_{1}s_{2};s_{3}s_{4}}
	&= L_{s_{1}s_{2};s_{3}s_{4}},
\end{align}
\begin{align}
	P_{s_{1}s_{2};s_{3}s_{4}} = \Gamma^{(1)\{3,-1\}}_{s_{1}s_{2};s_{3}s_{4}}
	&= -3A_{s_{1}s_{2};s_{3}s_{4}},\\
	Q_{s_{1}s_{2};s_{3}s_{4}} = \Gamma^{(1)\{1,-3\}}_{s_{1}s_{2};s_{3}s_{4}}
	&= P_{s_{1}s_{2};s_{3}s_{4}},\\
	R_{s_{1}s_{2};s_{3}s_{4}} = \Gamma^{(1)\{4,-2\}}_{s_{1}s_{2};s_{3}s_{4}}
	&= -(U/8)s_1s_2s_3s_4\delta^2,\\
	\label{eq:RGIniLast}
	S_{s_{1}s_{2};s_{3}s_{4}} = \Gamma^{(1)\{2,-4\}}_{s_{1}s_{2};s_{3}s_{4}}
	&= R_{s_{1}s_{2};s_{3}s_{4}}
\end{align}
The second-order amplitude is derived from Eq.~(\ref{eq:GammaNDef}) with $n=2$
\begin{align}
	\notag &\Gamma_{s_1s_2;s_4s_3}^{(2)}(\bm{k},-\bm{k};\bm{p},-\bm{p})
	= -L\sum_{s}\{U_{s_{1}s_{2};ss}U_{ss;s_{3}s_{4}}
	+H_{s_{1}s_{2};ss}J_{ss;s_{3}s_{4}}\\
	\notag &+ [V_{s_{1}s_{2};ss}V_{ss;s_{3}s_{4}}+F_{s_{1}s_{2};ss}G_{ss;s_{3}s_{4}}
	+J_{s_{1}s_{2};ss}H_{ss;s_{3}s_{4}}]e^{i\theta_{\bm{p}}}e^{-i\theta_{\bm{k}}}
	+[W_{s_{1}s_{2};ss}W_{ss;s_{3}s_{4}}+G_{s_{1}s_{2};ss}F_{ss;s_{3}s_{4}}]
	e^{2i\theta_{\bm{p}}}e^{-2i\theta_{\bm{k}}}\\
	\notag &+A_{s_{1}s_{2};ss}V_{ss;s_{3}s_{4}}e^{i\theta_{\bm{p}}}e^{i\theta_{\bm{k}}}
	 +V_{s_{1}s_{2};ss}B_{ss;s_{3}s_{4}}e^{-i\theta_{\bm{p}}}e^{-i\theta_{\bm{k}}}
	+[V_{s_{1}s_{2};ss}F_{ss;s_{3}s_{4}}+F_{s_{1}s_{2};ss}W_{ss;s_{3}s_{4}}]
	e^{2i\theta_{\bm{p}}}e^{-i\theta_{\bm{k}}}\\
	\notag &+[W_{s_{1}s_{2};ss}G_{ss;s_{3}s_{4}}+G_{s_{1}s_{2};ss}V_{ss;s_{3}s_{4}}]
	e^{i\theta_{\bm{p}}}e^{-2i\theta_{\bm{k}}}
	+[H_{s_{1}s_{2};ss}V_{ss;s_{3}s_{4}}+U_{s_{1}s_{2};ss}H_{ss;s_{3}s_{4}}]
	e^{i\theta_{\bm{p}}}\\
	\notag &+[V_{s_{1}s_{2};ss}J_{ss;s_{3}s_{4}}+J_{s_{1}s_{2};ss}U_{ss;s_{3}s_{4}}]
	e^{-i\theta_{\bm{k}}}
	+[H_{s_{1}s_{2};ss}F_{ss;s_{3}s_{4}}+U_{s_{1}s_{2};ss}L_{ss;s_{3}s_{4}}
	+L_{s_{1}s_{2};ss}W_{ss;s_{3}s_{4}}]e^{2i\theta_{\bm{p}}}\\
	\notag &+[W_{s_{1}s_{2};ss}M_{ss;s_{3}s_{4}}+G_{s_{1}s_{2};ss}J_{ss;s_{3}s_{4}}
	+M_{s_{1}s_{2};ss}U_{ss;s_{3}s_{4}}]e^{-2i\theta_{\bm{k}}}
	+V_{s_{1}s_{2};ss}P_{ss;s_{3}s_{4}}e^{3i\theta_{\bm{p}}}e^{-i\theta_{\bm{k}}}\\
	&+Q_{s_{1}s_{2};ss}V_{ss;s_{3}s_{4}}e^{i\theta_{\bm{p}}}e^{-3i\theta_{\bm{k}}}
	+W_{s_{1}s_{2};ss}R_{ss;s_{3}s_{4}}e^{4i\theta_{\bm{p}}}e^{-2i\theta_{\bm{k}}}
	+S_{s_{1}s_{2};ss}W_{ss;s_{3}s_{4}}e^{2i\theta_{\bm{p}}}e^{-4i\theta_{\bm{k}}}\}
	+\mc{O}(\delta^3),\label{gamma2xx}
\end{align}
where $L=(m/2\pi)\ln(\Lambda/T)$.
The second-order amplitude contains the same combinations of the harmonics
$e^{im\theta_\bm{p}}$ and $e^{in\theta_\bm{k}}$ as the first-order amplitude, which proves the group property.
The RG flow equations are obtained by replacing the left-hand side of Eq.~(\ref{gamma2xx}) by the bare amplitude, letting the coefficients $U\dots S$ to depend on $L$, and differentiating with respect to $L$:
\begin{align}
	\label{eq:RGXXfirst}
	-\frac{d}{dL}U_{s_{1}s_{2};s_{3}s_{4}}(L) &=
	 U_{s_{1}s_{2};ss}(L)U_{ss;s_{3}s_{4}}(L)+H_{s_{1}s_{2};ss}(L)J_{ss;s_{3}s_{4}}(L),\\
	-\frac{d}{dL}V_{s_{1}s_{2};s_{3}s_{4}}(L) &= V_{s_{1}s_{2};ss}(L)V_{ss;s_{3}s_{4}}(L)
	 +F_{s_{1}s_{2};ss}(L)G_{ss;s_{3}s_{4}}(L)+J_{s_{1}s_{2};ss}(L)H_{ss;s_{3}s_{4}}(L),\\
	-\frac{d}{dL}W_{s_{1}s_{2};s_{3}s_{4}}(L) &=
	 W_{s_{1}s_{2};ss}(L)W_{ss;s_{3}s_{4}}(L)+G_{s_{1}s_{2};ss}(L)F_{ss;s_{3}s_{4}}(L),\\
	-\frac{d}{dL}A_{s_{1}s_{2};s_{3}s_{4}}(L) &= A_{s_{1}s_{2};ss}(L)V_{ss;s_{3}s_{4}}(L),\\
	-\frac{d}{dL}B_{s_{1}s_{2};s_{3}s_{4}}(L) &= V_{s_{1}s_{2};ss}(L)B_{ss;s_{3}s_{4}}(L),\\
	-\frac{d}{dL}F_{s_{1}s_{2};s_{3}s_{4}}(L) &=
	 V_{s_{1}s_{2};ss}(L)F_{ss;s_{3}s_{4}}(L)+F_{s_{1}s_{2};ss}(L)W_{ss;s_{3}s_{4}}(L),\\
	-\frac{d}{dL}G_{s_{1}s_{2};s_{3}s_{4}}(L) &=
	 W_{s_{1}s_{2};ss}(L)G_{ss;s_{3}s_{4}}(L)+G_{s_{1}s_{2};ss}(L)V_{ss;s_{3}s_{4}}(L),\\
	-\frac{d}{dL}H_{s_{1}s_{2};s_{3}s_{4}}(L) &=
	 H_{s_{1}s_{2};ss}(L)V_{ss;s_{3}s_{4}}(L)+U_{s_{1}s_{2};ss}(L)H_{ss;s_{3}s_{4}}(L),\\
	-\frac{d}{dL}J_{s_{1}s_{2};s_{3}s_{4}}(L) &=
	 V_{s_{1}s_{2};ss}(L)J_{ss;s_{3}s_{4}}(L)+J_{s_{1}s_{2};ss}(L)U_{ss;s_{3}s_{4}}(L),\\
	-\frac{d}{dL}L_{s_{1}s_{2};s_{3}s_{4}}(L) &= H_{s_{1}s_{2};ss}(L)F_{ss;s_{3}s_{4}}(L)
	 +U_{s_{1}s_{2};ss}(L)L_{ss;s_{3}s_{4}}(L)+L_{s_{1}s_{2};ss}(L)W_{ss;s_{3}s_{4}}(L),\\
	-\frac{d}{dL}M_{s_{1}s_{2};s_{3}s_{4}}(L) &= W_{s_{1}s_{2};ss}(L)M_{ss;s_{3}s_{4}}(L)
	 +G_{s_{1}s_{2};ss}(L)J_{ss;s_{3}s_{4}}(L)+M_{s_{1}s_{2};ss}(L)U_{ss;s_{3}s_{4}}(L),\\
	-\frac{d}{dL}P_{s_{1}s_{2};s_{3}s_{4}}(L) &= V_{s_{1}s_{2};ss}(L)P_{ss;s_{3}s_{4}}(L),\\
	-\frac{d}{dL}Q_{s_{1}s_{2};s_{3}s_{4}}(L) &= Q_{s_{1}s_{2};ss}(L)V_{ss;s_{3}s_{4}}(L),\\
	-\frac{d}{dL}R_{s_{1}s_{2};s_{3}s_{4}}(L) &= W_{s_{1}s_{2};ss}(L)R_{ss;s_{3}s_{4}}(L),\\
	\label{eq:RGXXlast}
	-\frac{d}{dL}S_{s_{1}s_{2};s_{3}s_{4}}(L) &= S_{s_{1}s_{2};ss}(L)W_{ss;s_{3}s_{4}}(L),
\end{align}
where summation over the repeated index $s$ is implied. The initial conditions are given by $X_{s_1,s_2;s_3,s_4}(0)=X_{s_1,s_2;s_3,s_4}$ with $X=U\dots S$.
Since it is very difficult to solve this system of differential equations analytically, a new approach is required. In what follows, we will determine $U(L),\dots,S(L)$ for a few lowest orders
in the \lq\lq RG time\rq\rq\/  $L$ and then make a guess for a form of an arbitrary-order term. The RG equations will then provide a necessary check as to whether our guess, based on the perturbative calculation, gives a  correct answer.

A few lowest order amplitudes can be derived perturbatively from Eq.~(\ref{eq:GammaNDef}), copied here for the reader's convenience
\begin{equation}
	\Gamma_{s_1s_2;s_4s_3}^{(j)}(\bm{k},-\bm{k};\bm{p},-\bm{p})(L)
	= -L\sum_{s}\int_0^{2\pi}\frac{d\theta_l}{2\pi}
	\Gamma_{s_1s_2;ss}^{(j-1)}(\bm{k},-\bm{k};\bm{l},-\bm{l})
	\Gamma_{ss;s_4s_3}^{(1)}(\bm{l},-\bm{l};\bm{p},-\bm{p})
\end{equation}
with $j\geq2$ standing for order of the perturbation theory. Since the scattering amplitudes depend on angles $\theta_\bm{k}$ and $\theta_\bm{p}$, they can be decomposed order by order into the Fourier series
\begin{equation}
	\Gamma_{s_{1}s_{2};s_{3}s_{4}}^{(j)}(\bm{k,-k};\bm{p,-p})
	= \sum_{m,n=-\infty}^{\infty}\Gamma^{(j)\{m,n\}}_{s_{1}s_{2};s_{3}s_{4}}
	e^{mi\theta_{\bm{p}}}e^{ni\theta_{\bm{k}}},
\end{equation}
where the coefficients in front of $e^{im\theta_\bm{p}}e^{in\theta_\bm{k}}$ are determined using the orthogonality property
\begin{equation}
	\Gamma_{s_1s_2;s_4s_3}^{(j)\{m,n\}}(\bm{k},-\bm{k};\bm{p},-\bm{p})
	= \int_{0}^{2\pi}\frac{d\theta_{\bm{k}}}{2\pi}\int_{0}^{2\pi}\frac{d\theta_{\bm{p}}}{2\pi}
	\Gamma_{s_1s_2;s_4s_3}^{(j)}(\bm{k},-\bm{k};\bm{p},-\bm{p})
	e^{-im\theta_\bm{p}}e^{-in\theta_\bm{k}}.
\end{equation}
Resumming the coefficients of $e^{im\theta_\bm{p}}e^{in\theta_\bm{k}}$ to infinite order (with $m$ and $n$ being fixed)
\begin{equation}
	\Gamma_{s_1s_2;s_4s_3}^{(\infty)\{m,n\}}(\bm{k},-\bm{k};\bm{p},-\bm{p})
	=\sum_{j=1}^{\infty}\Gamma_{s_1s_2;s_4s_3}^{(j)\{m,n\}}(\bm{k},-\bm{k};\bm{p},-\bm{p})
\end{equation}
we can find the renormalized amplitudes. For each combination of partial harmonics, which
occurs to second order in the magnetic field, we derive explicitly the scattering amplitudes up to seventh order in the Cooper channel renormalization parameter $UL$ and then make a guess for general $j$-th order amplitude.
The final result is obtained by resumming these amplitudes to infinite order and then substituted
into the RG flow equations to check the correctness of our guess. In all cases, the guess turns out to be correct. All \emph{nonzero} RG charges as well as their large $L$ limits are listed below. We begin with the $n=m=0$ and $n=m=1$ harmonics, given by
\begin{align}
	U_{ss;ss}(L) &= \frac{U}{2}\frac{1}{2+UL}+\frac{U}{8}\delta^2
	= \frac{U}{8}\delta^2 + \mc{O}(\ln^{-1}T),\\
	U_{ss;-s-s}(L) &= \frac{U}{2}\frac{1}{2+UL}-\frac{U}{8}\delta^2
	= -\frac{U}{8}\delta^2 + \mc{O}(\ln^{-1}T),\\
	U_{s-s;-ss}(L) &= \frac{U}{2}\frac{1}{2+UL}-\frac{U}{8}\frac{1}{1+UL}\delta^2
	=  \mc{O}(\ln^{-1}T),\\
	U_{s-s;s-s}(L) &= \frac{U}{2}\frac{1}{2+UL}+\frac{U}{8}\frac{1}{1+UL}\delta^2
	=  \mc{O}(\ln^{-1}T),\\
	U_{\sigma(\pm\mp;\mp\mp)}(L) &= \frac{U}{2}\frac{1}{2+UL} = \mc{O}(\ln^{-1}T),
\end{align}
\begin{align}
	V_{ss;ss}(L) &= \frac{U}{2}\frac{1}{1+UL}-\frac{U}{4}\frac{1}{(1+UL)^2}\delta^2
	= \mc{O}(\ln^{-1}T),\\
	V_{ss;-s-s}(L) &= -\frac{U}{2}\frac{1}{1+UL}+\frac{U}{4}\frac{1}{(1+UL)^2}\delta^2
	= \mc{O}(\ln^{-1}T),\\
	V_{s-s;-ss}(L) &= -\frac{U}{2}+\frac{U}{4}(1+UL)\delta^2
	= - \frac{U}{2} + \mc{O}(U^2), \label{eq:Vgrow1} \\
	V_{s-s;s-s}(L) &= \frac{U}{2}-\frac{U}{4}(1+UL)\delta^2
	= \frac{U}{2} + \mc{O}(U^2). \label{eq:Vgrow2}
\end{align}
An important remark should be made at this point: in addition to
amplitudes which flow either to zero or to finite values at low temperatures, there are also
amplitudes which grow logarithmically at low temperatures, namely, the amplitudes in Eqs.(\ref{eq:Vgrow1}) and (\ref{eq:Vgrow2}). This peculiar feature, which occurs only in the presence of both the SOI and in-plane magnetic field,
may indicate a phase transition below certain field-dependent temperature
or it may be an artifact of the expansion to lowest order in $\delta^2$.
 In the derivation of the spin susceptibility that follows in App.~\ref{app:ChiXXRen}, we assume that the electron gas is far above
 the temperature below which the instability becomes important,
  i.e., that $UL\ll1/\delta^2$, so that the effect of the instability can be neglected but the nonperturbative regime of Cooper renormalization, where $1\ll UL\ll 1/\delta^2$, can still be accessed.

The remaining harmonics are
\begin{align}
	W_{ss;ss}(L) &= U_{ss;ss}(L),\\
	W_{ss;-s-s}(L) &= U_{ss;-s-s}(L),\\
	W_{s-s;-ss}(L) &= U_{s-s;-ss}(L),\\
	W_{s-s;s-s}(L) &= U_{s-s;s-s}(L),\\
	W_{\sigma(\pm\mp;\mp\mp)}(L) &= -U_{\sigma(\pm\mp;\mp\mp)}(L),
\end{align}
\begin{align}
	A_{ss;ss}(L) &= -A_{ss;-s-s}(L) = \frac{U}{16}\frac{1}{1+UL}\delta^2 = \mc{O}(\ln^{-1}T),\\
	A_{s-s;-ss}(L) &= -A_{s-s;s-s}(L) = -\frac{U}{16}\delta^2
\end{align}
\begin{equation}
	B_{s_1s_2;s_3s_4}(L)=A_{s_1s_2;s_3s_4}(L).
\end{equation}
\begin{align}
	F_{\pm\mp;\mp\mp}(L) &= -\frac{U}{4}i\delta,\\
	F_{\mp\pm;\mp\mp}(L) &= \frac{U}{4}i\delta,\\
	F_{\mp\mp;\pm\mp}(L) &= -\frac{U}{4}\frac{1}{1+UL}i\delta = \mc{O}(\ln^{-1}T),\\
	F_{\mp\mp;\mp\pm}(L) &= \frac{U}{4}\frac{1}{1+UL}i\delta = \mc{O}(\ln^{-1}T),
\end{align}
\begin{align}
	G_{\pm\mp;\mp\mp}(L) &= \frac{U}{4}\frac{1}{1+UL}i\delta = \mc{O}(\ln^{-1}T),\\
	G_{\mp\pm;\mp\mp}(L) &= -\frac{U}{4}\frac{1}{1+UL}i\delta = \mc{O}(\ln^{-1}T),\\
	G_{\mp\mp;\pm\mp}(L) &= \frac{U}{4}i\delta,\\
	G_{\mp\mp;\mp\pm}(L) &= -\frac{U}{4}i\delta,
\end{align}
\begin{equation}
	H_{s_1s_2;s_3s_4}(L)=G_{s_1s_2;s_3s_4}(L)
\end{equation}
\begin{equation}
	J_{s_1s_2;s_3s_4}(L)=F_{s_1s_2;s_3s_4}(L)
\end{equation}
\begin{align}
	L_{ss;ss}(L) &= \frac{U}{2}\frac{1}{(2+UL)^2}+\frac{U}{8}\delta^2
	= \frac{U}{8}\delta^2 + \mc{O}(\ln^{-1}T),\\
	L_{ss;-s-s}(L) &= \frac{U}{2}\frac{1}{(2+UL)^2}-\frac{U}{8}\delta^2
	= -\frac{U}{8}\delta^2 + \mc{O}(\ln^{-1}T),\\
	L_{s-s;-ss}(L) &= \frac{U}{8}\frac{(4+3UL)UL}{(1+UL)(2+UL)^2}\delta^2 = \mc{O}(\ln^{-1}T),\\
	L_{s-s;s-s}(L) &= \frac{U}{8}\frac{8+12UL+5U^2L^2}{(1+UL)(2+UL)^2}\delta^2 = \mc{O}(\ln^{-1}T),\\
	L_{\pm\mp;\mp\mp}(L) &= L_{\mp\pm;\mp\mp}(L) = -\frac{U}{2}\frac{1+UL}{(2+UL)^2}\delta^2
	= \mc{O}(\ln^{-1}T),\\
	L_{\mp\mp;\pm\mp}(L) &= L_{\mp\mp;\mp\pm}(L) = -\frac{U}{2}\frac{1}{(2+UL)^2}\delta^2
	= \mc{O}(\ln^{-1}T).
\end{align}
\begin{align}
	M_{ss;ss}(L) &= L_{ss;ss}(L),\\
	M_{ss;-s-s}(L) &= L_{ss;-s-s}(L),\\
	M_{s-s;-ss}(L) &= L_{s-s;-ss}(L),\\
	M_{s-s;s-s}(L) &= L_{s-s;s-s}(L),\\
	M_{\pm\mp;\mp\mp}(L) &= L_{\mp\pm;\mp\mp}(L) = -\frac{U}{2}\frac{1}{(2+UL)^2}\delta^2
	= \mc{O}(\ln^{-1}T),\\
	M_{\mp\mp;\pm\mp}(L) &= L_{\mp\mp;\mp\pm}(L) = -\frac{U}{2}\frac{1+UL}{(2+UL)^2}\delta^2
	= \mc{O}(\ln^{-1}T),
\end{align}
\begin{equation}
	P_{s_1s_2;s_3s_4}(L)=Q_{s_1s_2;s_3s_4}(L)=-3A_{s_1s_2;s_3s_4}(L),
\end{equation}
\begin{align}
	R_{ss;ss}(L) = R_{ss;-s-s}(L) = R_{s-s;-ss}&(L) = R_{s-s;s-s}(L)
	= -\frac{U}{4}\frac{1}{2+UL}\delta^2 = \mc{O}(\ln^{-1}T),\\
	R_{\sigma(\pm\mp;\mp\mp)}(L) &= \frac{U}{4}\frac{1}{2+UL}\delta^2 = \mc{O}(\ln^{-1}T),
\end{align}
\begin{equation}
	S_{s_1s_2;s_3s_4}(L)=R_{s_1s_2;s_3s_4}(L).
\end{equation}
It can be readily verified that all the amplitudes satisfy RG equations (\ref{eq:RGXXfirst})--(\ref{eq:RGXXlast}) with initial conditions (\ref{eq:RGIniFirst})--(\ref{eq:RGIniLast} up to $\mc{O}(\delta^3)$ accuracy.

Finally, the renormalized scattering amplitude is given by
\begin{align}
	\notag \Gamma_{s_{1}s_{2};s_{3}s_{4}}&(\bm{k,-k};\bm{p,-p})
	= U_{s_{1}s_{2};s_{3}s_{4}}(L)
	+ V_{s_{1}s_{2};s_{3}s_{4}}(L)e^{i\theta_{\bm{p}}}e^{-i\theta_{\bm{k}}}
	+ W_{s_{1}s_{2};s_{3}s_{4}}(L)e^{2i\theta_{\bm{p}}}e^{-2i\theta_{\bm{k}}}\\
	\notag &+ A_{s_{1}s_{2};s_{3}s_{4}}(L)e^{i\theta_{\bm{p}}}e^{i\theta_{\bm{k}}}
	+ B_{s_{1}s_{2};s_{3}s_{4}}(L)e^{-i\theta_{\bm{p}}}e^{-i\theta_{\bm{k}}}
	+ F_{s_{1}s_{2};s_{3}s_{4}}(L)e^{2i\theta_{\bm{p}}}e^{-i\theta_{\bm{k}}}
	+ G_{s_{1}s_{2};s_{3}s_{4}}(L)e^{i\theta_{\bm{p}}}e^{-2i\theta_{\bm{k}}}\\
	\notag &+ H_{s_{1}s_{2};s_{3}s_{4}}(L)e^{i\theta_{\bm{p}}}
	+ J_{s_{1}s_{2};s_{3}s_{4}}(L)e^{-i\theta_{\bm{k}}}
	+ L_{s_{1}s_{2};s_{3}s_{4}}(L)e^{2i\theta_{\bm{p}}}
	+ M_{s_{1}s_{2};s_{3}s_{4}}(L)e^{-2i\theta_{\bm{k}}}\\
	&+ P_{s_{1}s_{2};s_{3}s_{4}}(L)e^{3i\theta_{\bm{p}}}e^{-i\theta_{\bm{k}}}
	+ Q_{s_{1}s_{2};s_{3}s_{4}}(L)e^{i\theta_{\bm{p}}}e^{-3i\theta_{\bm{k}}}
	+ R_{s_{1}s_{2};s_{3}s_{4}}(L)e^{4i\theta_{\bm{p}}}e^{-2i\theta_{\bm{k}}}
	+ S_{s_{1}s_{2};s_{3}s_{4}}(L)e^{2i\theta_{\bm{p}}}e^{-4i\theta_{\bm{k}}}.
\end{align}

\subsubsection{\label{app:ChiXXRen}Renormalization of $\chi_{xx}$}

As for the transverse-field case, the free energy for the in-plane magnetic field is found by replacing the bare interaction $U$ in Eq.~(\ref{eq:Xi2XX}) by the renormalized vertex $\Gamma$
\begin{equation}
	\delta\Xi _{xx}=-\frac{1}{4}\int_{0}^{2\pi}\frac{d\theta_\bm{k}}{2\pi}
	T\sum_{\Omega}{\sum_{\{s_{i}\}}}\int_{0}^{\infty}\frac{qdq}{2\pi}
	\Gamma_{s_1s_4;s_3s_2}(\mathbf{k},\mathbf{-k};\mathbf{-k},\mathbf{k})
	\Gamma_{s_3s_2;s_1s_4}(\mathbf{-k},\mathbf{k};\mathbf{k},\mathbf{-k})
	\Pi _{s_{1}s_{2}}^{+\bm{k}_F}\Pi _{s_{3}s_{4}}^{-\bm{k}_F},
\end{equation}
where $\Pi _{ss^{\prime }}^{\pm \mathbf{k}_{F}}$ given by Eq.~(\ref{eq:PiAngleDep}) depends on the direction of the electron momentum with respect to the magnetic field.

A general formula for $\delta\Xi_{xx}$ is very complicated; however, in the regime of strong Cooper renormalization, i.e. for $1\ll UL\ll 1/\delta^2$,  there are only a few partial amplitudes which survive the downward renormalization. Keeping only these partial amplitudes in $\Gamma$, we obtain for the thermodynamic potential
\begin{align}
	\notag \delta\Xi _{xx}= &- \frac{U^2}{16}\int_0^{2\pi}\frac{d\theta_\bm{k}}{2\pi}
	T\sum_{\Omega}\int\frac{qdq}{2\pi}[(\Pi_{-+}^{+\bm{k}_F}\Pi_{-+}^{-\bm{k}_F}
	+\Pi_{+-}^{+\bm{k}_F}\Pi_{+-}^{-\bm{k}_F}-2\Pi_0^2)+4\Pi_0^2]\\
	&- \frac{\Delta^2}{\alpha^2k_F^2}\frac{U^2}{8}
	T\sum_{\Omega}\int\frac{qdq}{2\pi}[\Pi_0(\Pi_{-+}+\Pi_{+-}-2\Pi_0)+\Pi_0^2],
\end{align}
where the angular dependence of the polarization bubbles in the second line was neglected because of an overall factor $\Delta^2$ originating from the scattering amplitudes, and the angular integral in those terms was readily performed. On the other hand, the field dependence in the first term is exclusively due to the bubbles, hence the angular integration has to be carried out last. The integrals over the momentum and frequency yield
\begin{equation}
	\delta\Xi _{xx}= - \frac{T^3}{8\pi v_F^2}\left(\frac{mU}{4\pi}\right)^2\bigg[
	\int_0^{2\pi}\frac{d\theta_\bm{k}}{2\pi}
	\mc{F}\left(\frac{\bar\Delta_{\bm{k}_F}+\bar\Delta_{-\bm{k}_F}}{T}\right)
	+ 2\frac{\Delta^2}{\alpha^2k_F^2}\mc{F}\left(\frac{|\alpha|k_F}{T}\right)\bigg].
\end{equation}
Expanding $\mc{F}(x)\approx x^3/3$ for $x\gg1$ and differentiating with respect to the field twice, we obtain for the nonanalytic part of the spin susceptibility
\begin{equation}\label{eq:chixxfull}
	\delta\chi _{xx}= \frac13\chi_0\left(\frac{mU}{4\pi}\right)^2\frac{|\alpha|k_F}{E_F}.
\end{equation}
Somewhat unexpectedly, the fully renormalized result (\ref{eq:chixxfull}) coincides with the leading (first) term in the second-order result (\ref{eq:ChiXX2ndAlpha}).
The formally subleading but $T$-dependent $T/2E_F$ term in Eq.~(\ref{eq:ChiXX2ndAlpha}) does not show up in the fully renormalized result, which implies that, at best, it is of order $T/UL\propto T/\ln T$ for large but finite $UL$. Hence follows the result for $\delta\chi_{xx}$ presented in the main text, Eq.~(\ref{eq:chixxfullmain}).

\end{widetext}

\end{document}